\documentclass{llncs}
\usepackage[table]{xcolor}    

\usepackage{multirow}

\def\tmp#1\CT@column@color\CT@row@color#2!#3#4{%
  \def\@classz{#3\@addtopreamble{#1\CT@row@color\CT@column@color#2}#4}}

\usepackage{eufrak}
\usepackage{mathtools}
\usepackage{xparse}
\usepackage[english]{babel}
\usepackage[utf8]{inputenc}
\usepackage{amsmath,amssymb} 
\usepackage{graphicx}
\usepackage{color}
\usepackage{wrapfig}
\usepackage[colorinlistoftodos]{todonotes}

\usepackage{array}

\usepackage{extpfeil}

\usepackage{program}
\usepackage{proof}
\usepackage[normalem]{ulem}
\usepackage[symbol]{footmisc}

\usepackage{url}

\usepackage{listings}

\lstdefinelanguage{custom}
{
morekeywords={public, void, name},
sensitive=false,
morecomment=[l]{//},
morecomment=[s]{/*}{*/},
morestring=[b]",
}

\newcommand\TRUE{\textsf{true}}
\newcommand\FALSE{\textsf{false}}

\newcommand\x{x}
\newcommand\y{y}

\newcommand{\act}{\alpha}

\newenvironment{itemize*}%
  {\begin{itemize}%
    \setlength{\itemsep}{0.0in}%
    \setlength{\topsep}{0.0in}%
    \setlength{\parskip}{0.0in}}%
  {\end{itemize}}
\newenvironment{enumerate*}%
  {\begin{enumerate}%
    \setlength{\itemsep}{0.0in}%
    \setlength{\topsep}{0.0in}%
    \setlength{\parskip}{0.0in}}%
  {\end{enumerate}}

\makeatletter
{\catcode`\/\active\catcode`\_\active\catcode`\.\active
\gdef\URLslash{\futurelet\next\@@URLslash}%
\gdef\@@URLslash{\ifx\next\URLslash\char`\/\else\slash\fi}%
\gdef\URLdot{\char`\.\penalty\exhyphenpenalty}%
\gdef\URLprepare{\catcode`\/\active\catcode`\_\active\catcode`\.\active
        \let/\URLslash\let.\URLdot\def~{\char`\~}\def_{\char`\_}}}%
\def\URL{\bgroup\URLprepare\realURL}%
\def\realURL#1{\tt #1\egroup}%

\definecolor{light-gray}{gray}{0.85}
\newcommand\red[1]{{\color{red} #1}}

\newcommand\ignore[1]{}

\newif \ifDRAFT
\DRAFTtrue
\ifDRAFT

\newcommand\removed[1]{{\textcolor{red}{\sout{#1}}}}
\newcommand\esays[1]{}
\newcommand\osays[1]{\todo[color=red!40]{{\bf O says:} #1}}
\newcommand\ksays[1]{\todo[color=orange!40]{{\bf K says:} #1}}
\else

\newcommand\removed[1]{}
\newcommand\esays[1]{}
\newcommand\osays[1]{}
\newcommand\ksays[1]{}
\fi

\newcommand\BLIND[1]{}

\newcommand\SHORT[1]{}

\newcommand\LLNCS[1]{}

\newcommand\figboxS[1]{\noindent{\ \\\fbox{\begin{minipage}{6.9cm}
#1
\end{minipage}
}}}

\newcommand\ie{{\it i.e.}}
\newcommand\eg{{\it e.g.}}
\newcommand\smartpar[1]{\smallskip\noindent {\bf #1}\hspace{0.1in}}

\newcommand\Tool{{\sc Servois}}

\newcommand\mv[1]{\textsf{val}_{#1}}
\newcommand\ms[1]{\textsf{sendr}_{#1}}
\newcommand\bn[1]{\textsf{bk}_{#1}}
\newcommand\rd[1]{\textsf{rnd}_{#1}}
\newcommand\mvA{\mv{1}}
\newcommand\msA{\ms{1}}
\newcommand\bnA{\bn{1}}
\newcommand\rdA{\rd{1}}
\newcommand\mvB{\mv{2}}
\newcommand\msB{\ms{2}}
\newcommand\bnB{\bn{2}}
\newcommand\rdB{\rd{2}}
\newcommand\mvN{\mv{}}
\newcommand\msN{\ms{}}
\newcommand\bnN{\bn{}}
\newcommand\rdN{\rd{}}

\newcommand\Lbowtie{\hat{\bowtie}}
\newcommand\Lrm{\hat{\triangleright}}
\newcommand\LSigma{\hat{\Sigma}}
\newcommand\Lsigma{\hat{\sigma}}

\newcommand\LGammaOf[1]{(\!|\!] #1 [\!|\!)}

\newcommand\dom[1]{\text{\bf dom}(#1)}
\newcommand\NLbowtie{\mathrlap{\backslash}\Lbowtie}

\newcommand\hatvarphi{\tilde{\varphi}}
\newcommand\Choose{{\sc Choose}}

\newcommand\arxivOnly[1]{#1}
\newcommand\tacasOnly[1]{}
\newif \ifARXIV
\ARXIVtrue

\begin{document}
\title{Automatic Generation of \\ Precise and Useful Commutativity Conditions}
\subtitle{\arxivOnly{(Extended Version)}}

\newcommand\MyThanks{Koskinen was previously supported in part by NSF CCF Award \#1421126 at New York University.}

\author{Kshitij Bansal \inst{1}\thanks{This work was partially supported by NSF award \#1228768. Author was at New York University when part of the work was completed.} \and Eric Koskinen \inst{2}\thanks{Support in part by NSF CCF Award \#1421126, and CCF Award \#1618542. Some of the research was done while author was at IBM Research.} \and Omer Tripp \inst{1}\thanks{Some of the research was done while author was at IBM Research.}}
\institute{Google, Inc.
  \and Stevens Institute of Technology}

\maketitle

\begin{abstract}
Reasoning about commutativity between data-structure operations is an important problem with applications including parallelizing compilers, optimistic parallelization and, more recently, Ethereum smart contracts. There have been research results on automatic generation of commutativity conditions, yet we are unaware of any fully automated technique to generate conditions that are both sound and effective.

We have designed such a technique, driven by an algorithm that iteratively refines a conservative approximation of the commutativity (and non-commutativity) condition for a pair of methods into an increasingly precise version. The algorithm terminates if/when the entire state space has been considered, and can be aborted at any time to obtain a partial yet sound commutativity condition.
We have generalized our work to left-/right-movers~\cite{lmrm} and proved relative completeness. 
We describe aspects of our technique that lead to \emph{useful} commutativity conditions, including how predicates are selected during refinement and heuristics that impact the output shape of the condition. \red{\tacasOnly{AE. arxiv. sp}}

We have implemented our technique in a prototype open-source tool \Tool. Our algorithm produces quantifier-free queries that are dispatched to a back-end SMT solver. We evaluate \Tool\ through two case studies: (i) We synthesize commutativity conditions for a range of data structures including Set, HashTable, Accumulator, Counter, and Stack. (ii) We consider an Ethereum smart contract called \textsf{BlockKing}, and show that \Tool\ can detect serious concurrency-related vulnerabilities and guide developers to construct robust and efficient implementations.

\bigskip
\arxivOnly{\emph{Note: This is an extended version of our paper, which
  appears in TACAS 2018~\cite{tacas18}.}}


\end{abstract}

\newcommand\Pre[1]{\text{\it Pre}_{#1}}
\newcommand\Post[1]{\text{\it Post}_{#1}} 

\section{Introduction}

Reasoning about the conditions under which data-structure operations commute is an important problem. The ability to derive sound yet effective commutativity conditions unlocks the potential of multicore architectures, including parallelizing compilers~\cite{rinard,DBLP:conf/oopsla/TrippYFS11}, speculative execution (\eg\ transactional memory~\cite{ppopp08}), peephole partial-order reduction~\cite{peephole}, futures, etc. Another important application domain that has emerged recently is Ethereum~\cite{ethereum} smart contracts: efficient  execution of such contracts hinges on exploiting their commutativity~\cite{Dickerson:2017:ACS:3087801.3087835}
and block-wise concurrency can lead to vulnerabilities~\cite{sergeyhobor}.
Intuitively, commutativity is an important property because linearizable data-structure operations that commute can
be executed concurrently: their effects do not interfere with
each other in an observable way. When using a linearizable HashTable, for example, knowledge that  \texttt{put(x,'a')} commutes with \texttt{get(y)} provided that
$\texttt{x}\neq \texttt{y}$ enables significant parallelization opportunities.  
Indeed, it’s important for the commutativity condition to be sufficiently granular so that parallelism can be exploited effectively~\cite{DBLP:journals/tocs/ClementsKZMK15}. At the same time, to make safe use of a commutativity condition, it must be sound~\cite{DBLP:conf/popl/KoskinenPH10,DBLP:conf/pldi/KoskinenP15}.
Achieving both of these goals using manual reasoning is burdensome and error prone.

In light of that, researchers have investigated ways of verifying user-provided commutativity conditions~\cite{KR:PLDI11} as well as synthesizing such conditions automatically, \eg\ based on random interpretation~\cite{aleen}, profiling
\cite{TrippJanus} or sampling \cite{DBLP:conf/cav/GehrDV15}. None of these approaches, however,
 meet the goal of computing a commutativity condition that is both \textit{sound} and \textit{granular} in a \textit{fully automated} manner.

 In this paper, we present a refinement-based technique for synthesizing commutativity conditions.  Our technique builds on
 well-known
 descriptions and representations of abstract data types (ADTs) in terms of logical
$(\Pre{m},\Post{m})$ 
specifications~\cite{Hoare02,modula,misra,Barnett,Meyer92,Leino06} 
for each method $m$. 
Our algorithm
iteratively relaxes under-approximations
of the commutativity \emph{and} non-commutativity conditions of methods $m$ and $n$, starting from {\sf false}, into 
increasingly precise versions. At each step, we conjunctively subdivide
the symbolic state space into regions, searching for areas where $m$ and $n$
commute and where they don’t. Counterexamples to both the positive side
and the negative side are used in the next symbolic subdivision.
Throughout this recursive process, we accumulate the
commutativity condition as a growing disjunction of these regions.
The output of our procedure is a logical formula
$\varphi_m^n$ which specifies when method $m$ commutes with method $n$.
We have proven that the algorithm is sound, and can also be aborted at any time to obtain a
partial, yet useful~\cite{TrippJanus,ppopp08}, commutativity condition. 
We show that, under certain conditions,
termination is guaranteed (relative completeness).

We address several challenges that arise in using an iterative refinement
  approach to generating precise and useful commutativity conditions.
First, we show how to pose the commutativity question
  in a way that does not introduce additional quantifiers.
We also show how to generate the predicate vocabulary for expressing the condition
$\varphi_m^n$, as well as how to choose the predicates throughout the refinement loop.
A further question that we address is how predicate selection impacts the conciseness and readability of the generated commutativity conditions.
Finally, we have generalized our algorithm to left-/right-movers~\cite{lmrm}, a more
precise version of commutativity.

We have implemented our approach as the \Tool\ tool, whose code and documentation are available online~\cite{servoishomepage}. \Tool\ is built on top of the CVC4 SMT solver~\cite{cvc4}.
We evaluate \Tool\ through two case studies. First, we generate commutativity conditions for a collection of popular data structures, including Set, HashTable, Accumulator, Counter, and Stack. The conditions 
typically combine multiple theories, such as sets, integers, arrays, etc. We show the conditions to be comparable in granularity to manually specified conditions~\cite{KR:PLDI11}. Second, we consider \textsf{BlockKing}~\cite{sergeyhobor}, an Ethereum smart contract, with its known vulnerability. We demonstrate how a developer can be guided by \Tool\ to create a more robust implementation.

\smartpar{Contributions.}
In summary, this paper makes the following contributions:
\begin{itemize}
\item The first sound and precise technique to
  automatically generate commutativity
  conditions (Sec.~\ref{sec:alg}).

\item Proof of
  soundness and relative completeness (Sec.~\ref{sec:alg}).

\item An implementation that takes an abstract
  code specification and automatically generates
  commutativity conditions using an SMT solver  
  (Sec.~\ref{sec:impl}). 

\item A novel technique for selecting refinement predicates that
  improves scalability and the simplicity of the generated formulae (Sec.~\ref{sec:impl}).

\item Demonstrated efficacy for several key data structures (Sec.~\ref{subsec:ds}) as well as the \textsf{BlockKing} Ethereum smart contract~\cite{sergeyhobor}.
  (Sec.~\ref{subsec:blockking}).
\end{itemize}

\noindent
\arxivOnly{This is an extended version of our paper~\cite{tacas18}.}
\tacasOnly{An extended version of this paper can be found in \red{arXiv}~\cite{arxiv}.}


\smartpar{Related work.}
The closest to our contribution in this paper is a recent technique by Gehr \emph{et al.}~\cite{DBLP:conf/cav/GehrDV15} for learning, or inference, of commutativity conditions based on black-box sampling. They draw concrete arguments, extract relevant predicates from the sampled set of examples, and then search for a formula over the predicates. There are no soundness or completeness guarantees.

Both Aleen and Clark~\cite{aleen} and Tripp \emph{et al.}~\cite{TrippJanus} identify sequences of actions that commute (via random interpretation and dynamic analysis, respectively). However, neither technique yields an explicit commutativity condition.
Kulkarni et al. \cite{KNPSP:PLDI11} point out that 
varying degrees of commutativity specification precision are useful.
Kim and Rinard \cite{KR:PLDI11} use Jahob to
verify manually specified commutativity conditions 
of several different linked data structures.
Commutativity specifications are also
found in dynamic analysis techniques~\cite{DBLP:conf/pldi/DimitrovRVK14}.
More distantly related is work on
synthesis of programs~\cite{DBLP:conf/pldi/Solar-LezamaJB08}
and of synchronization~\cite{DBLP:conf/popl/VechevYY10,DBLP:conf/pldi/VechevY08}.


\newcommand\preds{{\cal P}}
\newcommand\Set{\textsf{Set}}
\newcommand\Contents{S}
\newcommand\prop{\varphi}
\newcommand\hatprop{\hatvarphi}
\newcommand\cexC{\chi_\text{c}}
\newcommand\cexNC{\chi_\text{nc}}
\newcommand\trieq{\stackrel{\triangle}{=}}

\section{Example}\label{sec:overview}

Specifying commutativity conditions is generally nontrivial, more
importantly it is easy to miss subtle corner cases.  Additionally, it
has to be done pairwise for all methods.
For ease of illustration, we will focus on the relatively simple
\Set\ ADT, whose state consists of a single set
$\Contents$ that stores an unordered collection of unique elements.
Let us consider one pair of operations:
%
	(i) {\tt contains($\x$)/bool}, a side-effect-free check whether the element $\x$ is in $\Contents$; and
	(ii) {\tt add($\y$)/bool} adds $\y$ to $\Contents$ if it is not already there and returns \TRUE, or otherwise returns \FALSE.
%
{\tt add} and {\tt contains} clearly commute if they refer to different elements in the set. There is another case that is less obvious:
{\tt add} and {\tt contains} commute if they refer to the same element $e$, as long as in the pre-state $e \in \Contents$. In this case, under both orders of execution, {\tt add} and {\tt contains} leave the set unmodified and return \FALSE\ and \TRUE, respectively.
The algorithm we describe in this paper takes 3.6s to automatically produce a
precise logical formula $\prop$ that captures this commutativity condition, \ie\ the disjunction of the two cases above:
$\prop \equiv \x\neq\y \vee (\x = \y \wedge \x
\in \Contents)$.
The algorithm also generates the conditions under which the methods 
\emph{do not} commute: $\hatprop \equiv\x=\y \wedge x\notin \Contents$.
These are precise, since $\prop$ is the negation of $\hatprop$.

A more complicated commutativity condition generated by our tool,
\Tool, for \textsf{BlockKing}
(Sec.~\ref{subsec:blockking}) is for method
\textsf{enter($\mvA$,$\msA$,$\bnA$...)} and completed in
1.4s. It does not commute with itself
\textsf{enter($\mvB$,$\msB$,$\bnB$...)} \emph{iff}:
\[\begin{array}{c}
\bigvee \left\{
  \begin{array}{l}
    \mvA \geq 50 \wedge \mvB \geq 50 \wedge \msA \neq \msB\\
    \mvA \geq 50 \wedge \mvB \geq 50 \wedge \msA = \msB \wedge \mvA \neq \mvB\\
    \mvA \geq 50 \wedge \mvB \geq 50 \wedge \msA = \msB \wedge \mvA = \mvB \wedge \bnA \neq \bnB\\
  \end{array} \right.
\end{array}
    \]
This disjunction enumerates the non-commutativity cases and, as discussed in Sec.~\ref{subsec:blockking}, directly identifies a vulnerability.

Capturing precise conditions such as these by hand, and doing so for
many pairs of operations, is tedious and error prone. This paper
instead presents a way to automate this.
Our algorithm recursively subdivides the state space via predicates
until, at the base case, regions are found that are either entirely
commutative or else entirely non-commutative.
Returning to our \Set\ example, the conditions we incrementally
generate are denoted $\prop$ and $\hatprop$, respectively.  The
following diagram illustrates how our algorithm proceeds to generate
the commutativity conditions for {\tt add} and {\tt contains}.
\begin{center}
		\includegraphics[width=4.8in]{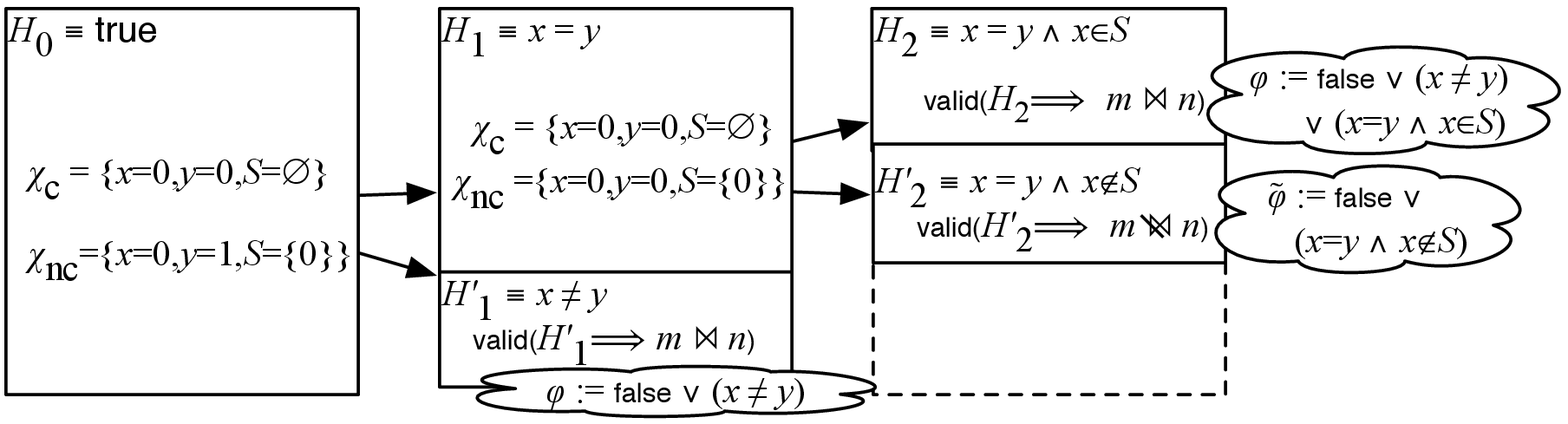}
\end{center}
In this diagram, each subsequent panel depicts a partitioning of the
state space into regions of commutativity ($\prop$) or non-commutativity
($\hatprop$). The counterexamples $\cexC,\cexNC$ give values for the
arguments $x$, $y$ and the current state $\Contents$.

We denote by $H$ the logical formula that describes the
current state space at a given recursive call. 
We begin with $H_0=\TRUE$, $\prop=\FALSE$, and $\hatprop=\FALSE$. 
There are three cases for a given $H$:
({\it i}) $H$ describes a precondition for $m$ and $n$
in which they \emph{always} commute;
({\it ii}) $H$ describes a precondition for $m$ and $n$ 
in which they \emph{never} commute; or
({\it iii}) neither of the above.
The latter case drives the algorithm to subdivide the region
by choosing a new predicate. 

We now detail the run of this refinement loop on our earlier
\Set\ example.  We elaborate on the other 
challenges that arise in later sections.
At each step of the algorithm, we determine which case we are in
via carefully designed validity queries to an SMT solver (Sec.~\ref{sec:noAlt}). For $H_0$,
  it returns the
  commutativity counterexample: $\cexC = \{ \x=0,\y=0,S=\emptyset \}$ as well as the non-commutativity counterexample $\cexNC = \{ \x=0,\y=1,S=\{0\} \}$.
Since, therefore, $H_0=\TRUE$ is neither a commutativity nor a non-commutativity
condition, we must refine $H_0$ into regions (or stronger conditions). In
particular, we would like to perform a \emph{useful} subdivision:
Divide $H_0$ into an $H_1$ that allows $\cexC$ but disallows $\cexNC$,
and an $H'_1$ that allows $\cexNC$ but not $\cexC$. 
So we must choose a predicate $p$ 
(from a suitable set of predicates $\preds$, discussed later),
such that $H_0 \wedge p \Rightarrow \cexC$ while $H_0 \wedge \neg p \Rightarrow \cexNC$ (or vice versa).
The predicate $\x=\y$ satisfies this property.
The algorithm then makes the next two recursive calls, adding $p$ as a
conjunct to $H$, as  shown in the
second column of the diagram above: one with 
$H_1 \equiv \TRUE \wedge \x=\y$ and
one with
$H'_1 \equiv \TRUE \wedge \x\neq\y$.
Taking the $H'_1$ case, our algorithm makes another SMT query and finds
that $\x\neq \y$ implies that {\tt add} always commutes with {\tt
  contains}. At this point, it can update the commutativity condition
$\prop$, letting $\prop := \prop \vee H'_1$, adding this $H'_1$ region
to the growing disjunction.
On the other hand, $H_1$ is neither a sufficient commutativity nor
a sufficient non-commutativity condition, and so our algorithm, again, produces the
respective counterexamples:
$\cexC = \{ \x=0,\y=0,\Contents=\emptyset \}$ and
$\cexNC = \{ \x=0,\y=0,\Contents=\{0\} \}$.
In this case, our algorithm selects the predicate $\x\in \Contents$,
and makes two further recursive calls: one with 
$H_2 \equiv \x=\y \wedge \x\in\Contents$ and
another with
$H'_2 \equiv \x=\y \wedge \x\notin\Contents$.
In this case, it finds that $H_2$ is a sufficiently strong precondition for
commutativity, while $H'_2$  is a strong enough precondition for
non-commutativity. Consequently, $H_2$ is added as a new conjunct to
$\prop$, yielding
$\prop \equiv \x\neq y \vee (\x=\y \wedge \x\in\Contents)$. Similarly, 
$\hatprop$ is updated to be:
$\hatprop \equiv (\x=\y \wedge \x\notin\Contents)$.
No further recursive calls are made so the algorithm terminates and
we have obtained a precise
(complete) commutativity/non-commutativity specification:
$\prop \vee \hatprop$ is valid (Lem.~\ref{lemma:term}).

\smartpar{Challenges \& outline.}
While the algorithm outlined so far is a relatively
standard refinement, the above generated conditions were not immediate.
We now discuss challenges involved in
generating sound \emph{and} useful conditions.

(Sec.~\ref{sec:noAlt}) A first question is how to pose the
underlying commutativity queries for each subsequent $H$ in
a way that avoids the introduction of additional quantifiers, so that
we can remain in fragments for which the solver has complete decision
procedures.
Thus, if the data structure can be encoded using theories that are
decidable, then the queries we pose to the SMT solver are guaranteed
to be decidable as well. $\Pre{m}/\Post{m}$ specifications that are
partial would introduce quantifier alternation, but we show how this
can be avoided by, instead, transforming them into total
specifications.

(Sec.~\ref{sec:alg})
  We have proved that our algorithm is sound even if aborted or 
the ADT description involves undecidable theories. We further show that
termination implies
completeness, and specify broad conditions that imply termination.

(Sec.~\ref{sec:impl})
Another challenge is to prioritize predicates during the refinement
loop. This choice impacts not only the algorithm's performance, but
also the quality/conciseness of the resulting conditions. Our
choice of next predicate $p$ is governed by two requirements. First,
for progress, $p$/$\neg p$ must eliminate the counterexamples to
commutativity/non-commutativity due to the last iteration
. This may still leave multiple choices, and we
propose two heuristics -- called \emph{simple} and \emph{poke}---with
different trade-offs to break ties. 

(Sec.~\ref{sec:eval}) We conclude with an evaluation on a range of popular data structures 
and a case study on boosting the security of an
Ethereum smart contract.


\newcommand\sem[1]{[\![ #1 ]\!]}
\newcommand\opeq{\preccurlyeq} 
\newcommand\methA{m}
\newcommand\methB{n}
\newcommand\TT{\mathfrak{T}}
\newcommand\LTT{\hat{\mathfrak{T}}}

\newcommand\sigmat[1]{\sigma_{#1}}
\newcommand\sigmatt[2]{\sigma^{#1}_{#2}}
\newcommand\sigmaM{\sigmat{m}}
\newcommand\sigmaMN{\sigmatt{m}{n}}
\newcommand\sigmaN{\sigmat{n}}
\newcommand\sigmaNM{\sigmatt{n}{m}}
\newcommand\Xs{\bar{x}}
\newcommand\Ys{\bar{y}}
\newcommand\Rms{\bar{r}}
\newcommand\Rns{\bar{s}}
\newcommand\As{\bar{a}}
\newcommand\Bs{\bar{b}}
\newcommand\Us{\bar{u}}
\newcommand\Vs{\bar{v}}

\newcommand\GammaOf[1]{(\!| #1 |\!)}

\section{Preliminaries}
\label{sec:qf}
\vspace{-10pt}
\smartpar{States, actions, methods.}
We will work with a state space $\Sigma$, with decidable
equality and a set of \emph{actions} $A$. For each $\act\in A$, we have a
transition function
$\GammaOf{\act} : \Sigma \rightharpoondown \Sigma$.
We denote a single transition as $\sigma\xrightarrow{\act}\sigma'$.
We assume that each such action arc completes in finite time.
Let $\TT\equiv(\Sigma,A,\GammaOf{\bullet})$.
%
  We say that two \emph{actions} $\act_1$ and $\act_2$ \emph{commute}~\cite{DBLP:conf/pldi/DimitrovRVK14}, denoted
  $\act_1 \bowtie \act_2$, provided that
  $\GammaOf{\act_1} \circ \GammaOf{\act_2} = \GammaOf{\act_2} \circ
  \GammaOf{\act_1}$.
%
Note that $\bowtie$ is with respect to
$\TT=(\Sigma,A,\GammaOf{\bullet})$.
Our formalism, implementation, and evaluation all extend to a more
fine-grained notion of commutativity: an asymmetric version called
left-movers and right-movers~\cite{lmrm}, where a method commutes in
one direction and not the other.
\tacasOnly{Details can be found in~\cite{arxiv}.}
\arxivOnly{We return to this in Sec.~\ref{sec:lmrm}.}
Also, in our evaluation (Sec.~\ref{sec:eval}) we
show left-/right-mover conditions that were generated by our
implementation.


%
An action $\act \in A$ is of the form
$m(\Xs)/\Rms$, where $m$, $\Xs$ and $\Rms$ are called a \emph{method},
\emph{arguments} and \emph{return values} respectively. As a convention, for 
actions corresponding to a method $n$, we use $\Ys$ for
arguments and $\Rns$ for return values.
The set of methods will be finite,
inducing a finite partitioning of $A$. We refer to an
action, say $m(\As)/\Vs$, as \emph{corresponding} to method $m$
(where $\As$ and $\Vs$ are vectors of values). The
set of actions corresponding to a method $m$, denoted $A_m$,
might be infinite as arguments and return values may be from an
infinite domain. 
\begin{definition}\label{def:Mcommute} Methods $m$ and $n$ \emph{commute}, denoted $m\ \bowtie\ n$ provided that
$ \forall \Xs\ \Ys\ \Rms\ \Rns.\;\; m(\Xs)/\Rms \bowtie n(\Ys)/\Rns$.
\end{definition}
The quantification $\forall \Xs \Rms$ above means 
$\forall m(\Xs)/\Rms \in A_m$, i.e., all
vectors of arguments and return values that constitute an action in $A_m$.


\smartpar{Abstract specifications.}
We symbolically describe the actions of a method $m$
as pre-condition $\Pre{m}$ and post-condition
$\Post{m}$. Pre-conditions are logical
formulae over method arguments and the initial state:
$\sem{\Pre{m}} : \Xs \rightarrow \Sigma \rightarrow \mathbb{B}$.
Post-conditions are over method arguments, and return
values, initial state and final state:
$\sem{\Post{m}} : \Xs \rightarrow \Rms \rightarrow \Sigma
                       \rightarrow \Sigma \rightarrow \mathbb{B}$.
%
Given $(\Pre{m},\Post{m})$ for every method $m$,
we define a transition system $\TT=(\Sigma,A,\GammaOf{\bullet})$ such that
$\sigma \xrightarrow{m(\As)/\Vs} \sigma'$ \emph{iff}
$\sem{\Pre{m}}\ \As\ \sigma$ and
$\sem{\Post{m}}\ \As\ \Vs\ \sigma\ \sigma'$.

Since our approach works on deterministic transition systems, we have
implemented an SMT-based check (Sec.~\ref{sec:eval}) that ensures the input
transition system is deterministic.  Deterministic specifications were
sufficient in our examples.
This is unsurprising given the inherent difficulty of creating efficient concurrent implementations of nondeterministic operations, whose effects are hard to characterize.
Reducing
nondeterministic data-structure methods to deterministic ones through
symbolic partial
determinization~\cite{AbadiLamport,CookKoskinenPOPL11} is left as future work.

\smartpar{Logical commutativity formulae.}
We will generate a commutativity condition for methods
$m$ and $n$ as logical
formulae over initial states and the arguments/return values of the methods. We denote a logical commutativity formula as $\varphi$ and assume a decidable
interpretation of formulae:
$\sem{\varphi} : (\sigma,\Xs,\Ys,\Rms,\Rns) \rightarrow \mathbb{B}$.
%
(We tuple the arguments for brevity.)
The first argument is the initial state.  Commutativity \emph{post}-
and \emph{mid}-conditions can also be written~\cite{KR:PLDI11} but
here, for simplicity, we focus on commutativity \emph{pre}-conditions.
We may write $\sem{\varphi}$ as 
$\varphi$ when it is clear from context that $\varphi$ is meant to be interpreted.

We say that $\varphi_m^n$ is a \emph{sound commutativity condition}, and
$\hat{\varphi}_m^n$ a sound \emph{non}-commutativity condition resp., for $m$ and $n$
provided that
\[\begin{array}{l}
\forall \sigma \Xs \Ys \Rms \Rns.\
\sem{\varphi_m^n}\ \sigma\ \Xs\ \Ys\ \Rms\ \Rns
\Rightarrow m(\Xs)/\Rms\ \bowtie\ n(\Ys)/\Rns, \text{ and}\\
\forall \sigma \Xs \Ys \Rms \Rns.\
\sem{\hat{\varphi}_m^n}\ \sigma\ \Xs\ \Ys\ \Rms\ \Rns 
\Rightarrow \neg(m(\Xs)/\Rms\ \bowtie\ n(\Ys)/\Rns), \text{ resp.}
\end{array}\]


\newcommand\Err{\textsf{err}}

\section{Commutativity without quantifier alternation}
\label{sec:noAlt}
Def.~\ref{def:Mcommute} requires showing equivalence between
different compositions of potentially partial functions. That
is,
$
\GammaOf{\act_1} \circ \GammaOf{\act_2} = 
\GammaOf{\act_2} \circ \GammaOf{\act_1}
$ if and only if:
\[\begin{array}{c}
  \forall \sigma_0\ \sigma_1\ \sigma_{12}.\ 
     \GammaOf{\act_1}\sigma_0 = \sigma_1 \wedge
     \GammaOf{\act_2}\sigma_1 = \sigma_{12}
  \;\; \Rightarrow\;\;
     \exists \sigma_3.\ \GammaOf{\act_2}\sigma_0 = \sigma_3
        \wedge \GammaOf{\act_1}\sigma_3 = \sigma_{12}\\
\text{(\emph{and a symmetric  case for the other direction})}
\end{array}\]
Even when the transition relation can be expressed in a decidable
theory, because of $\forall \exists$ quantifier alternation in the
above encoding (which is undecidable in general), any procedure
requiring such a check would be incomplete. SMT solvers are
particularly poor at handling such constraints.


We observe that when the transition system is specified as
  $Pre_m$ and $Post_m$ conditions, and the $Post_m$ condition is
  \emph{consistent} with $Pre_m$, then it is possible to avoid
  quantifier alternation. By consistent we mean that whenever $Pre_m$
  holds, there is always some state and return value for which $Post_m$
  holds. 
\[\forall \As\ \sigma.\;\;\; \Pre{m}(\As,\sigma) = \TRUE \;\Rightarrow\;
\exists \sigma'\ \Rms.\ \Post{m}(\As,\Rms,\sigma,\sigma').\]
This assumption holds for all of the specifications in
  the examples we considered (Sec.~\ref{sec:eval}).
This allows us to perform a simple transformation
on transition
systems to a lifted domain, and enforce a definition of commutativity in the
lifted domain $m\ \Lbowtie\ n$ that is equivalent to
Def.~\ref{def:Mcommute}. This new definition requires only
\emph{universal} quantification, and as such, is better suited to
SMT-backed algorithms (Sec.~\ref{sec:alg}).


\begin{definition}[Lifted transition function]\label{def:lifted}
For $\TT=(\Sigma,A,\GammaOf{\bullet})$, we lift $\TT$ to
$\LTT=(\LSigma,A,\LGammaOf{\bullet})$
where $\LSigma = \Sigma \cup \{ \Err{} \}$, $\Err{} \notin \Sigma$, and 
$\LGammaOf{\act} : \LSigma \rightarrow \LSigma$, as:
$$
   \LGammaOf{\act} \hat{\sigma} \equiv
      \begin{cases}
         \Err{} & \text{if } \hat{\sigma} = \Err{} \red{remove?}\\
         \GammaOf{\act} \Lsigma & \text{if }  \Lsigma \in
         \dom{\GammaOf{\act}}\\
         \Err{} & \text{otherwise}
       \end{cases}
$$
\end{definition}

\noindent
Intuitively, $\LGammaOf{\act}$ wraps $\GammaOf{\act}$ so that 
$\Err{}$ loops back to $\Err{}$, and the (potentially partial)
$\GammaOf{\act}$ is made to be total by mapping elements to $\Err{}$
when they are undefined in $\GammaOf{\act}$.
It is not necessary to lift the actions (or, indeed,
  the methods), but only the states and transition function.
Once lifted, for a given state $\Lsigma_0$, the question of
\emph{some} successor state becomes equivalent to
\emph{all} successor states because there is exactly one successor
state.


\newcommand\LPre[1]{\widehat{\text{\it Pre}}_{#1}}
\newcommand\LPost[1]{\widehat{\text{\it Post}}_{#1}} 
\newcommand\err{\textsf{err}}
\newcommand\nerr{\textsf{nerr}}

\smartpar{Abstraction.} Pre-/post-conditions
$(\Pre{m},\Post{m})$ 
are suitable for specifications of potentially partial transition
systems. One can translate these into a new pair
$(\LPre{m},\LPost{m})$ that induces a corresponding lifted transition
system that is total and remains deterministic.
These lifted specifications have types over
lifted state spaces:
$   \sem{\LPre{m}} : \Xs \rightarrow \LSigma
\rightarrow \mathbb{B}$
and
$ \sem{\LPost{m}} :
    \Xs \rightarrow \Rms \rightarrow \LSigma \rightarrow \LSigma
    \rightarrow \mathbb{B}$.
    Our implementation performs this lifting via translation
    denoted {\sc Lift} from 
$(\Pre{m},\Post{m})$ to:
\[\begin{array}{rl}
   \LPre{m}(\Xs,\Lsigma) &\equiv\;\;\TRUE\\
   \LPost{m}(\Xs,\Rms,\Lsigma,\Lsigma') &\equiv
    \;\bigvee \begin{cases}
       \Lsigma=\Err{} \wedge \Lsigma'=\Err{}\\
       \Lsigma \neq \Err{} \wedge\Pre{m}(\Xs,\Lsigma) \wedge
                \Lsigma'\neq \Err{} \wedge \Post{m}(\Xs,\Rms,\Lsigma,\Lsigma')\\
       \Lsigma \neq \Err{} \wedge \neg \Pre{m}(\Xs,\Lsigma) \wedge \Lsigma'=\Err{}
    \end{cases}
\end{array}\]
(We abuse notation, giving $\Lsigma$ as an argument to $\Pre{m}$,
etc.) It is easy
to see that the lifted transition system induced by this translation
($\LSigma,\LGammaOf{\bullet}$) is of the form given in Def.~\ref{def:lifted}.
\tacasOnly{In~\cite{arxiv}, }
\arxivOnly{In Apx.~\ref{yml:counterauto}, }
we show how our tool transforms a counter
specification into an equivalent
lifted version that is total.
%


We use the notation $\Lbowtie$ to mean $\bowtie$ but over lifted
transition system $\LTT$. Since $\Lbowtie$ is over
total, determinsitic transition functions, 
$ \act_1\ \Lbowtie\ \act_2 $ is equivalent to:
\begin{equation}
    \forall \Lsigma_0.\; \Lsigma_0 \neq \Err{} \;\Rightarrow\;
       \LGammaOf{\act_2}\ \LGammaOf{\act_1}\ \Lsigma_0 =
       \LGammaOf{\act_1}\ \LGammaOf{\act_2}\ \Lsigma_0 \label{eqn:liftedcommu}
\end{equation}
The equivalence above is in terms of state equality. Importantly, this is a
universally quantified formula that translates to a ground
satisfiability check in an SMT solver (modulo the theories used to model the data structure).
In our refinement algorithm (Sec.~\ref{sec:alg}), we will use this format to check
whether candidate logical formulae describe commutative subregions.

\begin{lemma}\label{lemma:foo}
  $m\ \bowtie\ n \text{ if and only if } m\ \Lbowtie\ n$.
\ifARXIV
\begin{proof} Follows from classical reasoning, functional
  extensionality and case analysis on
  totality-vs-partiality.
\end{proof}
\else
(\emph{All proofs in~\cite{arxiv}.})
\fi
\end{lemma}


\section{Right-/Left-movers}
\label{sec:lmrm}

We now describe how the formalism
presented thus far can be extend to a more fine-grained notion of
commutativity: an asymmetric version called left-movers and
right-movers~\cite{lmrm}, where a method commutes in one direction and
not the other.

\begin{definition}[Action right-mover~\cite{lmrm}]
  We say that an action $\act_1$ \emph{moves to the right of} action $\act_2$ commute, denoted
  $\act_1 \triangleright \act_2$, provided that
  $\GammaOf{\act_2} \circ \GammaOf{\act_1} \subseteq \GammaOf{\act_1} \circ
  \GammaOf{\act_2}$.
\end{definition}

\noindent
Note that left-movers can be defined as right-movers, but with
arguments swapped.

\begin{definition}[Method right-mover]
For $m$ and $n$, 
$$
m\ \triangleright\ n \;\;\equiv\;\;
 \forall \Xs\ \Ys\ \Rms\ \Rns.\;\; m(\Xs)/\Rms \triangleright n(\Ys)/\Rns
$$
\end{definition}

\newcommand\propRM{\vec{\Psi}_m^n}

A \emph{logical right-mover condition} denoted $\propRM$ has the
same type as a commutativity condition and, again $\sem{\propRM}$
denotes interpretations of $\propRM$. Moreover, we say that $\propRM$
is a right-mover condition for $m$ and $n$ provided that
$\forall \sigma_0\ \Xs\ \Ys\ \Rms\ \Rns.\
\sem{\propRM}\ \sigma_0\ (m(\Xs)/\Rms)\ (n(\Ys)/\Rns) = \TRUE
\Rightarrow m\ \triangleright\ n
$ and similar for a \emph{non}-right-mover condition.

\smartpar{Checking whether $H_m^n \Rightarrow m\ \Lrm\ n$.}
After performing the lifting transformation, 
we again are able to reduce the question of whether a
formula $H_m^n$ is a right-mover condition to a validity check that
does not introduce quantifier alternation.
\[ \begin{array}{l}
\textsf{valid}\\
\;\left(
 \begin{array}{l}
  \forall \Lsigma_0\ \Xs\ \Ys\ \Rms\ \Rns.\\
   \;\; \varphi_m^n(\Lsigma_0,\Xs,\Ys,\Rms,\Rns) \;\Rightarrow\\
   \;\; \Lsigma_0 \neq \Err{} \;\Rightarrow\\
   \;\; \LGammaOf{n(\Ys)/\Rns}\ \LGammaOf{m(\Xs)/\Rms}\ \Lsigma_0 \neq \Err{} \;\Rightarrow\\
   \;\; \LGammaOf{n(\Ys)/\Rns}\ \LGammaOf{m(\Xs)/\Rms}\ \Lsigma_0 =
        \LGammaOf{m(\Xs)/\Rms}\ \LGammaOf{n(\Ys)/\Rns}\ \Lsigma_0.
\end{array} \right) \end{array} \]
Notice that this is a generalization of the validity check for 
commutativity.


\newcommand\Halg{H} 
\section{Iterative refinement}
\label{sec:alg}

We now present an iterative
refinement strategy that, when given a
lifted abstract transition system, generates the commutativity and the non-commutativity
conditions. We then discuss
soundness and relative completeness and, in Secs.~\ref{sec:impl} and~\ref{sec:eval},
challenges in generating precise \emph{and}
useful commutativity conditions.

The refinement algorithm symbolically searches
the state space for regions where the operations commute (or do not
commute) in a conjunctive manner, adding on one predicate at a
time. We add each subregion $H$ (described conjunctively) in which
commutativity always holds to a growing disjunctive description of the
commutativity condition $\prop$, and each subregion $H$ in which commutativity
never holds to a growing disjunctive description of the
non-commutativity condition $\hatprop$.

\ifARXIV
\begin{figure}
\else
\begin{wrapfigure}[21]{r}{7.1cm}
\fi
  \centering
\figboxS{\begin{program}[style=sf,number=true]
{\sc Re}\tab $\text{\sc fine}^m_n(\Halg, \preds)$ \{\\
   if\tab\ valid($\Halg \;\Rightarrow\; m\ \Lbowtie\ n$) then\label{ln:valid1}\\
      $\varphi$ := $\varphi \vee \Halg$;\untab\\
   el\tab se if valid($\Halg \;\Rightarrow\; m\ \NLbowtie\ n$) then\label{ln:valid2}\\
      $\hatvarphi$ := $\hatvarphi \vee \Halg$;\untab\\
  el\tab se\\
     let $(\cexC,\cexNC)$ = counterexs. to $\Lbowtie$ and $\NLbowtie$ \\
     let\tab\ $p$ = \Choose($\Halg,\preds,\cexC,\cexNC$) in \label{ln:choose}\\
        {\sc Refine}$^m_n$($\Halg\wedge p$, $\preds\setminus\{p\}$);\\
        {\sc Refine}$^m_n$($\Halg\wedge \neg p$, $\preds\setminus\{p\}$);\untab \untab \untab\\
\}\\
main \{\; \tab $\varphi$ := $\FALSE$;\;\; $\hatvarphi$ := $\FALSE$;\\
   try \{ $\text{\sc Refine}^m_n(\TRUE, \preds)$; \}\\
   catch (InterruptedExn e) \{ skip; \} \\ 
   return($\varphi,\hatvarphi$);  \} \untab\\
\end{program}}
\caption{\label{fig:alg} Algorithm for generating 
  commutativity $\varphi$ and non-commutativity
  $\hatvarphi$.} 
\ifARXIV
\end{figure}
\else
\end{wrapfigure}
\fi
The algorithm in Fig.~\ref{fig:alg} begins by setting $\varphi=\FALSE$ and
$\hatvarphi=\FALSE$. 
{\sc Refine} begins a symbolic binary search through the state space
$H$, starting from the entire state: $H=\TRUE$.
It also may use a collection of predicates $\preds$ (discussed later).
At each iteration, {\sc Refine} checks whether the current $H$ represents a
region of space for which $m$ and $n$ always commute:
$H \Rightarrow m\ \Lbowtie\ n$ (described below). If so, $H$ can be
disjunctively added to $\varphi$.
It may, instead be the case that  $H$ represents a
region of space for which $m$ and $n$ never commute:
$H \Rightarrow m\ \NLbowtie\ n$. If so, $H$ can be
disjunctively added to $\hatvarphi$.
If neither of these cases hold, we have two counterexamples.
$\cexC$ is the counterexample to commutativity, returned if the validity check on
Line~\ref{ln:valid1} fails.
$\cexNC$ is the counterexample to \emph{non}-commutativity, returned
if the validity check on Line~\ref{ln:valid2} fails.

We now need to subdivide $H$ into two regions. This is accomplished by
selecting a new predicate $p$ via the \Choose\  method. For now,
let the method \Choose\  and the choice of predicate vocabulary $\preds$
be parametric. {\sc Refine} is sound regardless of
the behavior of \Choose. Below we give
the conditions on \Choose\  that ensure relative completeness,
and in Sec.~\ref{sec:eval} we discuss our particular strategy.
Regardless of what $p$ is returned by \Choose, two recursive
calls are made to {\sc Refine}, one with argument $H \wedge p$, and
the other with argument $H\wedge\neg p$. The algorithm is 
exponential in the number of predicates. In Sec.~\ref{sec:impl}
we discuss prioritizing predicates.

The refinement
algorithm generates commutativity conditions 
in disjunctive normal form.
Hence, any finite
logical formula can be represented.
This logical language is more expressive than previous commutativity logics that,
because they were designed for run-time purposes, 
were restricted to conjunctions of inequalities~\cite{KNPSP:PLDI11}
and boolean combinations of predicates over finite domains~\cite{DBLP:conf/pldi/DimitrovRVK14}.

\smartpar{Checking a candidate $H_m^n$.}
Our algorithm involves checking whether $(H_m^n \Rightarrow
m\ \Lbowtie\ n)$ or $(H_m^n \Rightarrow m\ \NLbowtie\ n)$.
As shown in Sec.~\ref{sec:noAlt},
we can check whether $H_m^n$ specifies
conditions under which $m\ \bowtie\ n$ via an SMT query that does not
introduce quantifier alternation.  For brevity, we define:
\[ \begin{array}{l}\textsf{valid}(H_m^n \;\Rightarrow\;m\
     \Lbowtie\ n) \;\equiv\;
\textsf{valid}\!\!\left(
 \begin{array}{l}
  \forall \Lsigma_0\ \Xs\ \Ys\ \Rms\ \Rns.\;\;\;
        H_m^n(\Lsigma_0,\Xs,\Ys,\Rms,\Rns) \;\Rightarrow\; \\
   \;\;\;\;\;
      m(\Xs)/\Rms\;\;n(\Ys)/\Rns\;\;\Lsigma_0 = 
      n(\Ys)/\Rns\;\;m(\Xs)/\Rms\;\;\Lsigma_0 
 \end{array} \right) \end{array} \]
Above we assume as a black box an SMT solver
providing \textsf{valid}. 
Here we have lifted the universal quantification within $\Lbowtie$
outside the implication.

We can similarly check whether $H_m^n$ is a condition under which $m$
and $n$ \emph{do not} commute. First, we define negative analogs
of commutativity:
\[\begin{array}{rl}
 \act_1\ \NLbowtie\ \act_2 \;\equiv&
    \forall \Lsigma_0.\; \Lsigma_0 \neq \Err{} \;\Rightarrow\;
        \LGammaOf{\act_2}\ \LGammaOf{\act_1}\ \Lsigma_0 \neq
        \LGammaOf{\act_1}\ \LGammaOf{\act_2}\ \Lsigma_0\\
m\ \NLbowtie\ n \;\equiv&
    \forall \Xs\ \Ys\ \Rms\ \Rns.\;
     m(\Xs)/\Rms\ \NLbowtie\ n(\Ys)/\Rns 
\end{array}\]
We thus define a check for when
$\varphi_m^n$ is a \emph{non}-commutativity condition with:
\[ \begin{array}{l}\textsf{valid}(H_m^n \;\Rightarrow\;m\
     \NLbowtie\ n) \;\equiv\;
\textsf{valid}\!\!\left(
 \begin{array}{l}
  \forall \Lsigma_0\ \Xs\ \Ys\ \Rms\ \Rns. 
   \; H_m^n(\Lsigma_0,\Xs,\Ys,\Rms,\Rns) \;\Rightarrow\; \Lsigma_0 \neq \Err{} \;\Rightarrow\;\\
   \;\;\;\;\; 
      m(\Xs)/\Rms\;\;n(\Ys)/\Rns\;\;\Lsigma_0 \neq 
      n(\Ys)/\Rns\;\;m(\Xs)/\Rms\;\;\Lsigma_0 
 \end{array} \right) \end{array} \]





\begin{theorem}[Soundness]\label{thm:sound}
For each $\text{\sc Refine}^m_n$ iteration:
$\varphi \Rightarrow m\ \Lbowtie\ n$, and
$\hatvarphi \Rightarrow m\ \NLbowtie\ n$.
\ifARXIV
\smartpar{Theorem~\ref{thm:sound}.}
\begin{proof}\normalfont
  By induction. Initially, $\FALSE$ is a suitable condition
  for when commutativity holds.  $\FALSE$ is also a suitable condition
  under which commutativity does not hold. At each iteration,
  $\varphi$ or $\hatvarphi$ may be updated (not both, but for
  soundness this does not matter). Consider $\varphi$.  It must also
  be the case that $(\varphi\vee H)\Rightarrow m\ \Lbowtie\ n$ because
  we know that $\varphi\Rightarrow m\ \Lbowtie\ n$ (from the previous
  iteration) and that $H\Rightarrow m\ \Lbowtie\ n$ (from the
  \textsf{valid} check at Line~\ref{ln:valid1}). Analogous reasoning
  for $\hatvarphi$.
\end{proof}
\fi
\end{theorem}

\noindent
\tacasOnly{All proofs available in~\cite{arxiv}.}
Soundness holds regardless of what
\Choose\ returns
and even when the theories used to
model the underlying data-structure are incomplete.
Next we show termination implies completeness:

\begin{lemma}\label{lemma:term} If {\sc Refine}$^m_n$ terminates,
 then $\prop \vee \hatvarphi$.
 \ifARXIV
 \begin{proof}\normalfont
  The recursive calls of the {\sc Refine} algorithm induce a
  \emph{binary} tree $T$, where nodes are labeled by the conjunction of
  predicates. If {\sc Refine} terminates, then $T$ is finite, and each
  node is labeled with a finite conjunction $p_0 \wedge...\wedge p_n$.

  \emph{Claim.} The disjunction of all leaf node labels is valid.
  \emph{Pf.} By induction on the tree. Base case: a single-node
  tree has label $\TRUE$. Inductive case: for every new node 
  created, labeled with a new conjunct $...\wedge p$, there is a sibling node with
  label $...\wedge \neg p$.

Each leaf node of tree $T$, labeled with conjunction
$\gamma$, arises from {\sc Refine} reaching a base case where, by
construction, the conjunction $\gamma$ is disjunctively added to
either $\prop$ or $\hatvarphi$.
Since {\sc Refine} terminates, \emph{all} conjunctions are added to
either $\prop$ or $\hatvarphi$, and thus $\prop\vee\hatvarphi$ must be valid.
\end{proof}
 \fi
\end{lemma}


\begin{theorem}[Conditions for Termination]\label{thm:rc} {\sc Refine}$^m_n$ terminates if
1. {\bf (expressiveness)} the state space $\Sigma$ is partitionable
  into a finite set of regions $\Sigma_1,...,\Sigma_N$, each described
  by a finite conjunction of predicates $\psi_i$, such that 
either
  $\psi_i\Rightarrow m\ \Lbowtie\ n$
or
$\psi_i\Rightarrow m\ \NLbowtie\ n$;
and 2.
{\bf (fairness)} for every $p\in\preds$,
  \Choose\ eventually picks $p$
(note that this does not imply that $\preds$ is finite),
%
%
  \ifARXIV
  \begin{proof}\normalfont
By contradiction. As in the proof for Lemma \ref{lemma:term},
we represent the algorithm's execution as a binary tree $T$, induced by the recursive {\sc Refine} calls, whose nodes are labeled by the conjunction of
predicates (Lines 9 and 10 in Algorithm \ref{fig:alg}). 
Assume there exists an infinite path along $T$, and let 
its respective labels be 
$\pi = p_0, p_0\wedge p_1, p_0\wedge p_1\wedge p_2,...$.

\emph{Claim.} There is no finite prefix of $\pi$ that contains all the predicates $\psi_i$. 
\emph{Pf.} Had there been such a prefix $\varpi$, by the expressiveness assumption the running condition $H$ would satisfy one of the validity checks at lines 2 and 4 within, or immediately after, $\varpi$. This is because $H$ would be equal to, or stronger than, the conjunction of the predicates $\psi_i$.
This would have made $\pi$ finite, as $\pi$ is extended only if both of the validity checks fail, where we assume $\pi$ is infinite.

By the above claim, at least one of the predicates $\psi_i$ is not contained in any finite prefix of $\pi$. This contradicts the fairness assumption, whereby any predicate $p \in \preds$ is chosen after finitely many \Choose\ invocations (provided the algorithm hasn't terminated). 
\end{proof}
\fi
\end{theorem}

\noindent
Note that while these conditions ensure termination, the bound on the
number of iterations depends on the predicate language and behavior of
\Choose.


\section{The \Tool\ tool and practical considerations}
\label{sec:impl}
\vspace{-10pt}

\newcommand{\specfont}[1]{\textsf{#1}}

\smartpar{Input.}
We use an input specification language building on YAML (which has
parser and printer support for all common programming languages) with
SMTLIB as the logical language. This can
be automatically generated relatively easily, thus enabling the
integration with other
tools~\cite{Hoare02,modula,misra,Barnett,Meyer92,Leino06}.
\tacasOnly{In~\cite{arxiv}, }
\arxivOnly{In Apx.~\ref{yml:counter}, }
we show the Counter ADT specification, which was
derived from the $\Pre{}$ and $\Post{}$ conditions used in earlier
work~\cite{KR:PLDI11}.
The states of a transition system describing an ADT
are encoded as list of variables (each as a name/type pair),
and each method 
specification requires a list of argument types,
return type, and $\Pre{}$/$\Post{}$ conditions.
Again, the Counter example can be seen in
\tacasOnly{\cite{arxiv}}.
\arxivOnly{\ref{yml:counter}}.

\smartpar{Implementation.} 
We have developed the open-source \Tool\ tool~\cite{servois}, which
implements {\sc Refine}, {\sc Lift}, predicate generation, and
a method for selecting predicates ({\sc Choose}) discussed below.
\Tool\ uses CVC4~\cite{cvc4} as a backend SMT solver. 
\Tool\ begins by performing some pre-processing on the input
transition system. It checks that 
the transition system is deterministic. Next, in case the
transition system is partial, \Tool\ performs the 
{\sc Lift} transformation (Sec.~\ref{sec:noAlt}).
An example of {\sc Lift} applied to Counter is
\tacasOnly{in~\cite{arxiv}.}
\arxivOnly{in Apx.~\ref{yml:counterauto}.}

Next, \Tool\ automatically generates the predicate language
({\sc PGen}) in addition to user-provided hints.
If the predicate vocabulary is not
sufficiently expressive, then the algorithm would not be able to
converge on precise commutativity and non-commutativity conditions
(Sec.~\ref{sec:alg}).
We generate predicates by using terms and operators that appear in the
specification, and generating well-typed atoms not trivially true or
false.
%
%
%
%
As we demonstrate in Sec.~\ref{sec:eval},
this strategy works well in practice. Intuitively,
$\Pre{}$ and $\Post{}$ formulas suffice to
express the footprint of an operation. So, the atoms comprising
them are an effective vocabulary to express when operations do or do
not interfere.


\smartpar{Predicate selection ({\sc \Choose}).}
Even though the number of computed predicates is relatively small,
since our algorithm is exponential in number of predicates it is
essential to be able to identify \emph{relevant} predicates for the
algorithm.
To this end, in addition to filtering trivial predicates,
we prioritize predicates based on the \emph{two}
counterexamples generated by the validity checks in {\sc Refine}.
Predicates that distinguish between the given counter examples are
tried first (call these \emph{distinguishing} predicates).
\Choose\ must return a predicate such that $\cexC
\Rightarrow\; H \wedge p$ and $\cexNC \Rightarrow\; H \wedge \neg p$.
This guarantees progress on both recursive calls.
When combined with a heuristic to favor less complex atoms,
this ensured timely termination on our examples.
We refer to this as the \emph{simple} heuristic.
%


Though this produced precise conditions, they were not always very
concise, which is
desirable for human understanding, and inspection purposes. We thus
introduced a new heuristic
which significantly improves the \emph{qualitative} aspect of our
algorithm.
%
We found that doing a lookahead (recurse on each predicate one level
deep, or \emph{poke}) and computing the number of distinguishing
predicates for the two branches as a good indicator of
importance of the predicate. More precisely, we pick the
predicate with lowest sum of remaining number of distinguishing
predicates by the two calls.
%
%
As an aside, those familiar with decision tree learning, might see
a connection with the notion of entropy gain.
This requires more calls to the SMT solver at each call, but it cuts
down the total number of branches to be explored.
Also, all individual queries were relatively simple for CVC4.
The heuristic
converges much faster to the relevant predicates, and produces
smaller, concise conditions.
%


\newcommand{\CCandtwo}[2]{#1 \wedge #2}
\newcommand{\CCandthree}[3]{#1 \wedge #2 \wedge #3}
\newcommand{\CCandfour}[4]{#1 \wedge #2 \wedge #3 \wedge #4}
\newcommand{\CCequal}[2]{#1 = #2}
\newcommand{\CCplus}[2]{#1 + #2}
\newcommand{\CCminus}[2]{#1 - #2}
\newcommand{\CCsetminus}[2]{#1\setminus#2}
\newcommand{\CCmember}[2]{#1 \in #2}
\newcommand{\CCnot}[1]{\neg(#1)}
\newcommand{\CCsingleton}[1]{\{#1\}}
\newcommand{\CCortwo}[2]{[#1] \vee [#2]}
\newcommand{\CCorthree}[3]{[#1] \vee [#2] \vee [#3]}
\newcommand{\CCorfour}[4]{[#1] \vee [#2] \vee [#3] \vee [#4]}
\newcommand{\CCselect}[2]{#1[#2]}
\newcommand{\CCstore}[3]{#1[#2\leftarrow#3]}
\newcommand{\CCvar}[1]{#1}
\newcommand{\CCrow}[1]{\hline%
  \multirow{-6}{*}{\rotatebox[origin=c]{90}{{\bf #1}}} & }
\newcommand{\CCqueries}[1]{#1}
\newcommand{\CCtime}[1]{#1}
\newcommand{\CCmethod}[1]{\texttt{#1}}
\newcommand{\CCbowtie}{$\bowtie$}
\newcommand{\CCleftmover}{$\lhd$}
\newcommand{\CCrightmover}{$\rhd$}
\newcommand{\mydots}{...}

\definecolor{Gray}{gray}{0.85}

\newcolumntype{a}{>{\columncolor{Gray}}c}

\begin{figure*}[!htp]
  \scalebox{0.9}{  
    \begin{tabular}{c|rcl|r|r|p{2.5in}|}
\cline{2-2}\cline{4-7}
\multicolumn{1}{c|}{} &
\multicolumn{1}{c|}{$m(\Xs)$} &
\multicolumn{1}{c|}{} &
\multicolumn{1}{c|}{$n(\Ys)$} &
\multicolumn{1}{c|}{\bf Simple} &
\multicolumn{1}{c|}{\bf Poke} &
\multicolumn{1}{c|}{\bf $\varphi^m_n$ (Poke)}\\ 
\cline{2-2}\cline{4-7}
\multicolumn{4}{c}{} & \multicolumn{1}{r}{\scriptsize Qs (time)} & \multicolumn{1}{r}{\scriptsize Qs (time)}
\input{resultstable}
\end{tabular}
}
\caption{\label{fig:res}Automatically generated commutativity
  conditions ($\prop^m_n$).
  Right-moverness (\CCrightmover) conditions identical
  for a pair of methods denoted by
  \CCbowtie. \textbf{Qs} denotes number of SMT queries. Running
  time in seconds.
  Longer conditions have been truncated, see
  \cite{B16}.
}
\end{figure*}

\section{Case studies}
\label{sec:eval}

\smartpar{Common Data-Structures.}
We applied \Tool\ to Set, HashTable, Accumulator, Counter, and Stack.
The generated commutativity conditions for these data structures
typically combine multiple theories, such as sets, integers and arrays.
%
We used the quantifier-free integer theory in SMTLIB to encode the
abstract state and contracts for the Counter and Accumulator ADTs. For
Set, the theory of finite sets~\cite{BRBT16} for tracking elements along with
integers to track size; for HashTable, finite sets to track keys, and
arrays for the HashMap itself. For Stack, we observed that for the
purpose of pairwise commutativity it is sufficient to track the
behavior of boundedly many top elements. Since
two operations can \emph{at most} either pop the top two elements or
push two elements, tracking four elements is sufficient.
%
All evaluation data is available on our website~\cite{servoishomepage}.

Depending on the pair of methods, the number of predicates generated
by {\sc PGen} were (count after filtering in parentheses):
Counter:
25-25 (12-12)%
, Accumulator:
1-20 (0-20)%
, Set:
17-55 (17-34)%
, HashTable:
18-36 (6-36)%
, Stack:
41-61 (41-42)%
. We did not provide any hints to the algorithm for
this case study.
On all our examples, the \emph{simple} heuristic terminated with
precise commutativity conditions.  In Fig.~\ref{fig:res}, we give the
number of solver queries and total time (in paren.) consumed by this
heuristic.  The experiments were run on a 2.53 GHz Intel Core 2 Duo
machine with 8 GB RAM.  The conditions in Fig.\ref{fig:res} are those
generated by the \emph{poke} heuristic, and interested reader may
compare them with the simple heuristic in~\cite{B16}.
On the theoretical side, our \Choose\ implementation is fair
(satisfies condition 2 of Thm.~\ref{thm:rc}, as Lines 9-10 of the
algorithm remove from $\mathcal{P}$ the predicate being tried).  From
our experiments we conclude that our choice of predicates satisfies
condition 1 of Thm.~\ref{thm:rc}.

Although our algorithm is sound, we manually validated the implementation
of \Tool\ by examining its output and comparing the generated
commutativity conditions with those reported by prior studies. 
In the case of Accumulator and Counter, our commutativity
conditions were identical to those given in~\cite{KR:PLDI11}. For the
Set data structure, the work of~\cite{KR:PLDI11} used a less precise
Set abstraction, so we instead validated against the conditions
of~\cite{KNPSP:PLDI11}.
As for HashTable, we validated that our conditions match those
by Dimitrov {\it et al.}~\cite{DBLP:conf/pldi/DimitrovRVK14}.

\newcommand{\sort}[1]{\texttt{#1}}


\begin{figure}[t]
  \footnotesize
  \begin{program}[style=sf,number=true]
int warrior, warriorGold, warriorBlock, callback\_result, king, kingBlock; \label{ln:vars}
voi\tab d enter(int $\mvN$, int $\msN$, \textcolor{gray}{int $\bnN$}, \textcolor{gray}{int $\rdN$}) \{
  if ($\mvN$ $<$ 50)  \{ send($\msN$,$\mvN$); return; \} \label{ln:lt}
  warrior = $\msN$; warriorGold = $\mvN$; warriorBlock = \textcolor{gray}{$\bnN$} // write global variables
  rpc\_call("random number generator",\_\_callback,res); \label{ln:rpc}
  // $\boxed{\text{Another call to enter() can execute while waiting for RPC}}$
  fun\tab ction \_\_callback(int res\_RN) \{ \label{ln:callback}

  // $\boxed{\text{Most recent writer to warrior now reaps benefit of every callback}}$
  if \tab(modFun(warriorBlock)  == res\_RN) \{
    king = warrior; kingBlock = warriorBlock; // winner \label{ln:kingBlock} \} \} \}
\end{program}
  \caption{Simplified code for \textsf{BlockKing} in a C-like language. }
  \label{fig:blockking}
\end{figure}

\smartpar{The \textsf{BlockKing} Ethereum smart contract.}
%
We further validated our approach by examining a real-world situation
in which non-commutativity opens the door for attacks that exploit
interleavings. We examined ``smart contracts’’, which are programs
written in the Solidity programming language~\cite{solidity} and
executed on the Ethereum blockchain~\cite{ethereum}.
Eliding many details, smart contracts are like objects, and blockchain
participants can invoke methods on these objects.
Although the initial intuition is that smart contracts are executed
sequentially, practitioners and academics~\cite{sergeyhobor}
are increasingly realizing that the blockchain is a concurrent
environment due to the fact the execution of one actor's smart
  contract can be split across multiple blocks, with other actors'
smart contracts interleaved.
Therefore, the execution model of the blockchain has been compared to that of concurrent objects~\cite{sergeyhobor}. 
Unfortunately, many smart contracts are not written with this in mind,
and attackers can exploit interleavings to their benefit.

As an example, we study the \textsf{BlockKing} smart
contract. Fig.~\ref{fig:blockking} provides a simplification of its
description, as discussed in~\cite{sergeyhobor}. This is a simple game
in which the players---each identified by an address
$\msN$---participate by making calls to \textsf{BlockKing.enter()},
sending money $\mvN$ to the contract. (The grey variables are external
input that we have lifted to be parameters. $\bnN$ reflects the
caller's current block number and $\rdN$ is the outcome of a random
number generation, described shortly.) The variables on
Line~\ref{ln:vars} are globals, writable in any call to
\textsf{enter}. On Line~\ref{ln:lt} there is a trivial case when the
caller hasn’t put enough value into the game, and the money is simply
returned. Otherwise, the caller stores their address and value into
the shared state. A random number is then generated and, since this
requires complex algorithms, it is done via a remote procedure call to
a third-party on Line~\ref{ln:rpc}, with a callback method provided on
Line~\ref{ln:callback}. If the randomly generated number is equal to a
modulus of the current block number, then the caller is the winner,
and \textsf{warrior}'s (caller's) details are stored to \textsf{king}
and \textsf{kingBlock} on Line~\ref{ln:kingBlock}.

Since random number generation is done via an RPC,
players' invocations of \textsf{enter} can be
interleaved. Moreover, these calls all write $\msN$ and $\mvN$ to
shared variables, so the RPC callback will always roll the dice for
whomever most recently wrote to \textsf{warriorBlock}. An attacker can
use this to leverage other players' investments to increase
his/her own chance to win.

We now explore how \Tool~can aid a programmer in developing a more
secure implementation. We observe that, as in traditional parallel
programming contexts, if smart contracts are commutative then these
interleavings are not problematic. Otherwise, there is cause for
concern. To this end, we translated the \textsf{BlockKing} game into
\Tool~format (see Apx.~\ref{yml:blockking}). \Tool\ took
1.4s (on machine with 2.4 GHz Intel Core i5 processor and 8 GB RAM) and
yielded the following \emph{non-commutativity} condition for two calls
to \texttt{enter}:
\[\begin{array}{c}
\texttt{enter}(\mvA,\msA,\bnA,\rdA)
\;\NLbowtie\;
\texttt{enter}(\mvB,\msB,\bnB,\rdB)
\;\;\;\;\;\;\;\;\Leftrightarrow\\
\bigvee \left\{
  \begin{array}{l}
    \mvA \geq 50 \wedge \mvB \geq 50 \wedge \msA \neq \msB\\
    \mvA \geq 50 \wedge \mvB \geq 50 \wedge \msA = \msB \wedge \mvA \neq \mvB\\
    \mvA \geq 50 \wedge \mvB \geq 50 \wedge \msA = \msB \wedge \mvA = \mvB \wedge \bnA \neq \bnB\\
  \end{array} \right.
\end{array}
    \]
This disjunction effectively enumerates cases under which they
contract calls \emph{do not} commute. Of particular note is the first
disjunct. From this first disjunct, whenever $\msA \neq \msB$, the
calls will not commute. Since in practice $\msA$ will always be
different from $\msB$ (two different callers) and $\mvA \geq 50 \wedge
\mvB \geq 50$ is the non-trivial case, the operations will almost
never commute. This should be immediate cause for concern to the
developer.

A commutative version of \textsf{BlockKing} would mean that there are
no interleavings to be concerned about. Indeed, a simple way to
improve commutativity is for each player to write their respective
$\msN$ and $\mvN$ to distinct shared state, perhaps via a hashtable
keyed on $\msN$. To this end, we created a new version
\textsf{enter\_fixed}\arxivOnly{, shown in Fig.~\ref{fig:blockkingfixed}. (YML versions of these two programs can be found in Appendix~\ref{apx:blockking}
and~\ref{apx:blockking_fixed}.)}\tacasOnly{ (see~\cite{arxiv})}. \Tool\ generated the following
\emph{non-}commutativity condition after 3.5s.

\[\begin{array}{c}
\texttt{enter\_fixed}(\mvA,\msA,\bnA,\rdA)
\;\NLbowtie\;
\texttt{enter\_fixed}(\mvB,\msB,\bnB,\rdB)
\;\;\;\;\;\;\text{\emph{iff}}\\
\bigvee \left\{
  \begin{array}{l}
        \mvA \geq 50 \wedge
         \mvB \geq 50 \wedge 
         \mvA = \mvB \wedge
         \bnA \neq \bnB \wedge
         \msA = \msB  \\      
    \mvA \geq 50 \wedge
         \mvB \geq 50 \wedge
         \mvA \neq \mvB \wedge
         \msA = \msB \\         
\mvA \geq 50 \wedge
\mvB \geq 50 \wedge
\textsf{md}(\bnB) = \rdB \wedge   
\textsf{md}(\bnA) = \rdA \wedge   
    \msA \neq \msB 
  \end{array} \right.
    \end{array}\]
In the above non-commutativity condition, $\textsf{md}$ is shorthand for \textsf{modFun}. In the first two disjuncts above, $\msA=\msB$ which is, again, a case that will not occur in practice.
All that remains is the third disjunct where
$\textsf{md}(\bnB) = \rdB$ and $\textsf{md}(\bnA) = \rdA$. This corresponds to the case where \emph{both} players have won. In this case, it is acceptable for the operations to not commute, because whomever won more recently will store their address/block to the shared \textsf{king}/\textsf{kingBlock}.


In summary, if we assume that $\msA \neq \msB$, the non-commutativity
of the original version is $\mvA \geq 50 \vee \mvB \geq 50$ (very
strong). By contrast, the non-commutativity of the fixed version is
$\mvA \geq 50 \wedge \mvB \geq 50 \wedge \textsf{md}(\bnB) = \rdB \wedge
\textsf{md}(\bnA) = \rdA$.
We have thus demonstrated that the commutativity (and non-commutativity)
conditions generated by \Tool\ can help developers understand the model of
interference between two concurrent calls.

\begin{figure}
  \begin{program}[style=sf,number=true]
str\tab uct storage \{
  int warrior;
  int warriorGold;
  int warriorBlock;
  int res; \untab
  \};
  $\;$\\
hashtable[int,struct storage] scratch = ...;
int king, kingBlock;
$\;$\\
voi\tab d enter\_fixed(int $\mvN$, int $\msN$, \textcolor{gray}{int $\bnN$}, \textcolor{gray}{int $\rdN$}) \{
  if ($\mvN$ $<$ 50)  \{ send($\msN$,$\mvN$); return; \}
  scratch[$\msN$].warrior = $\msN$;
  scratch[$\msN$].warriorGold = $\mvN$;
  scratch[$\msN$].warriorBlock = \textcolor{gray}{$\bnN$};
  // callback generates the random number in scratch[$\msN$]
  rpc\_call("random number generator",\_\_callback,scratch[$\msN$].res);
  fun\tab ction \_\_callback() \{
  if \tab(modFun(scratch[$\msN$].warriorBlock)  == scratch[$\msN$].res) \{
    king = scratch[$\msN$].warrior; // winner
    kingBlock = scratch[$\msN$].warriorBlock; \} \} \}
\end{program}
  \caption{\label{fig:blockkingfixed} Our fixed version of \textsf{BlockKing}
    in a C-like language.}
\end{figure}

\section{Conclusions and future work}

This paper demonstrates that it is possible to automatically generate
sound and effective commutativity conditions, a task that has so far been done manually or without soundness.
Our commutativity conditions are applicable in a variety of
contexts including transactional boosting~\cite{ppopp08}, open nested
transactions~\cite{DBLP:conf/ppopp/NiMAHHMSS07}, and other
non-transactional concurrency paradigms such as race
detection~\cite{DBLP:conf/pldi/DimitrovRVK14},
parallelizing compilers~\cite{rinard,DBLP:conf/oopsla/TrippYFS11},
and, as we show, robustness of Ethereum smart contracts~\cite{sergeyhobor}.
It has been shown that understanding
the commutativity of data-structure operations provides a key avenue
to improved performance~\cite{DBLP:journals/tocs/ClementsKZMK15} or ease of verification~\cite{DBLP:conf/popl/KoskinenPH10,DBLP:conf/pldi/KoskinenP15}.

This work opens several avenues of future research.
For instance, leveraging the internal state of the SMT solver
(beyond counterexamples) in order to generate new predicates~\cite{HBG04};
automatically building abstract representation or making inferences
such as one we made for the stack example; and
exploring strategies to compute commutativity conditions directly
from the program's code, without the need for an intermediate abstract
representation~\cite{DBLP:conf/oopsla/TrippYFS11}.

\vfill
\pagebreak
{\footnotesize\linespread{0.9}

}

\vfill
\pagebreak
\appendix
\section{Data Structure Representations}
\label{apx:ds}
\lstset{numbers=left,numberstyle=\tiny,numbersep=5pt,language=Lisp,
  stringstyle=\ttfamily\small,basicstyle=\ttfamily\footnotesize,
  showstringspaces=false,breaklines,frame=none}
\renewcommand{\CCandtwo}[2]{#1 $\wedge$ #2}
\renewcommand{\CCandthree}[3]{#1 $\wedge$ #2 $\wedge$ #3}
\renewcommand{\CCandfour}[4]{#1 $\wedge$ #2 $\wedge$ #3 $\wedge$ #4}
\renewcommand{\CCequal}[2]{#1 = #2}
\renewcommand{\CCplus}[2]{#1 $+$ #2}
\renewcommand{\CCminus}[2]{#1 $-$ #2}
\renewcommand{\CCsetminus}[2]{#1$\setminus$#2}
\renewcommand{\CCmember}[2]{#1 $\in$ #2}
\renewcommand{\CCnot}[1]{$\neg$(#1)}
\renewcommand{\CCsingleton}[1]{\{#1\}}
\renewcommand{\CCortwo}[2]{[#1] $\vee$ [#2]}
\renewcommand{\CCorthree}[3]{[#1] $\vee$ [#2] $\vee$ [#3]}
\renewcommand{\CCorfour}[4]{[#1] $\vee$ [#2] $\vee$ [#3] $\vee$ [#4]}
\renewcommand{\CCselect}[2]{#1[#2]}
\renewcommand{\CCstore}[3]{#1[#2=#3]}
\renewcommand{\CCvar}[1]{\texttt{#1}}

\subsection{Counter}
\scriptsize
\label{yml:counter}
{\ttfamily\noindent
\textcolor[HTML]{0000C0}{\char35{}\enskip{}Counter\enskip{}data\enskip{}structure's\enskip{}abstract\enskip{}definition}\\
\\
\textcolor[HTML]{008080}{name}\textcolor[HTML]{C000C0}{:}\textcolor[HTML]{000000}{\enskip{}}\textcolor[HTML]{000000}{counter}\\
\\
\textcolor[HTML]{008080}{state}\textcolor[HTML]{C000C0}{:}\\
\textcolor[HTML]{000000}{\enskip{}\enskip{}}\textcolor[HTML]{AF5F00}{-\enskip{}}\textcolor[HTML]{008080}{name}\textcolor[HTML]{C000C0}{:}\textcolor[HTML]{000000}{\enskip{}}\textcolor[HTML]{000000}{contents}\\
\textcolor[HTML]{000000}{\enskip{}\enskip{}\enskip{}\enskip{}}\textcolor[HTML]{008080}{type}\textcolor[HTML]{C000C0}{:}\textcolor[HTML]{000000}{\enskip{}}\textcolor[HTML]{000000}{Int}\\
\\
\textcolor[HTML]{008080}{states\char95{}equal}\textcolor[HTML]{C000C0}{:}\\
\textcolor[HTML]{000000}{\enskip{}\enskip{}}\textcolor[HTML]{008080}{definition}\textcolor[HTML]{C000C0}{:}\textcolor[HTML]{000000}{\enskip{}}\textcolor[HTML]{000000}{(=}\textcolor[HTML]{000000}{\enskip{}}\textcolor[HTML]{000000}{contents\char95{}1}\textcolor[HTML]{000000}{\enskip{}}\textcolor[HTML]{000000}{contents\char95{}2)}\\
\\
\textcolor[HTML]{008080}{methods}\textcolor[HTML]{C000C0}{:}\\
\textcolor[HTML]{000000}{\enskip{}\enskip{}}\textcolor[HTML]{AF5F00}{-\enskip{}}\textcolor[HTML]{008080}{name}\textcolor[HTML]{C000C0}{:}\textcolor[HTML]{000000}{\enskip{}}\textcolor[HTML]{000000}{increment}\\
\textcolor[HTML]{000000}{\enskip{}\enskip{}\enskip{}\enskip{}}\textcolor[HTML]{008080}{args}\textcolor[HTML]{C000C0}{:}\textcolor[HTML]{000000}{\enskip{}}\textcolor[HTML]{C000C0}{\char91{}\char93{}}\\
\textcolor[HTML]{000000}{\enskip{}\enskip{}\enskip{}\enskip{}}\textcolor[HTML]{008080}{return}\textcolor[HTML]{C000C0}{:}\\
\textcolor[HTML]{000000}{\enskip{}\enskip{}\enskip{}\enskip{}\enskip{}\enskip{}}\textcolor[HTML]{AF5F00}{-\enskip{}}\textcolor[HTML]{008080}{name}\textcolor[HTML]{C000C0}{:}\textcolor[HTML]{000000}{\enskip{}}\textcolor[HTML]{000000}{result}\\
\textcolor[HTML]{000000}{\enskip{}\enskip{}\enskip{}\enskip{}\enskip{}\enskip{}\enskip{}\enskip{}}\textcolor[HTML]{008080}{type}\textcolor[HTML]{C000C0}{:}\textcolor[HTML]{000000}{\enskip{}}\textcolor[HTML]{000000}{Bool}\\
\textcolor[HTML]{000000}{\enskip{}\enskip{}\enskip{}\enskip{}}\textcolor[HTML]{008080}{requires}\textcolor[HTML]{C000C0}{:}\textcolor[HTML]{000000}{\enskip{}|}\\
\textcolor[HTML]{000000}{\enskip{}\enskip{}\enskip{}\enskip{}\enskip{}\enskip{}}\textcolor[HTML]{000000}{(>=}\textcolor[HTML]{000000}{\enskip{}}\textcolor[HTML]{000000}{contents}\textcolor[HTML]{000000}{\enskip{}}\textcolor[HTML]{000000}{0)}\\
\textcolor[HTML]{000000}{\enskip{}\enskip{}\enskip{}\enskip{}}\textcolor[HTML]{008080}{ensures}\textcolor[HTML]{C000C0}{:}\textcolor[HTML]{000000}{\enskip{}|}\\
\textcolor[HTML]{000000}{\enskip{}\enskip{}\enskip{}\enskip{}\enskip{}\enskip{}}\textcolor[HTML]{000000}{(and}\textcolor[HTML]{000000}{\enskip{}}\textcolor[HTML]{000000}{(=}\textcolor[HTML]{000000}{\enskip{}}\textcolor[HTML]{000000}{contents\char95{}new}\textcolor[HTML]{000000}{\enskip{}}\textcolor[HTML]{000000}{(+}\textcolor[HTML]{000000}{\enskip{}}\textcolor[HTML]{000000}{contents}\textcolor[HTML]{000000}{\enskip{}}\textcolor[HTML]{000000}{1))}\\
\textcolor[HTML]{000000}{\enskip{}\enskip{}\enskip{}\enskip{}\enskip{}\enskip{}\enskip{}\enskip{}\enskip{}\enskip{}\enskip{}}\textcolor[HTML]{000000}{(=}\textcolor[HTML]{000000}{\enskip{}}\textcolor[HTML]{000000}{result}\textcolor[HTML]{000000}{\enskip{}}\textcolor[HTML]{C00000}{true}\textcolor[HTML]{000000}{))}\\
\textcolor[HTML]{000000}{\enskip{}\enskip{}\enskip{}\enskip{}}\textcolor[HTML]{008080}{terms}\textcolor[HTML]{C000C0}{:}\\
\textcolor[HTML]{000000}{\enskip{}\enskip{}\enskip{}\enskip{}\enskip{}\enskip{}}\textcolor[HTML]{008080}{Int}\textcolor[HTML]{C000C0}{:}\textcolor[HTML]{000000}{\enskip{}}\textcolor[HTML]{C000C0}{\char91{}}\textcolor[HTML]{000000}{contents}\textcolor[HTML]{000000}{,\enskip{}}\textcolor[HTML]{C00000}{1}\textcolor[HTML]{000000}{,\enskip{}}\textcolor[HTML]{000000}{(+}\textcolor[HTML]{000000}{\enskip{}}\textcolor[HTML]{000000}{contents}\textcolor[HTML]{000000}{\enskip{}}\textcolor[HTML]{000000}{1)}\textcolor[HTML]{C000C0}{\char93{}}\\
\textcolor[HTML]{000000}{\enskip{}\enskip{}}\textcolor[HTML]{AF5F00}{-\enskip{}}\textcolor[HTML]{008080}{name}\textcolor[HTML]{C000C0}{:}\textcolor[HTML]{000000}{\enskip{}}\textcolor[HTML]{000000}{decrement}\\
\textcolor[HTML]{000000}{\enskip{}\enskip{}\enskip{}\enskip{}}\textcolor[HTML]{008080}{args}\textcolor[HTML]{C000C0}{:}\textcolor[HTML]{000000}{\enskip{}}\textcolor[HTML]{C000C0}{\char91{}\char93{}}\\
\textcolor[HTML]{000000}{\enskip{}\enskip{}\enskip{}\enskip{}}\textcolor[HTML]{008080}{return}\textcolor[HTML]{C000C0}{:}\\
\textcolor[HTML]{000000}{\enskip{}\enskip{}\enskip{}\enskip{}\enskip{}\enskip{}}\textcolor[HTML]{AF5F00}{-\enskip{}}\textcolor[HTML]{008080}{name}\textcolor[HTML]{C000C0}{:}\textcolor[HTML]{000000}{\enskip{}}\textcolor[HTML]{000000}{result}\\
\textcolor[HTML]{000000}{\enskip{}\enskip{}\enskip{}\enskip{}\enskip{}\enskip{}\enskip{}\enskip{}}\textcolor[HTML]{008080}{type}\textcolor[HTML]{C000C0}{:}\textcolor[HTML]{000000}{\enskip{}}\textcolor[HTML]{000000}{Bool}\\
\textcolor[HTML]{000000}{\enskip{}\enskip{}\enskip{}\enskip{}}\textcolor[HTML]{008080}{requires}\textcolor[HTML]{C000C0}{:}\textcolor[HTML]{000000}{\enskip{}|}\\
\textcolor[HTML]{000000}{\enskip{}\enskip{}\enskip{}\enskip{}\enskip{}\enskip{}}\textcolor[HTML]{000000}{(>=}\textcolor[HTML]{000000}{\enskip{}}\textcolor[HTML]{000000}{contents}\textcolor[HTML]{000000}{\enskip{}}\textcolor[HTML]{000000}{1)}\\
\textcolor[HTML]{000000}{\enskip{}\enskip{}\enskip{}\enskip{}}\textcolor[HTML]{008080}{ensures}\textcolor[HTML]{C000C0}{:}\textcolor[HTML]{000000}{\enskip{}|}\\
\textcolor[HTML]{000000}{\enskip{}\enskip{}\enskip{}\enskip{}\enskip{}\enskip{}}\textcolor[HTML]{000000}{(and}\textcolor[HTML]{000000}{\enskip{}}\textcolor[HTML]{000000}{(=}\textcolor[HTML]{000000}{\enskip{}}\textcolor[HTML]{000000}{contents\char95{}new}\textcolor[HTML]{000000}{\enskip{}}\textcolor[HTML]{000000}{(-}\textcolor[HTML]{000000}{\enskip{}}\textcolor[HTML]{000000}{contents}\textcolor[HTML]{000000}{\enskip{}}\textcolor[HTML]{000000}{1))}\\
\textcolor[HTML]{000000}{\enskip{}\enskip{}\enskip{}\enskip{}\enskip{}\enskip{}\enskip{}\enskip{}\enskip{}\enskip{}\enskip{}}\textcolor[HTML]{000000}{(=}\textcolor[HTML]{000000}{\enskip{}}\textcolor[HTML]{000000}{result}\textcolor[HTML]{000000}{\enskip{}}\textcolor[HTML]{C00000}{true}\textcolor[HTML]{000000}{))}\\
\textcolor[HTML]{000000}{\enskip{}\enskip{}\enskip{}\enskip{}}\textcolor[HTML]{008080}{terms}\textcolor[HTML]{C000C0}{:}\\
\textcolor[HTML]{000000}{\enskip{}\enskip{}\enskip{}\enskip{}\enskip{}\enskip{}}\textcolor[HTML]{008080}{Int}\textcolor[HTML]{C000C0}{:}\textcolor[HTML]{000000}{\enskip{}}\textcolor[HTML]{C000C0}{\char91{}}\textcolor[HTML]{000000}{contents}\textcolor[HTML]{000000}{,\enskip{}}\textcolor[HTML]{C00000}{1}\textcolor[HTML]{000000}{,\enskip{}}\textcolor[HTML]{000000}{(-}\textcolor[HTML]{000000}{\enskip{}}\textcolor[HTML]{000000}{contents}\textcolor[HTML]{000000}{\enskip{}}\textcolor[HTML]{000000}{1)}\textcolor[HTML]{000000}{,\enskip{}}\textcolor[HTML]{C00000}{0}\textcolor[HTML]{C000C0}{\char93{}}\\
\textcolor[HTML]{000000}{\enskip{}\enskip{}}\textcolor[HTML]{AF5F00}{-\enskip{}}\textcolor[HTML]{008080}{name}\textcolor[HTML]{C000C0}{:}\textcolor[HTML]{000000}{\enskip{}}\textcolor[HTML]{000000}{reset}\\
\textcolor[HTML]{000000}{\enskip{}\enskip{}\enskip{}\enskip{}}\textcolor[HTML]{008080}{args}\textcolor[HTML]{C000C0}{:}\textcolor[HTML]{000000}{\enskip{}}\textcolor[HTML]{C000C0}{\char91{}\char93{}}\\
\textcolor[HTML]{000000}{\enskip{}\enskip{}\enskip{}\enskip{}}\textcolor[HTML]{008080}{return}\textcolor[HTML]{C000C0}{:}\\
\textcolor[HTML]{000000}{\enskip{}\enskip{}\enskip{}\enskip{}\enskip{}\enskip{}}\textcolor[HTML]{AF5F00}{-\enskip{}}\textcolor[HTML]{008080}{name}\textcolor[HTML]{C000C0}{:}\textcolor[HTML]{000000}{\enskip{}}\textcolor[HTML]{000000}{result}\\
\textcolor[HTML]{000000}{\enskip{}\enskip{}\enskip{}\enskip{}\enskip{}\enskip{}\enskip{}\enskip{}}\textcolor[HTML]{008080}{type}\textcolor[HTML]{C000C0}{:}\textcolor[HTML]{000000}{\enskip{}}\textcolor[HTML]{000000}{Bool}\\
\textcolor[HTML]{000000}{\enskip{}\enskip{}\enskip{}\enskip{}}\textcolor[HTML]{008080}{requires}\textcolor[HTML]{C000C0}{:}\textcolor[HTML]{000000}{\enskip{}|}\\
\textcolor[HTML]{000000}{\enskip{}\enskip{}\enskip{}\enskip{}\enskip{}\enskip{}}\textcolor[HTML]{000000}{(>=}\textcolor[HTML]{000000}{\enskip{}}\textcolor[HTML]{000000}{contents}\textcolor[HTML]{000000}{\enskip{}}\textcolor[HTML]{000000}{0)}\\
\textcolor[HTML]{000000}{\enskip{}\enskip{}\enskip{}\enskip{}}\textcolor[HTML]{008080}{ensures}\textcolor[HTML]{C000C0}{:}\textcolor[HTML]{000000}{\enskip{}|}\\
\textcolor[HTML]{000000}{\enskip{}\enskip{}\enskip{}\enskip{}\enskip{}\enskip{}}\textcolor[HTML]{000000}{(and}\textcolor[HTML]{000000}{\enskip{}}\textcolor[HTML]{000000}{(=}\textcolor[HTML]{000000}{\enskip{}}\textcolor[HTML]{000000}{contents\char95{}new}\textcolor[HTML]{000000}{\enskip{}}\textcolor[HTML]{000000}{0)}\\
\textcolor[HTML]{000000}{\enskip{}\enskip{}\enskip{}\enskip{}\enskip{}\enskip{}\enskip{}\enskip{}\enskip{}\enskip{}\enskip{}}\textcolor[HTML]{000000}{(=}\textcolor[HTML]{000000}{\enskip{}}\textcolor[HTML]{000000}{result}\textcolor[HTML]{000000}{\enskip{}}\textcolor[HTML]{C00000}{true}\textcolor[HTML]{000000}{))}\\
\textcolor[HTML]{000000}{\enskip{}\enskip{}\enskip{}\enskip{}}\textcolor[HTML]{008080}{terms}\textcolor[HTML]{C000C0}{:}\\
\textcolor[HTML]{000000}{\enskip{}\enskip{}\enskip{}\enskip{}\enskip{}\enskip{}}\textcolor[HTML]{008080}{Int}\textcolor[HTML]{C000C0}{:}\textcolor[HTML]{000000}{\enskip{}}\textcolor[HTML]{C000C0}{\char91{}}\textcolor[HTML]{000000}{contents}\textcolor[HTML]{000000}{,\enskip{}}\textcolor[HTML]{C00000}{0}\textcolor[HTML]{C000C0}{\char93{}}\\
\textcolor[HTML]{000000}{\enskip{}\enskip{}}\textcolor[HTML]{AF5F00}{-\enskip{}}\textcolor[HTML]{008080}{name}\textcolor[HTML]{C000C0}{:}\textcolor[HTML]{000000}{\enskip{}}\textcolor[HTML]{000000}{zero}\\
\textcolor[HTML]{000000}{\enskip{}\enskip{}\enskip{}\enskip{}}\textcolor[HTML]{008080}{args}\textcolor[HTML]{C000C0}{:}\textcolor[HTML]{000000}{\enskip{}}\textcolor[HTML]{C000C0}{\char91{}\char93{}}\\
\textcolor[HTML]{000000}{\enskip{}\enskip{}\enskip{}\enskip{}}\textcolor[HTML]{008080}{return}\textcolor[HTML]{C000C0}{:}\\
\textcolor[HTML]{000000}{\enskip{}\enskip{}\enskip{}\enskip{}\enskip{}\enskip{}}\textcolor[HTML]{AF5F00}{-\enskip{}}\textcolor[HTML]{008080}{name}\textcolor[HTML]{C000C0}{:}\textcolor[HTML]{000000}{\enskip{}}\textcolor[HTML]{000000}{result}\\
\textcolor[HTML]{000000}{\enskip{}\enskip{}\enskip{}\enskip{}\enskip{}\enskip{}\enskip{}\enskip{}}\textcolor[HTML]{008080}{type}\textcolor[HTML]{C000C0}{:}\textcolor[HTML]{000000}{\enskip{}}\textcolor[HTML]{000000}{Bool}\\
\textcolor[HTML]{000000}{\enskip{}\enskip{}\enskip{}\enskip{}}\textcolor[HTML]{008080}{requires}\textcolor[HTML]{C000C0}{:}\textcolor[HTML]{000000}{\enskip{}|}\\
\textcolor[HTML]{000000}{\enskip{}\enskip{}\enskip{}\enskip{}\enskip{}\enskip{}}\textcolor[HTML]{000000}{(>=}\textcolor[HTML]{000000}{\enskip{}}\textcolor[HTML]{000000}{contents}\textcolor[HTML]{000000}{\enskip{}}\textcolor[HTML]{000000}{0)}\\
\textcolor[HTML]{000000}{\enskip{}\enskip{}\enskip{}\enskip{}}\textcolor[HTML]{008080}{ensures}\textcolor[HTML]{C000C0}{:}\textcolor[HTML]{000000}{\enskip{}|}\\
\textcolor[HTML]{000000}{\enskip{}\enskip{}\enskip{}\enskip{}\enskip{}\enskip{}}\textcolor[HTML]{000000}{(and}\textcolor[HTML]{000000}{\enskip{}}\textcolor[HTML]{000000}{(=}\textcolor[HTML]{000000}{\enskip{}}\textcolor[HTML]{000000}{contents\char95{}new}\textcolor[HTML]{000000}{\enskip{}}\textcolor[HTML]{000000}{contents)}\\
\textcolor[HTML]{000000}{\enskip{}\enskip{}\enskip{}\enskip{}\enskip{}\enskip{}\enskip{}\enskip{}\enskip{}\enskip{}\enskip{}}\textcolor[HTML]{000000}{(=}\textcolor[HTML]{000000}{\enskip{}}\textcolor[HTML]{000000}{result}\textcolor[HTML]{000000}{\enskip{}}\textcolor[HTML]{000000}{(=}\textcolor[HTML]{000000}{\enskip{}}\textcolor[HTML]{000000}{contents}\textcolor[HTML]{000000}{\enskip{}}\textcolor[HTML]{000000}{0)))}\\
\textcolor[HTML]{000000}{\enskip{}\enskip{}\enskip{}\enskip{}}\textcolor[HTML]{008080}{terms}\textcolor[HTML]{C000C0}{:}\\
\textcolor[HTML]{000000}{\enskip{}\enskip{}\enskip{}\enskip{}\enskip{}\enskip{}}\textcolor[HTML]{008080}{Int}\textcolor[HTML]{C000C0}{:}\textcolor[HTML]{000000}{\enskip{}}\textcolor[HTML]{C000C0}{\char91{}}\textcolor[HTML]{000000}{contents}\textcolor[HTML]{000000}{,\enskip{}}\textcolor[HTML]{C00000}{0}\textcolor[HTML]{C000C0}{\char93{}}\\
\\
\textcolor[HTML]{008080}{predicates}\textcolor[HTML]{C000C0}{:}\\
\textcolor[HTML]{000000}{\enskip{}\enskip{}}\textcolor[HTML]{AF5F00}{-\enskip{}}\textcolor[HTML]{008080}{name}\textcolor[HTML]{C000C0}{:}\textcolor[HTML]{000000}{\enskip{}}\textcolor[HTML]{C00000}{"}\textcolor[HTML]{C00000}{=}\textcolor[HTML]{C00000}{"}\\
\textcolor[HTML]{000000}{\enskip{}\enskip{}\enskip{}\enskip{}}\textcolor[HTML]{008080}{type}\textcolor[HTML]{C000C0}{:}\textcolor[HTML]{000000}{\enskip{}}\textcolor[HTML]{C000C0}{\char91{}}\textcolor[HTML]{000000}{Int}\textcolor[HTML]{000000}{,\enskip{}}\textcolor[HTML]{000000}{Int}\textcolor[HTML]{C000C0}{\char93{}}\\

}

\begin{itemize}
\item \CCmethod{decrement} \CCbowtie\ \CCmethod{decrement}

 Simple:

true

 Poke:

true

\item \CCmethod{increment} \CCrightmover\ \CCmethod{decrement}

 Simple:

[\CCequal{\CCvar{1}}{\CCvar{contents}}]

 $\vee$ [\CCnot{\CCequal{\CCvar{1}}{\CCvar{contents}}} $\wedge$ \CCnot{\CCequal{\CCvar{0}}{\CCvar{contents}}}]

 Poke:

\CCnot{\CCequal{\CCvar{0}}{\CCvar{contents}}}

\item \CCmethod{decrement} \CCrightmover\ \CCmethod{increment}

 Simple:

true

 Poke:

true

\item \CCmethod{decrement} \CCbowtie\ \CCmethod{reset}

 Simple:

false

 Poke:

false

\item \CCmethod{decrement} \CCbowtie\ \CCmethod{zero}

 Simple:

\CCnot{\CCequal{\CCvar{1}}{\CCvar{contents}}}

 Poke:

\CCnot{\CCequal{\CCvar{1}}{\CCvar{contents}}}

\item \CCmethod{increment} \CCbowtie\ \CCmethod{increment}

 Simple:

true

 Poke:

true

\item \CCmethod{increment} \CCbowtie\ \CCmethod{reset}

 Simple:

false

 Poke:

false

\item \CCmethod{increment} \CCbowtie\ \CCmethod{zero}

 Simple:

[\CCequal{\CCvar{1}}{\CCvar{contents}}]

 $\vee$ [\CCnot{\CCequal{\CCvar{1}}{\CCvar{contents}}} $\wedge$ \CCnot{\CCequal{\CCvar{0}}{\CCvar{contents}}}]

 Poke:

\CCnot{\CCequal{\CCvar{0}}{\CCvar{contents}}}

\item \CCmethod{reset} \CCbowtie\ \CCmethod{reset}

 Simple:

true

 Poke:

true

\item \CCmethod{reset} \CCbowtie\ \CCmethod{zero}

 Simple:

\CCnot{\CCequal{\CCvar{1}}{\CCvar{contents}}} $\wedge$ \CCequal{\CCvar{0}}{\CCvar{contents}}

 Poke:

\CCequal{\CCvar{0}}{\CCvar{contents}}

\item \CCmethod{zero} \CCbowtie\ \CCmethod{zero}

 Simple:

true

 Poke:

true

\end{itemize}

\subsection{Counter (lifted, auto-generated)}
\scriptsize
\label{yml:counterauto}
{\ttfamily\noindent
\textcolor[HTML]{008080}{methods}\textcolor[HTML]{C000C0}{:}\\
\textcolor[HTML]{AF5F00}{-\enskip{}}\textcolor[HTML]{008080}{args}\textcolor[HTML]{C000C0}{:}\textcolor[HTML]{000000}{\enskip{}}\textcolor[HTML]{C000C0}{\char91{}\char93{}}\\
\textcolor[HTML]{000000}{\enskip{}\enskip{}}\textcolor[HTML]{008080}{ensures}\textcolor[HTML]{C000C0}{:}\textcolor[HTML]{000000}{\enskip{}}\textcolor[HTML]{C00000}{"}\textcolor[HTML]{C00000}{(or\enskip{}(and\enskip{}err\enskip{}err\char95{}new)}\textcolor[HTML]{C000C0}{\char92{}n}\textcolor[HTML]{C00000}{\enskip{}\enskip{}\enskip{}\enskip{}(and\enskip{}(not\enskip{}err)\enskip{}(not\enskip{}err\char95{}new)\enskip{}(>=\enskip{}contents\enskip{}0)}\textcolor[HTML]{C000C0}{\char92{}n\char92{}}\\
\textcolor[HTML]{C00000}{\enskip{}\enskip{}\enskip{}\enskip{}}\textcolor[HTML]{C000C0}{\char92{}\enskip{}}\textcolor[HTML]{C00000}{(and\enskip{}(=\enskip{}contents\char95{}new\enskip{}(+\enskip{}contents\enskip{}1))}\textcolor[HTML]{C000C0}{\char92{}n}\textcolor[HTML]{C00000}{\enskip{}\enskip{}\enskip{}\enskip{}\enskip{}(=\enskip{}result\enskip{}true))}\textcolor[HTML]{C000C0}{\char92{}n}\textcolor[HTML]{C00000}{)}\textcolor[HTML]{C000C0}{\char92{}n}\textcolor[HTML]{C00000}{\enskip{}\enskip{}\enskip{}\enskip{}(and\enskip{}(not}\textcolor[HTML]{C000C0}{\char92{}}\\
\textcolor[HTML]{C00000}{\enskip{}\enskip{}\enskip{}\enskip{}}\textcolor[HTML]{C000C0}{\char92{}\enskip{}}\textcolor[HTML]{C00000}{err)\enskip{}err\char95{}new\enskip{}(not\enskip{}(>=\enskip{}contents\enskip{}0)}\textcolor[HTML]{C000C0}{\char92{}n}\textcolor[HTML]{C00000}{)))}\textcolor[HTML]{C00000}{"}\\
\textcolor[HTML]{000000}{\enskip{}\enskip{}}\textcolor[HTML]{008080}{name}\textcolor[HTML]{C000C0}{:}\textcolor[HTML]{000000}{\enskip{}}\textcolor[HTML]{000000}{increment}\\
\textcolor[HTML]{000000}{\enskip{}\enskip{}}\textcolor[HTML]{008080}{requires}\textcolor[HTML]{C000C0}{:}\textcolor[HTML]{000000}{\enskip{}}\textcolor[HTML]{C00000}{'}\textcolor[HTML]{C00000}{true}\textcolor[HTML]{C00000}{'}\\
\textcolor[HTML]{000000}{\enskip{}\enskip{}}\textcolor[HTML]{008080}{return}\textcolor[HTML]{C000C0}{:}\\
\textcolor[HTML]{000000}{\enskip{}\enskip{}}\textcolor[HTML]{AF5F00}{-\enskip{}}\textcolor[HTML]{008080}{name}\textcolor[HTML]{C000C0}{:}\textcolor[HTML]{000000}{\enskip{}}\textcolor[HTML]{000000}{result}\\
\textcolor[HTML]{000000}{\enskip{}\enskip{}\enskip{}\enskip{}}\textcolor[HTML]{008080}{type}\textcolor[HTML]{C000C0}{:}\textcolor[HTML]{000000}{\enskip{}}\textcolor[HTML]{000000}{Bool}\\
\textcolor[HTML]{000000}{\enskip{}\enskip{}}\textcolor[HTML]{008080}{terms}\textcolor[HTML]{C000C0}{:}\\
\textcolor[HTML]{000000}{\enskip{}\enskip{}\enskip{}\enskip{}}\textcolor[HTML]{008080}{Int}\textcolor[HTML]{C000C0}{:}\\
\textcolor[HTML]{000000}{\enskip{}\enskip{}\enskip{}\enskip{}}\textcolor[HTML]{AF5F00}{-\enskip{}}\textcolor[HTML]{000000}{contents}\\
\textcolor[HTML]{000000}{\enskip{}\enskip{}\enskip{}\enskip{}}\textcolor[HTML]{AF5F00}{-\enskip{}}\textcolor[HTML]{C00000}{1}\\
\textcolor[HTML]{000000}{\enskip{}\enskip{}\enskip{}\enskip{}}\textcolor[HTML]{AF5F00}{-\enskip{}}\textcolor[HTML]{000000}{(+}\textcolor[HTML]{000000}{\enskip{}}\textcolor[HTML]{000000}{contents}\textcolor[HTML]{000000}{\enskip{}}\textcolor[HTML]{000000}{1)}\\
\textcolor[HTML]{AF5F00}{-\enskip{}}\textcolor[HTML]{008080}{args}\textcolor[HTML]{C000C0}{:}\textcolor[HTML]{000000}{\enskip{}}\textcolor[HTML]{C000C0}{\char91{}\char93{}}\\
\textcolor[HTML]{000000}{\enskip{}\enskip{}}\textcolor[HTML]{008080}{ensures}\textcolor[HTML]{C000C0}{:}\textcolor[HTML]{000000}{\enskip{}}\textcolor[HTML]{C00000}{"}\textcolor[HTML]{C00000}{(or\enskip{}(and\enskip{}err\enskip{}err\char95{}new)}\textcolor[HTML]{C000C0}{\char92{}n}\textcolor[HTML]{C00000}{\enskip{}\enskip{}\enskip{}\enskip{}(and\enskip{}(not\enskip{}err)\enskip{}(not\enskip{}err\char95{}new)\enskip{}(>=\enskip{}contents\enskip{}1)}\textcolor[HTML]{C000C0}{\char92{}n\char92{}}\\
\textcolor[HTML]{C00000}{\enskip{}\enskip{}\enskip{}\enskip{}}\textcolor[HTML]{C000C0}{\char92{}\enskip{}}\textcolor[HTML]{C00000}{(and\enskip{}(=\enskip{}contents\char95{}new\enskip{}(-\enskip{}contents\enskip{}1))}\textcolor[HTML]{C000C0}{\char92{}n}\textcolor[HTML]{C00000}{\enskip{}\enskip{}\enskip{}\enskip{}\enskip{}(=\enskip{}result\enskip{}true))}\textcolor[HTML]{C000C0}{\char92{}n}\textcolor[HTML]{C00000}{)}\textcolor[HTML]{C000C0}{\char92{}n}\textcolor[HTML]{C00000}{\enskip{}\enskip{}\enskip{}\enskip{}(and\enskip{}(not}\textcolor[HTML]{C000C0}{\char92{}}\\
\textcolor[HTML]{C00000}{\enskip{}\enskip{}\enskip{}\enskip{}}\textcolor[HTML]{C000C0}{\char92{}\enskip{}}\textcolor[HTML]{C00000}{err)\enskip{}err\char95{}new\enskip{}(not\enskip{}(>=\enskip{}contents\enskip{}1)}\textcolor[HTML]{C000C0}{\char92{}n}\textcolor[HTML]{C00000}{)))}\textcolor[HTML]{C00000}{"}\\
\textcolor[HTML]{000000}{\enskip{}\enskip{}}\textcolor[HTML]{008080}{name}\textcolor[HTML]{C000C0}{:}\textcolor[HTML]{000000}{\enskip{}}\textcolor[HTML]{000000}{decrement}\\
\textcolor[HTML]{000000}{\enskip{}\enskip{}}\textcolor[HTML]{008080}{requires}\textcolor[HTML]{C000C0}{:}\textcolor[HTML]{000000}{\enskip{}}\textcolor[HTML]{C00000}{'}\textcolor[HTML]{C00000}{true}\textcolor[HTML]{C00000}{'}\\
\textcolor[HTML]{000000}{\enskip{}\enskip{}}\textcolor[HTML]{008080}{return}\textcolor[HTML]{C000C0}{:}\\
\textcolor[HTML]{000000}{\enskip{}\enskip{}}\textcolor[HTML]{AF5F00}{-\enskip{}}\textcolor[HTML]{008080}{name}\textcolor[HTML]{C000C0}{:}\textcolor[HTML]{000000}{\enskip{}}\textcolor[HTML]{000000}{result}\\
\textcolor[HTML]{000000}{\enskip{}\enskip{}\enskip{}\enskip{}}\textcolor[HTML]{008080}{type}\textcolor[HTML]{C000C0}{:}\textcolor[HTML]{000000}{\enskip{}}\textcolor[HTML]{000000}{Bool}\\
\textcolor[HTML]{000000}{\enskip{}\enskip{}}\textcolor[HTML]{008080}{terms}\textcolor[HTML]{C000C0}{:}\\
\textcolor[HTML]{000000}{\enskip{}\enskip{}\enskip{}\enskip{}}\textcolor[HTML]{008080}{Int}\textcolor[HTML]{C000C0}{:}\\
\textcolor[HTML]{000000}{\enskip{}\enskip{}\enskip{}\enskip{}}\textcolor[HTML]{AF5F00}{-\enskip{}}\textcolor[HTML]{000000}{contents}\\
\textcolor[HTML]{000000}{\enskip{}\enskip{}\enskip{}\enskip{}}\textcolor[HTML]{AF5F00}{-\enskip{}}\textcolor[HTML]{C00000}{1}\\
\textcolor[HTML]{000000}{\enskip{}\enskip{}\enskip{}\enskip{}}\textcolor[HTML]{AF5F00}{-\enskip{}}\textcolor[HTML]{000000}{(-}\textcolor[HTML]{000000}{\enskip{}}\textcolor[HTML]{000000}{contents}\textcolor[HTML]{000000}{\enskip{}}\textcolor[HTML]{000000}{1)}\\
\textcolor[HTML]{000000}{\enskip{}\enskip{}\enskip{}\enskip{}}\textcolor[HTML]{AF5F00}{-\enskip{}}\textcolor[HTML]{C00000}{0}\\
\textcolor[HTML]{AF5F00}{-\enskip{}}\textcolor[HTML]{008080}{args}\textcolor[HTML]{C000C0}{:}\textcolor[HTML]{000000}{\enskip{}}\textcolor[HTML]{C000C0}{\char91{}\char93{}}\\
\textcolor[HTML]{000000}{\enskip{}\enskip{}}\textcolor[HTML]{008080}{ensures}\textcolor[HTML]{C000C0}{:}\textcolor[HTML]{000000}{\enskip{}}\textcolor[HTML]{C00000}{"}\textcolor[HTML]{C00000}{(or\enskip{}(and\enskip{}err\enskip{}err\char95{}new)}\textcolor[HTML]{C000C0}{\char92{}n}\textcolor[HTML]{C00000}{\enskip{}\enskip{}\enskip{}\enskip{}(and\enskip{}(not\enskip{}err)\enskip{}(not\enskip{}err\char95{}new)\enskip{}(>=\enskip{}contents\enskip{}0)}\textcolor[HTML]{C000C0}{\char92{}n\char92{}}\\
\textcolor[HTML]{C00000}{\enskip{}\enskip{}\enskip{}\enskip{}}\textcolor[HTML]{C000C0}{\char92{}\enskip{}}\textcolor[HTML]{C00000}{(and\enskip{}(=\enskip{}contents\char95{}new\enskip{}0)}\textcolor[HTML]{C000C0}{\char92{}n}\textcolor[HTML]{C00000}{\enskip{}\enskip{}\enskip{}\enskip{}\enskip{}(=\enskip{}result\enskip{}true))}\textcolor[HTML]{C000C0}{\char92{}n}\textcolor[HTML]{C00000}{)}\textcolor[HTML]{C000C0}{\char92{}n}\textcolor[HTML]{C00000}{\enskip{}\enskip{}\enskip{}\enskip{}(and\enskip{}(not\enskip{}err)\enskip{}err\char95{}new}\textcolor[HTML]{C000C0}{\char92{}}\\
\textcolor[HTML]{C00000}{\enskip{}\enskip{}\enskip{}\enskip{}}\textcolor[HTML]{C000C0}{\char92{}\enskip{}}\textcolor[HTML]{C00000}{(not\enskip{}(>=\enskip{}contents\enskip{}0)}\textcolor[HTML]{C000C0}{\char92{}n}\textcolor[HTML]{C00000}{)))}\textcolor[HTML]{C00000}{"}\\
\textcolor[HTML]{000000}{\enskip{}\enskip{}}\textcolor[HTML]{008080}{name}\textcolor[HTML]{C000C0}{:}\textcolor[HTML]{000000}{\enskip{}}\textcolor[HTML]{000000}{reset}\\
\textcolor[HTML]{000000}{\enskip{}\enskip{}}\textcolor[HTML]{008080}{requires}\textcolor[HTML]{C000C0}{:}\textcolor[HTML]{000000}{\enskip{}}\textcolor[HTML]{C00000}{'}\textcolor[HTML]{C00000}{true}\textcolor[HTML]{C00000}{'}\\
\textcolor[HTML]{000000}{\enskip{}\enskip{}}\textcolor[HTML]{008080}{return}\textcolor[HTML]{C000C0}{:}\\
\textcolor[HTML]{000000}{\enskip{}\enskip{}}\textcolor[HTML]{AF5F00}{-\enskip{}}\textcolor[HTML]{008080}{name}\textcolor[HTML]{C000C0}{:}\textcolor[HTML]{000000}{\enskip{}}\textcolor[HTML]{000000}{result}\\
\textcolor[HTML]{000000}{\enskip{}\enskip{}\enskip{}\enskip{}}\textcolor[HTML]{008080}{type}\textcolor[HTML]{C000C0}{:}\textcolor[HTML]{000000}{\enskip{}}\textcolor[HTML]{000000}{Bool}\\
\textcolor[HTML]{000000}{\enskip{}\enskip{}}\textcolor[HTML]{008080}{terms}\textcolor[HTML]{C000C0}{:}\\
\textcolor[HTML]{000000}{\enskip{}\enskip{}\enskip{}\enskip{}}\textcolor[HTML]{008080}{Int}\textcolor[HTML]{C000C0}{:}\\
\textcolor[HTML]{000000}{\enskip{}\enskip{}\enskip{}\enskip{}}\textcolor[HTML]{AF5F00}{-\enskip{}}\textcolor[HTML]{000000}{contents}\\
\textcolor[HTML]{000000}{\enskip{}\enskip{}\enskip{}\enskip{}}\textcolor[HTML]{AF5F00}{-\enskip{}}\textcolor[HTML]{C00000}{0}\\
\textcolor[HTML]{AF5F00}{-\enskip{}}\textcolor[HTML]{008080}{args}\textcolor[HTML]{C000C0}{:}\textcolor[HTML]{000000}{\enskip{}}\textcolor[HTML]{C000C0}{\char91{}\char93{}}\\
\textcolor[HTML]{000000}{\enskip{}\enskip{}}\textcolor[HTML]{008080}{ensures}\textcolor[HTML]{C000C0}{:}\textcolor[HTML]{000000}{\enskip{}}\textcolor[HTML]{C00000}{"}\textcolor[HTML]{C00000}{(or\enskip{}(and\enskip{}err\enskip{}err\char95{}new)}\textcolor[HTML]{C000C0}{\char92{}n}\textcolor[HTML]{C00000}{\enskip{}\enskip{}\enskip{}\enskip{}(and\enskip{}(not\enskip{}err)\enskip{}(not\enskip{}err\char95{}new)\enskip{}(>=\enskip{}contents\enskip{}0)}\textcolor[HTML]{C000C0}{\char92{}n\char92{}}\\
\textcolor[HTML]{C00000}{\enskip{}\enskip{}\enskip{}\enskip{}}\textcolor[HTML]{C000C0}{\char92{}\enskip{}}\textcolor[HTML]{C00000}{(and\enskip{}(=\enskip{}contents\char95{}new\enskip{}contents)}\textcolor[HTML]{C000C0}{\char92{}n}\textcolor[HTML]{C00000}{\enskip{}\enskip{}\enskip{}\enskip{}\enskip{}(=\enskip{}result\enskip{}(=\enskip{}contents\enskip{}0)))}\textcolor[HTML]{C000C0}{\char92{}n}\textcolor[HTML]{C00000}{)}\textcolor[HTML]{C000C0}{\char92{}n}\textcolor[HTML]{C00000}{\enskip{}\enskip{}\enskip{}\enskip{}(and}\textcolor[HTML]{C000C0}{\char92{}}\\
\textcolor[HTML]{C00000}{\enskip{}\enskip{}\enskip{}\enskip{}}\textcolor[HTML]{C000C0}{\char92{}\enskip{}}\textcolor[HTML]{C00000}{(not\enskip{}err)\enskip{}err\char95{}new\enskip{}(not\enskip{}(>=\enskip{}contents\enskip{}0)}\textcolor[HTML]{C000C0}{\char92{}n}\textcolor[HTML]{C00000}{)))}\textcolor[HTML]{C00000}{"}\\
\textcolor[HTML]{000000}{\enskip{}\enskip{}}\textcolor[HTML]{008080}{name}\textcolor[HTML]{C000C0}{:}\textcolor[HTML]{000000}{\enskip{}}\textcolor[HTML]{000000}{zero}\\
\textcolor[HTML]{000000}{\enskip{}\enskip{}}\textcolor[HTML]{008080}{requires}\textcolor[HTML]{C000C0}{:}\textcolor[HTML]{000000}{\enskip{}}\textcolor[HTML]{C00000}{'}\textcolor[HTML]{C00000}{true}\textcolor[HTML]{C00000}{'}\\
\textcolor[HTML]{000000}{\enskip{}\enskip{}}\textcolor[HTML]{008080}{return}\textcolor[HTML]{C000C0}{:}\\
\textcolor[HTML]{000000}{\enskip{}\enskip{}}\textcolor[HTML]{AF5F00}{-\enskip{}}\textcolor[HTML]{008080}{name}\textcolor[HTML]{C000C0}{:}\textcolor[HTML]{000000}{\enskip{}}\textcolor[HTML]{000000}{result}\\
\textcolor[HTML]{000000}{\enskip{}\enskip{}\enskip{}\enskip{}}\textcolor[HTML]{008080}{type}\textcolor[HTML]{C000C0}{:}\textcolor[HTML]{000000}{\enskip{}}\textcolor[HTML]{000000}{Bool}\\
\textcolor[HTML]{000000}{\enskip{}\enskip{}}\textcolor[HTML]{008080}{terms}\textcolor[HTML]{C000C0}{:}\\
\textcolor[HTML]{000000}{\enskip{}\enskip{}\enskip{}\enskip{}}\textcolor[HTML]{008080}{Int}\textcolor[HTML]{C000C0}{:}\\
\textcolor[HTML]{000000}{\enskip{}\enskip{}\enskip{}\enskip{}}\textcolor[HTML]{AF5F00}{-\enskip{}}\textcolor[HTML]{000000}{contents}\\
\textcolor[HTML]{000000}{\enskip{}\enskip{}\enskip{}\enskip{}}\textcolor[HTML]{AF5F00}{-\enskip{}}\textcolor[HTML]{C00000}{0}\\
\textcolor[HTML]{008080}{name}\textcolor[HTML]{C000C0}{:}\textcolor[HTML]{000000}{\enskip{}}\textcolor[HTML]{000000}{counter}\\
\textcolor[HTML]{008080}{predicates}\textcolor[HTML]{C000C0}{:}\\
\textcolor[HTML]{AF5F00}{-\enskip{}}\textcolor[HTML]{008080}{name}\textcolor[HTML]{C000C0}{:}\textcolor[HTML]{000000}{\enskip{}}\textcolor[HTML]{C00000}{'}\textcolor[HTML]{C00000}{=}\textcolor[HTML]{C00000}{'}\\
\textcolor[HTML]{000000}{\enskip{}\enskip{}}\textcolor[HTML]{008080}{type}\textcolor[HTML]{C000C0}{:}\\
\textcolor[HTML]{000000}{\enskip{}\enskip{}}\textcolor[HTML]{AF5F00}{-\enskip{}}\textcolor[HTML]{000000}{Int}\\
\textcolor[HTML]{000000}{\enskip{}\enskip{}}\textcolor[HTML]{AF5F00}{-\enskip{}}\textcolor[HTML]{000000}{Int}\\
\textcolor[HTML]{008080}{state}\textcolor[HTML]{C000C0}{:}\\
\textcolor[HTML]{AF5F00}{-\enskip{}}\textcolor[HTML]{008080}{name}\textcolor[HTML]{C000C0}{:}\textcolor[HTML]{000000}{\enskip{}}\textcolor[HTML]{000000}{contents}\\
\textcolor[HTML]{000000}{\enskip{}\enskip{}}\textcolor[HTML]{008080}{type}\textcolor[HTML]{C000C0}{:}\textcolor[HTML]{000000}{\enskip{}}\textcolor[HTML]{000000}{Int}\\
\textcolor[HTML]{AF5F00}{-\enskip{}}\textcolor[HTML]{008080}{name}\textcolor[HTML]{C000C0}{:}\textcolor[HTML]{000000}{\enskip{}}\textcolor[HTML]{000000}{err}\\
\textcolor[HTML]{000000}{\enskip{}\enskip{}}\textcolor[HTML]{008080}{type}\textcolor[HTML]{C000C0}{:}\textcolor[HTML]{000000}{\enskip{}}\textcolor[HTML]{000000}{Bool}\\
\textcolor[HTML]{008080}{states\char95{}equal}\textcolor[HTML]{C000C0}{:}\\
\textcolor[HTML]{000000}{\enskip{}\enskip{}}\textcolor[HTML]{008080}{definition}\textcolor[HTML]{C000C0}{:}\textcolor[HTML]{000000}{\enskip{}}\textcolor[HTML]{C00000}{'}\textcolor[HTML]{C00000}{(or\enskip{}(and\enskip{}err\char95{}1\enskip{}err\char95{}2)\enskip{}(and\enskip{}(not\enskip{}err\char95{}1)\enskip{}(not\enskip{}err\char95{}2)}\\
\\
\textcolor[HTML]{000000}{\enskip{}\enskip{}\enskip{}\enskip{}}\textcolor[HTML]{000000}{(=}\textcolor[HTML]{000000}{\enskip{}}\textcolor[HTML]{000000}{contents\char95{}1}\textcolor[HTML]{000000}{\enskip{}}\textcolor[HTML]{000000}{contents\char95{}2)}\\
\\
\textcolor[HTML]{000000}{\enskip{}\enskip{}\enskip{}\enskip{}}\textcolor[HTML]{000000}{))'}\\
\\

}

\subsection{Accumulator}
\scriptsize
\label{yml:accumulator}
{\ttfamily\noindent
\\
\textcolor[HTML]{0000C0}{\char35{}\enskip{}Accumulator\enskip{}abstract\enskip{}definition}\\
\\
\textcolor[HTML]{008080}{name}\textcolor[HTML]{C000C0}{:}\textcolor[HTML]{000000}{\enskip{}}\textcolor[HTML]{000000}{accumulator}\\
\\
\textcolor[HTML]{008080}{state}\textcolor[HTML]{C000C0}{:}\\
\textcolor[HTML]{000000}{\enskip{}\enskip{}}\textcolor[HTML]{AF5F00}{-\enskip{}}\textcolor[HTML]{008080}{name}\textcolor[HTML]{C000C0}{:}\textcolor[HTML]{000000}{\enskip{}}\textcolor[HTML]{000000}{contents}\\
\textcolor[HTML]{000000}{\enskip{}\enskip{}\enskip{}\enskip{}}\textcolor[HTML]{008080}{type}\textcolor[HTML]{C000C0}{:}\textcolor[HTML]{000000}{\enskip{}}\textcolor[HTML]{000000}{Int}\\
\\
\textcolor[HTML]{008080}{options}\textcolor[HTML]{C000C0}{:}\\
\textcolor[HTML]{000000}{\enskip{}\enskip{}}\\
\\
\textcolor[HTML]{008080}{states\char95{}equal}\textcolor[HTML]{C000C0}{:}\\
\textcolor[HTML]{000000}{\enskip{}\enskip{}}\textcolor[HTML]{008080}{definition}\textcolor[HTML]{C000C0}{:}\textcolor[HTML]{000000}{\enskip{}}\textcolor[HTML]{000000}{(=}\textcolor[HTML]{000000}{\enskip{}}\textcolor[HTML]{000000}{contents\char95{}1}\textcolor[HTML]{000000}{\enskip{}}\textcolor[HTML]{000000}{contents\char95{}2)}\\
\\
\textcolor[HTML]{008080}{methods}\textcolor[HTML]{C000C0}{:}\\
\textcolor[HTML]{000000}{\enskip{}\enskip{}}\textcolor[HTML]{AF5F00}{-\enskip{}}\textcolor[HTML]{008080}{name}\textcolor[HTML]{C000C0}{:}\textcolor[HTML]{000000}{\enskip{}}\textcolor[HTML]{000000}{increase}\\
\textcolor[HTML]{000000}{\enskip{}\enskip{}\enskip{}\enskip{}}\textcolor[HTML]{008080}{args}\textcolor[HTML]{C000C0}{:}\\
\textcolor[HTML]{000000}{\enskip{}\enskip{}\enskip{}\enskip{}\enskip{}\enskip{}}\textcolor[HTML]{AF5F00}{-\enskip{}}\textcolor[HTML]{008080}{name}\textcolor[HTML]{C000C0}{:}\textcolor[HTML]{000000}{\enskip{}}\textcolor[HTML]{000000}{n}\\
\textcolor[HTML]{000000}{\enskip{}\enskip{}\enskip{}\enskip{}\enskip{}\enskip{}\enskip{}\enskip{}}\textcolor[HTML]{008080}{type}\textcolor[HTML]{C000C0}{:}\textcolor[HTML]{000000}{\enskip{}}\textcolor[HTML]{000000}{Int}\\
\textcolor[HTML]{000000}{\enskip{}\enskip{}\enskip{}\enskip{}}\textcolor[HTML]{008080}{return}\textcolor[HTML]{C000C0}{:}\\
\textcolor[HTML]{000000}{\enskip{}\enskip{}\enskip{}\enskip{}\enskip{}\enskip{}}\textcolor[HTML]{AF5F00}{-\enskip{}}\textcolor[HTML]{008080}{name}\textcolor[HTML]{C000C0}{:}\textcolor[HTML]{000000}{\enskip{}}\textcolor[HTML]{000000}{result}\\
\textcolor[HTML]{000000}{\enskip{}\enskip{}\enskip{}\enskip{}\enskip{}\enskip{}\enskip{}\enskip{}}\textcolor[HTML]{008080}{type}\textcolor[HTML]{C000C0}{:}\textcolor[HTML]{000000}{\enskip{}}\textcolor[HTML]{000000}{Bool}\\
\textcolor[HTML]{000000}{\enskip{}\enskip{}\enskip{}\enskip{}}\textcolor[HTML]{008080}{requires}\textcolor[HTML]{C000C0}{:}\textcolor[HTML]{000000}{\enskip{}|}\\
\textcolor[HTML]{000000}{\enskip{}\enskip{}\enskip{}\enskip{}\enskip{}\enskip{}}\textcolor[HTML]{C00000}{true}\\
\textcolor[HTML]{000000}{\enskip{}\enskip{}\enskip{}\enskip{}}\textcolor[HTML]{008080}{ensures}\textcolor[HTML]{C000C0}{:}\textcolor[HTML]{000000}{\enskip{}|}\\
\textcolor[HTML]{000000}{\enskip{}\enskip{}\enskip{}\enskip{}\enskip{}\enskip{}}\textcolor[HTML]{000000}{(and}\textcolor[HTML]{000000}{\enskip{}}\textcolor[HTML]{000000}{(=}\textcolor[HTML]{000000}{\enskip{}}\textcolor[HTML]{000000}{contents\char95{}new}\textcolor[HTML]{000000}{\enskip{}}\textcolor[HTML]{000000}{(+}\textcolor[HTML]{000000}{\enskip{}}\textcolor[HTML]{000000}{contents}\textcolor[HTML]{000000}{\enskip{}}\textcolor[HTML]{000000}{n))}\\
\textcolor[HTML]{000000}{\enskip{}\enskip{}\enskip{}\enskip{}\enskip{}\enskip{}\enskip{}\enskip{}\enskip{}\enskip{}\enskip{}}\textcolor[HTML]{000000}{(=}\textcolor[HTML]{000000}{\enskip{}}\textcolor[HTML]{000000}{result}\textcolor[HTML]{000000}{\enskip{}}\textcolor[HTML]{C00000}{true}\textcolor[HTML]{000000}{))}\\
\textcolor[HTML]{000000}{\enskip{}\enskip{}\enskip{}\enskip{}}\textcolor[HTML]{008080}{terms}\textcolor[HTML]{C000C0}{:}\\
\textcolor[HTML]{000000}{\enskip{}\enskip{}\enskip{}\enskip{}\enskip{}\enskip{}}\textcolor[HTML]{008080}{Int}\textcolor[HTML]{C000C0}{:}\textcolor[HTML]{000000}{\enskip{}}\textcolor[HTML]{C000C0}{\char91{}}\textcolor[HTML]{000000}{\char36{}1}\textcolor[HTML]{000000}{,\enskip{}}\textcolor[HTML]{000000}{contents}\textcolor[HTML]{000000}{,\enskip{}}\textcolor[HTML]{000000}{(+}\textcolor[HTML]{000000}{\enskip{}}\textcolor[HTML]{000000}{contents}\textcolor[HTML]{000000}{\enskip{}}\textcolor[HTML]{000000}{\char36{}1)}\textcolor[HTML]{C000C0}{\char93{}}\\
\textcolor[HTML]{000000}{\enskip{}\enskip{}}\textcolor[HTML]{AF5F00}{-\enskip{}}\textcolor[HTML]{008080}{name}\textcolor[HTML]{C000C0}{:}\textcolor[HTML]{000000}{\enskip{}}\textcolor[HTML]{000000}{read}\\
\textcolor[HTML]{000000}{\enskip{}\enskip{}\enskip{}\enskip{}}\textcolor[HTML]{008080}{args}\textcolor[HTML]{C000C0}{:}\textcolor[HTML]{000000}{\enskip{}}\textcolor[HTML]{C000C0}{\char91{}\char93{}}\\
\textcolor[HTML]{000000}{\enskip{}\enskip{}\enskip{}\enskip{}}\textcolor[HTML]{008080}{return}\textcolor[HTML]{C000C0}{:}\\
\textcolor[HTML]{000000}{\enskip{}\enskip{}\enskip{}\enskip{}\enskip{}\enskip{}}\textcolor[HTML]{AF5F00}{-\enskip{}}\textcolor[HTML]{008080}{name}\textcolor[HTML]{C000C0}{:}\textcolor[HTML]{000000}{\enskip{}}\textcolor[HTML]{000000}{result}\\
\textcolor[HTML]{000000}{\enskip{}\enskip{}\enskip{}\enskip{}\enskip{}\enskip{}\enskip{}\enskip{}}\textcolor[HTML]{008080}{type}\textcolor[HTML]{C000C0}{:}\textcolor[HTML]{000000}{\enskip{}}\textcolor[HTML]{000000}{Int}\\
\textcolor[HTML]{000000}{\enskip{}\enskip{}\enskip{}\enskip{}}\textcolor[HTML]{008080}{requires}\textcolor[HTML]{C000C0}{:}\textcolor[HTML]{000000}{\enskip{}|}\\
\textcolor[HTML]{000000}{\enskip{}\enskip{}\enskip{}\enskip{}\enskip{}\enskip{}}\textcolor[HTML]{C00000}{true}\\
\textcolor[HTML]{000000}{\enskip{}\enskip{}\enskip{}\enskip{}}\textcolor[HTML]{008080}{ensures}\textcolor[HTML]{C000C0}{:}\textcolor[HTML]{000000}{\enskip{}|}\\
\textcolor[HTML]{000000}{\enskip{}\enskip{}\enskip{}\enskip{}\enskip{}\enskip{}}\textcolor[HTML]{000000}{(and}\textcolor[HTML]{000000}{\enskip{}}\textcolor[HTML]{000000}{(=}\textcolor[HTML]{000000}{\enskip{}}\textcolor[HTML]{000000}{contents\char95{}new}\textcolor[HTML]{000000}{\enskip{}}\textcolor[HTML]{000000}{contents)}\\
\textcolor[HTML]{000000}{\enskip{}\enskip{}\enskip{}\enskip{}\enskip{}\enskip{}\enskip{}\enskip{}\enskip{}\enskip{}\enskip{}}\textcolor[HTML]{000000}{(=}\textcolor[HTML]{000000}{\enskip{}}\textcolor[HTML]{000000}{result}\textcolor[HTML]{000000}{\enskip{}}\textcolor[HTML]{000000}{contents))}\\
\textcolor[HTML]{000000}{\enskip{}\enskip{}\enskip{}\enskip{}}\textcolor[HTML]{008080}{terms}\textcolor[HTML]{C000C0}{:}\\
\textcolor[HTML]{000000}{\enskip{}\enskip{}\enskip{}\enskip{}\enskip{}\enskip{}}\textcolor[HTML]{008080}{Int}\textcolor[HTML]{C000C0}{:}\textcolor[HTML]{000000}{\enskip{}}\textcolor[HTML]{C000C0}{\char91{}}\textcolor[HTML]{000000}{contents}\textcolor[HTML]{C000C0}{\char93{}}\\
\\
\textcolor[HTML]{008080}{predicates}\textcolor[HTML]{C000C0}{:}\\
\textcolor[HTML]{000000}{\enskip{}\enskip{}}\textcolor[HTML]{AF5F00}{-\enskip{}}\textcolor[HTML]{008080}{name}\textcolor[HTML]{C000C0}{:}\textcolor[HTML]{000000}{\enskip{}}\textcolor[HTML]{C00000}{"}\textcolor[HTML]{C00000}{=}\textcolor[HTML]{C00000}{"}\\
\textcolor[HTML]{000000}{\enskip{}\enskip{}\enskip{}\enskip{}}\textcolor[HTML]{008080}{type}\textcolor[HTML]{C000C0}{:}\textcolor[HTML]{000000}{\enskip{}}\textcolor[HTML]{C000C0}{\char91{}}\textcolor[HTML]{000000}{Int}\textcolor[HTML]{000000}{,\enskip{}}\textcolor[HTML]{000000}{Int}\textcolor[HTML]{C000C0}{\char93{}}\\

}

\begin{itemize}
\item \CCmethod{increase} \CCbowtie\ \CCmethod{increase}

 Simple:

true

 Poke:

true

\item \CCmethod{increase} \CCbowtie\ \CCmethod{read}

 Simple:

[\CCequal{\CCvar{x1}}{\CCvar{contents}} $\wedge$ \CCequal{\CCplus{\CCvar{contents}}{\CCvar{x1}}}{\CCvar{contents}}]

 $\vee$ [\CCnot{\CCequal{\CCvar{x1}}{\CCvar{contents}}} $\wedge$ \CCequal{\CCplus{\CCvar{contents}}{\CCvar{x1}}}{\CCvar{contents}}]

 Poke:

\CCequal{\CCplus{\CCvar{contents}}{\CCvar{x1}}}{\CCvar{contents}}

\item \CCmethod{read} \CCbowtie\ \CCmethod{read}

 Simple:

true

 Poke:

true

\end{itemize}

\subsection{Set}
\scriptsize
\label{yml:set}
{\ttfamily\noindent
\textcolor[HTML]{008080}{name}\textcolor[HTML]{C000C0}{:}\textcolor[HTML]{000000}{\enskip{}}\textcolor[HTML]{000000}{set}\\
\\
\textcolor[HTML]{008080}{preamble}\textcolor[HTML]{C000C0}{:}\textcolor[HTML]{000000}{\enskip{}|}\\
\textcolor[HTML]{000000}{\enskip{}\enskip{}}\textcolor[HTML]{000000}{(declare-sort}\textcolor[HTML]{000000}{\enskip{}}\textcolor[HTML]{000000}{E}\textcolor[HTML]{000000}{\enskip{}}\textcolor[HTML]{000000}{0)}\\
\\
\textcolor[HTML]{008080}{state}\textcolor[HTML]{C000C0}{:}\\
\textcolor[HTML]{000000}{\enskip{}\enskip{}}\textcolor[HTML]{AF5F00}{-\enskip{}}\textcolor[HTML]{008080}{name}\textcolor[HTML]{C000C0}{:}\textcolor[HTML]{000000}{\enskip{}}\textcolor[HTML]{000000}{S}\\
\textcolor[HTML]{000000}{\enskip{}\enskip{}\enskip{}\enskip{}}\textcolor[HTML]{008080}{type}\textcolor[HTML]{C000C0}{:}\textcolor[HTML]{000000}{\enskip{}}\textcolor[HTML]{000000}{(Set}\textcolor[HTML]{000000}{\enskip{}}\textcolor[HTML]{000000}{E)}\\
\textcolor[HTML]{000000}{\enskip{}\enskip{}}\textcolor[HTML]{AF5F00}{-\enskip{}}\textcolor[HTML]{008080}{name}\textcolor[HTML]{C000C0}{:}\textcolor[HTML]{000000}{\enskip{}}\textcolor[HTML]{000000}{size}\\
\textcolor[HTML]{000000}{\enskip{}\enskip{}\enskip{}\enskip{}}\textcolor[HTML]{008080}{type}\textcolor[HTML]{C000C0}{:}\textcolor[HTML]{000000}{\enskip{}}\textcolor[HTML]{000000}{Int}\\
\\
\textcolor[HTML]{008080}{states\char95{}equal}\textcolor[HTML]{C000C0}{:}\\
\textcolor[HTML]{000000}{\enskip{}\enskip{}}\textcolor[HTML]{008080}{definition}\textcolor[HTML]{C000C0}{:}\textcolor[HTML]{000000}{\enskip{}}\textcolor[HTML]{000000}{(and}\textcolor[HTML]{000000}{\enskip{}}\textcolor[HTML]{000000}{(=}\textcolor[HTML]{000000}{\enskip{}}\textcolor[HTML]{000000}{S\char95{}1}\textcolor[HTML]{000000}{\enskip{}}\textcolor[HTML]{000000}{S\char95{}2)}\textcolor[HTML]{000000}{\enskip{}}\textcolor[HTML]{000000}{(=}\textcolor[HTML]{000000}{\enskip{}}\textcolor[HTML]{000000}{size\char95{}1}\textcolor[HTML]{000000}{\enskip{}}\textcolor[HTML]{000000}{size\char95{}2))}\\
\\
\textcolor[HTML]{008080}{methods}\textcolor[HTML]{C000C0}{:}\\
\textcolor[HTML]{000000}{\enskip{}\enskip{}}\textcolor[HTML]{AF5F00}{-\enskip{}}\textcolor[HTML]{008080}{name}\textcolor[HTML]{C000C0}{:}\textcolor[HTML]{000000}{\enskip{}}\textcolor[HTML]{000000}{add}\\
\textcolor[HTML]{000000}{\enskip{}\enskip{}\enskip{}\enskip{}}\textcolor[HTML]{008080}{args}\textcolor[HTML]{C000C0}{:}\\
\textcolor[HTML]{000000}{\enskip{}\enskip{}\enskip{}\enskip{}\enskip{}\enskip{}}\textcolor[HTML]{AF5F00}{-\enskip{}}\textcolor[HTML]{008080}{name}\textcolor[HTML]{C000C0}{:}\textcolor[HTML]{000000}{\enskip{}}\textcolor[HTML]{000000}{v}\\
\textcolor[HTML]{000000}{\enskip{}\enskip{}\enskip{}\enskip{}\enskip{}\enskip{}\enskip{}\enskip{}}\textcolor[HTML]{008080}{type}\textcolor[HTML]{C000C0}{:}\textcolor[HTML]{000000}{\enskip{}}\textcolor[HTML]{000000}{E}\\
\textcolor[HTML]{000000}{\enskip{}\enskip{}\enskip{}\enskip{}}\textcolor[HTML]{008080}{return}\textcolor[HTML]{C000C0}{:}\\
\textcolor[HTML]{000000}{\enskip{}\enskip{}\enskip{}\enskip{}\enskip{}\enskip{}}\textcolor[HTML]{AF5F00}{-\enskip{}}\textcolor[HTML]{008080}{name}\textcolor[HTML]{C000C0}{:}\textcolor[HTML]{000000}{\enskip{}}\textcolor[HTML]{000000}{result}\\
\textcolor[HTML]{000000}{\enskip{}\enskip{}\enskip{}\enskip{}\enskip{}\enskip{}\enskip{}\enskip{}}\textcolor[HTML]{008080}{type}\textcolor[HTML]{C000C0}{:}\textcolor[HTML]{000000}{\enskip{}}\textcolor[HTML]{000000}{Bool}\\
\textcolor[HTML]{000000}{\enskip{}\enskip{}\enskip{}\enskip{}}\textcolor[HTML]{008080}{requires}\textcolor[HTML]{C000C0}{:}\textcolor[HTML]{000000}{\enskip{}|}\\
\textcolor[HTML]{000000}{\enskip{}\enskip{}\enskip{}\enskip{}\enskip{}\enskip{}}\textcolor[HTML]{C00000}{true}\\
\textcolor[HTML]{000000}{\enskip{}\enskip{}\enskip{}\enskip{}}\textcolor[HTML]{008080}{ensures}\textcolor[HTML]{C000C0}{:}\textcolor[HTML]{000000}{\enskip{}|}\\
\textcolor[HTML]{000000}{\enskip{}\enskip{}\enskip{}\enskip{}\enskip{}\enskip{}}\textcolor[HTML]{000000}{(ite}\textcolor[HTML]{000000}{\enskip{}}\textcolor[HTML]{000000}{(member}\textcolor[HTML]{000000}{\enskip{}}\textcolor[HTML]{000000}{v}\textcolor[HTML]{000000}{\enskip{}}\textcolor[HTML]{000000}{S)}\\
\textcolor[HTML]{000000}{\enskip{}\enskip{}\enskip{}\enskip{}\enskip{}\enskip{}\enskip{}\enskip{}\enskip{}\enskip{}\enskip{}}\textcolor[HTML]{000000}{(and}\textcolor[HTML]{000000}{\enskip{}}\textcolor[HTML]{000000}{(=}\textcolor[HTML]{000000}{\enskip{}}\textcolor[HTML]{000000}{S\char95{}new}\textcolor[HTML]{000000}{\enskip{}}\textcolor[HTML]{000000}{S)}\\
\textcolor[HTML]{000000}{\enskip{}\enskip{}\enskip{}\enskip{}\enskip{}\enskip{}\enskip{}\enskip{}\enskip{}\enskip{}\enskip{}\enskip{}\enskip{}\enskip{}\enskip{}\enskip{}}\textcolor[HTML]{000000}{(=}\textcolor[HTML]{000000}{\enskip{}}\textcolor[HTML]{000000}{size\char95{}new}\textcolor[HTML]{000000}{\enskip{}}\textcolor[HTML]{000000}{size)}\\
\textcolor[HTML]{000000}{\enskip{}\enskip{}\enskip{}\enskip{}\enskip{}\enskip{}\enskip{}\enskip{}\enskip{}\enskip{}\enskip{}\enskip{}\enskip{}\enskip{}\enskip{}\enskip{}}\textcolor[HTML]{000000}{(not}\textcolor[HTML]{000000}{\enskip{}}\textcolor[HTML]{000000}{result))}\\
\textcolor[HTML]{000000}{\enskip{}\enskip{}\enskip{}\enskip{}\enskip{}\enskip{}\enskip{}\enskip{}\enskip{}\enskip{}\enskip{}}\textcolor[HTML]{000000}{(and}\textcolor[HTML]{000000}{\enskip{}}\textcolor[HTML]{000000}{(=}\textcolor[HTML]{000000}{\enskip{}}\textcolor[HTML]{000000}{S\char95{}new}\textcolor[HTML]{000000}{\enskip{}}\textcolor[HTML]{000000}{(union}\textcolor[HTML]{000000}{\enskip{}}\textcolor[HTML]{000000}{S}\textcolor[HTML]{000000}{\enskip{}}\textcolor[HTML]{000000}{(singleton}\textcolor[HTML]{000000}{\enskip{}}\textcolor[HTML]{000000}{v)))}\\
\textcolor[HTML]{000000}{\enskip{}\enskip{}\enskip{}\enskip{}\enskip{}\enskip{}\enskip{}\enskip{}\enskip{}\enskip{}\enskip{}\enskip{}\enskip{}\enskip{}\enskip{}\enskip{}}\textcolor[HTML]{000000}{(=}\textcolor[HTML]{000000}{\enskip{}}\textcolor[HTML]{000000}{size\char95{}new}\textcolor[HTML]{000000}{\enskip{}}\textcolor[HTML]{000000}{(+}\textcolor[HTML]{000000}{\enskip{}}\textcolor[HTML]{000000}{size}\textcolor[HTML]{000000}{\enskip{}}\textcolor[HTML]{000000}{1))}\\
\textcolor[HTML]{000000}{\enskip{}\enskip{}\enskip{}\enskip{}\enskip{}\enskip{}\enskip{}\enskip{}\enskip{}\enskip{}\enskip{}\enskip{}\enskip{}\enskip{}\enskip{}\enskip{}}\textcolor[HTML]{000000}{result))}\\
\textcolor[HTML]{000000}{\enskip{}\enskip{}\enskip{}\enskip{}}\textcolor[HTML]{008080}{terms}\textcolor[HTML]{C000C0}{:}\\
\textcolor[HTML]{000000}{\enskip{}\enskip{}\enskip{}\enskip{}\enskip{}\enskip{}}\textcolor[HTML]{008080}{E}\textcolor[HTML]{C000C0}{:}\textcolor[HTML]{000000}{\enskip{}}\textcolor[HTML]{C000C0}{\char91{}}\textcolor[HTML]{000000}{\char36{}1}\textcolor[HTML]{C000C0}{\char93{}}\\
\textcolor[HTML]{000000}{\enskip{}\enskip{}\enskip{}\enskip{}\enskip{}\enskip{}}\textcolor[HTML]{008080}{Int}\textcolor[HTML]{C000C0}{:}\textcolor[HTML]{000000}{\enskip{}}\textcolor[HTML]{C000C0}{\char91{}}\textcolor[HTML]{000000}{size}\textcolor[HTML]{000000}{,\enskip{}}\textcolor[HTML]{C00000}{1}\textcolor[HTML]{000000}{,\enskip{}}\textcolor[HTML]{000000}{(+}\textcolor[HTML]{000000}{\enskip{}}\textcolor[HTML]{000000}{size}\textcolor[HTML]{000000}{\enskip{}}\textcolor[HTML]{000000}{1)}\textcolor[HTML]{C000C0}{\char93{}}\\
\textcolor[HTML]{000000}{\enskip{}\enskip{}\enskip{}\enskip{}\enskip{}\enskip{}}\textcolor[HTML]{000000}{(Set}\textcolor[HTML]{000000}{\enskip{}}\textcolor[HTML]{000000}{E)}\textcolor[HTML]{000000}{:\enskip{}}\textcolor[HTML]{C000C0}{\char91{}}\textcolor[HTML]{000000}{S}\textcolor[HTML]{000000}{,\enskip{}}\textcolor[HTML]{000000}{(singleton}\textcolor[HTML]{000000}{\enskip{}}\textcolor[HTML]{000000}{\char36{}1)}\textcolor[HTML]{000000}{,\enskip{}}\textcolor[HTML]{000000}{(union}\textcolor[HTML]{000000}{\enskip{}}\textcolor[HTML]{000000}{S}\textcolor[HTML]{000000}{\enskip{}}\textcolor[HTML]{000000}{(singleton}\textcolor[HTML]{000000}{\enskip{}}\textcolor[HTML]{000000}{\char36{}1))}\textcolor[HTML]{C000C0}{\char93{}}\\
\textcolor[HTML]{000000}{\enskip{}\enskip{}}\textcolor[HTML]{AF5F00}{-\enskip{}}\textcolor[HTML]{008080}{name}\textcolor[HTML]{C000C0}{:}\textcolor[HTML]{000000}{\enskip{}}\textcolor[HTML]{000000}{remove}\\
\textcolor[HTML]{000000}{\enskip{}\enskip{}\enskip{}\enskip{}}\textcolor[HTML]{008080}{args}\textcolor[HTML]{C000C0}{:}\\
\textcolor[HTML]{000000}{\enskip{}\enskip{}\enskip{}\enskip{}\enskip{}\enskip{}}\textcolor[HTML]{AF5F00}{-\enskip{}}\textcolor[HTML]{008080}{name}\textcolor[HTML]{C000C0}{:}\textcolor[HTML]{000000}{\enskip{}}\textcolor[HTML]{000000}{v}\\
\textcolor[HTML]{000000}{\enskip{}\enskip{}\enskip{}\enskip{}\enskip{}\enskip{}\enskip{}\enskip{}}\textcolor[HTML]{008080}{type}\textcolor[HTML]{C000C0}{:}\textcolor[HTML]{000000}{\enskip{}}\textcolor[HTML]{000000}{E}\\
\textcolor[HTML]{000000}{\enskip{}\enskip{}\enskip{}\enskip{}}\textcolor[HTML]{008080}{return}\textcolor[HTML]{C000C0}{:}\\
\textcolor[HTML]{000000}{\enskip{}\enskip{}\enskip{}\enskip{}\enskip{}\enskip{}}\textcolor[HTML]{AF5F00}{-\enskip{}}\textcolor[HTML]{008080}{name}\textcolor[HTML]{C000C0}{:}\textcolor[HTML]{000000}{\enskip{}}\textcolor[HTML]{000000}{result}\\
\textcolor[HTML]{000000}{\enskip{}\enskip{}\enskip{}\enskip{}\enskip{}\enskip{}\enskip{}\enskip{}}\textcolor[HTML]{008080}{type}\textcolor[HTML]{C000C0}{:}\textcolor[HTML]{000000}{\enskip{}}\textcolor[HTML]{000000}{Bool}\\
\textcolor[HTML]{000000}{\enskip{}\enskip{}\enskip{}\enskip{}}\textcolor[HTML]{008080}{requires}\textcolor[HTML]{C000C0}{:}\textcolor[HTML]{000000}{\enskip{}|}\\
\textcolor[HTML]{000000}{\enskip{}\enskip{}\enskip{}\enskip{}\enskip{}\enskip{}}\textcolor[HTML]{C00000}{true}\\
\textcolor[HTML]{000000}{\enskip{}\enskip{}\enskip{}\enskip{}}\textcolor[HTML]{008080}{ensures}\textcolor[HTML]{C000C0}{:}\textcolor[HTML]{000000}{\enskip{}|}\\
\textcolor[HTML]{000000}{\enskip{}\enskip{}\enskip{}\enskip{}\enskip{}\enskip{}}\textcolor[HTML]{000000}{(ite}\textcolor[HTML]{000000}{\enskip{}}\textcolor[HTML]{000000}{(member}\textcolor[HTML]{000000}{\enskip{}}\textcolor[HTML]{000000}{v}\textcolor[HTML]{000000}{\enskip{}}\textcolor[HTML]{000000}{S)}\\
\textcolor[HTML]{000000}{\enskip{}\enskip{}\enskip{}\enskip{}\enskip{}\enskip{}\enskip{}\enskip{}\enskip{}\enskip{}\enskip{}}\textcolor[HTML]{000000}{(and}\textcolor[HTML]{000000}{\enskip{}}\textcolor[HTML]{000000}{(=}\textcolor[HTML]{000000}{\enskip{}}\textcolor[HTML]{000000}{S\char95{}new}\textcolor[HTML]{000000}{\enskip{}}\textcolor[HTML]{000000}{(setminus}\textcolor[HTML]{000000}{\enskip{}}\textcolor[HTML]{000000}{S}\textcolor[HTML]{000000}{\enskip{}}\textcolor[HTML]{000000}{(singleton}\textcolor[HTML]{000000}{\enskip{}}\textcolor[HTML]{000000}{v)))}\\
\textcolor[HTML]{000000}{\enskip{}\enskip{}\enskip{}\enskip{}\enskip{}\enskip{}\enskip{}\enskip{}\enskip{}\enskip{}\enskip{}\enskip{}\enskip{}\enskip{}\enskip{}\enskip{}}\textcolor[HTML]{000000}{(=}\textcolor[HTML]{000000}{\enskip{}}\textcolor[HTML]{000000}{size\char95{}new}\textcolor[HTML]{000000}{\enskip{}}\textcolor[HTML]{000000}{(-}\textcolor[HTML]{000000}{\enskip{}}\textcolor[HTML]{000000}{size}\textcolor[HTML]{000000}{\enskip{}}\textcolor[HTML]{000000}{1))}\\
\textcolor[HTML]{000000}{\enskip{}\enskip{}\enskip{}\enskip{}\enskip{}\enskip{}\enskip{}\enskip{}\enskip{}\enskip{}\enskip{}\enskip{}\enskip{}\enskip{}\enskip{}\enskip{}}\textcolor[HTML]{000000}{result)}\\
\textcolor[HTML]{000000}{\enskip{}\enskip{}\enskip{}\enskip{}\enskip{}\enskip{}\enskip{}\enskip{}\enskip{}\enskip{}\enskip{}}\textcolor[HTML]{000000}{(and}\textcolor[HTML]{000000}{\enskip{}}\textcolor[HTML]{000000}{(=}\textcolor[HTML]{000000}{\enskip{}}\textcolor[HTML]{000000}{S\char95{}new}\textcolor[HTML]{000000}{\enskip{}}\textcolor[HTML]{000000}{S)}\\
\textcolor[HTML]{000000}{\enskip{}\enskip{}\enskip{}\enskip{}\enskip{}\enskip{}\enskip{}\enskip{}\enskip{}\enskip{}\enskip{}\enskip{}\enskip{}\enskip{}\enskip{}\enskip{}}\textcolor[HTML]{000000}{(=}\textcolor[HTML]{000000}{\enskip{}}\textcolor[HTML]{000000}{size\char95{}new}\textcolor[HTML]{000000}{\enskip{}}\textcolor[HTML]{000000}{size)}\\
\textcolor[HTML]{000000}{\enskip{}\enskip{}\enskip{}\enskip{}\enskip{}\enskip{}\enskip{}\enskip{}\enskip{}\enskip{}\enskip{}\enskip{}\enskip{}\enskip{}\enskip{}\enskip{}}\textcolor[HTML]{000000}{(not}\textcolor[HTML]{000000}{\enskip{}}\textcolor[HTML]{000000}{result)))}\\
\textcolor[HTML]{000000}{\enskip{}\enskip{}\enskip{}\enskip{}}\textcolor[HTML]{008080}{terms}\textcolor[HTML]{C000C0}{:}\\
\textcolor[HTML]{000000}{\enskip{}\enskip{}\enskip{}\enskip{}\enskip{}\enskip{}}\textcolor[HTML]{008080}{E}\textcolor[HTML]{C000C0}{:}\textcolor[HTML]{000000}{\enskip{}}\textcolor[HTML]{C000C0}{\char91{}}\textcolor[HTML]{000000}{\char36{}1}\textcolor[HTML]{C000C0}{\char93{}}\\
\textcolor[HTML]{000000}{\enskip{}\enskip{}\enskip{}\enskip{}\enskip{}\enskip{}}\textcolor[HTML]{008080}{Int}\textcolor[HTML]{C000C0}{:}\textcolor[HTML]{000000}{\enskip{}}\textcolor[HTML]{C000C0}{\char91{}}\textcolor[HTML]{000000}{size}\textcolor[HTML]{000000}{,\enskip{}}\textcolor[HTML]{C00000}{1}\textcolor[HTML]{000000}{,\enskip{}}\textcolor[HTML]{000000}{(-}\textcolor[HTML]{000000}{\enskip{}}\textcolor[HTML]{000000}{size}\textcolor[HTML]{000000}{\enskip{}}\textcolor[HTML]{000000}{1)}\textcolor[HTML]{C000C0}{\char93{}}\\
\textcolor[HTML]{000000}{\enskip{}\enskip{}\enskip{}\enskip{}\enskip{}\enskip{}}\textcolor[HTML]{000000}{(Set}\textcolor[HTML]{000000}{\enskip{}}\textcolor[HTML]{000000}{E)}\textcolor[HTML]{000000}{:\enskip{}}\textcolor[HTML]{C000C0}{\char91{}}\textcolor[HTML]{000000}{S}\textcolor[HTML]{000000}{,\enskip{}}\textcolor[HTML]{000000}{(singleton}\textcolor[HTML]{000000}{\enskip{}}\textcolor[HTML]{000000}{\char36{}1)}\textcolor[HTML]{000000}{,\enskip{}}\textcolor[HTML]{000000}{(setminus}\textcolor[HTML]{000000}{\enskip{}}\textcolor[HTML]{000000}{S}\textcolor[HTML]{000000}{\enskip{}}\textcolor[HTML]{000000}{(singleton}\textcolor[HTML]{000000}{\enskip{}}\textcolor[HTML]{000000}{\char36{}1))}\textcolor[HTML]{C000C0}{\char93{}}\\
\textcolor[HTML]{000000}{\enskip{}\enskip{}}\textcolor[HTML]{AF5F00}{-\enskip{}}\textcolor[HTML]{008080}{name}\textcolor[HTML]{C000C0}{:}\textcolor[HTML]{000000}{\enskip{}}\textcolor[HTML]{000000}{contains}\\
\textcolor[HTML]{000000}{\enskip{}\enskip{}\enskip{}\enskip{}}\textcolor[HTML]{008080}{args}\textcolor[HTML]{C000C0}{:}\\
\textcolor[HTML]{000000}{\enskip{}\enskip{}\enskip{}\enskip{}\enskip{}\enskip{}}\textcolor[HTML]{AF5F00}{-\enskip{}}\textcolor[HTML]{008080}{name}\textcolor[HTML]{C000C0}{:}\textcolor[HTML]{000000}{\enskip{}}\textcolor[HTML]{000000}{v}\\
\textcolor[HTML]{000000}{\enskip{}\enskip{}\enskip{}\enskip{}\enskip{}\enskip{}\enskip{}\enskip{}}\textcolor[HTML]{008080}{type}\textcolor[HTML]{C000C0}{:}\textcolor[HTML]{000000}{\enskip{}}\textcolor[HTML]{000000}{E}\\
\textcolor[HTML]{000000}{\enskip{}\enskip{}\enskip{}\enskip{}}\textcolor[HTML]{008080}{return}\textcolor[HTML]{C000C0}{:}\\
\textcolor[HTML]{000000}{\enskip{}\enskip{}\enskip{}\enskip{}\enskip{}\enskip{}}\textcolor[HTML]{AF5F00}{-\enskip{}}\textcolor[HTML]{008080}{name}\textcolor[HTML]{C000C0}{:}\textcolor[HTML]{000000}{\enskip{}}\textcolor[HTML]{000000}{result}\\
\textcolor[HTML]{000000}{\enskip{}\enskip{}\enskip{}\enskip{}\enskip{}\enskip{}\enskip{}\enskip{}}\textcolor[HTML]{008080}{type}\textcolor[HTML]{C000C0}{:}\textcolor[HTML]{000000}{\enskip{}}\textcolor[HTML]{000000}{Bool}\\
\textcolor[HTML]{000000}{\enskip{}\enskip{}\enskip{}\enskip{}}\textcolor[HTML]{008080}{requires}\textcolor[HTML]{C000C0}{:}\textcolor[HTML]{000000}{\enskip{}|}\\
\textcolor[HTML]{000000}{\enskip{}\enskip{}\enskip{}\enskip{}\enskip{}\enskip{}}\textcolor[HTML]{C00000}{true}\\
\textcolor[HTML]{000000}{\enskip{}\enskip{}\enskip{}\enskip{}}\textcolor[HTML]{008080}{ensures}\textcolor[HTML]{C000C0}{:}\textcolor[HTML]{000000}{\enskip{}|}\\
\textcolor[HTML]{000000}{\enskip{}\enskip{}\enskip{}\enskip{}\enskip{}\enskip{}}\textcolor[HTML]{000000}{(and}\textcolor[HTML]{000000}{\enskip{}}\textcolor[HTML]{000000}{(=}\textcolor[HTML]{000000}{\enskip{}}\textcolor[HTML]{000000}{S\char95{}new}\textcolor[HTML]{000000}{\enskip{}}\textcolor[HTML]{000000}{S)}\\
\textcolor[HTML]{000000}{\enskip{}\enskip{}\enskip{}\enskip{}\enskip{}\enskip{}\enskip{}\enskip{}\enskip{}\enskip{}\enskip{}}\textcolor[HTML]{000000}{(=}\textcolor[HTML]{000000}{\enskip{}}\textcolor[HTML]{000000}{size\char95{}new}\textcolor[HTML]{000000}{\enskip{}}\textcolor[HTML]{000000}{size)}\\
\textcolor[HTML]{000000}{\enskip{}\enskip{}\enskip{}\enskip{}\enskip{}\enskip{}\enskip{}\enskip{}\enskip{}\enskip{}\enskip{}}\textcolor[HTML]{000000}{(=}\textcolor[HTML]{000000}{\enskip{}}\textcolor[HTML]{000000}{(member}\textcolor[HTML]{000000}{\enskip{}}\textcolor[HTML]{000000}{v}\textcolor[HTML]{000000}{\enskip{}}\textcolor[HTML]{000000}{S)}\textcolor[HTML]{000000}{\enskip{}}\textcolor[HTML]{000000}{result))}\\
\textcolor[HTML]{000000}{\enskip{}\enskip{}\enskip{}\enskip{}}\textcolor[HTML]{008080}{terms}\textcolor[HTML]{C000C0}{:}\\
\textcolor[HTML]{000000}{\enskip{}\enskip{}\enskip{}\enskip{}\enskip{}\enskip{}}\textcolor[HTML]{008080}{E}\textcolor[HTML]{C000C0}{:}\textcolor[HTML]{000000}{\enskip{}}\textcolor[HTML]{C000C0}{\char91{}}\textcolor[HTML]{000000}{\char36{}1}\textcolor[HTML]{C000C0}{\char93{}}\\
\textcolor[HTML]{000000}{\enskip{}\enskip{}\enskip{}\enskip{}\enskip{}\enskip{}}\textcolor[HTML]{008080}{Int}\textcolor[HTML]{C000C0}{:}\textcolor[HTML]{000000}{\enskip{}}\textcolor[HTML]{C000C0}{\char91{}}\textcolor[HTML]{000000}{size}\textcolor[HTML]{C000C0}{\char93{}}\\
\textcolor[HTML]{000000}{\enskip{}\enskip{}\enskip{}\enskip{}\enskip{}\enskip{}}\textcolor[HTML]{000000}{(Set}\textcolor[HTML]{000000}{\enskip{}}\textcolor[HTML]{000000}{E)}\textcolor[HTML]{000000}{:\enskip{}}\textcolor[HTML]{C000C0}{\char91{}}\textcolor[HTML]{000000}{S}\textcolor[HTML]{000000}{,\enskip{}}\textcolor[HTML]{000000}{(singleton}\textcolor[HTML]{000000}{\enskip{}}\textcolor[HTML]{000000}{\char36{}1)}\textcolor[HTML]{000000}{,\enskip{}}\textcolor[HTML]{000000}{(setminus}\textcolor[HTML]{000000}{\enskip{}}\textcolor[HTML]{000000}{S}\textcolor[HTML]{000000}{\enskip{}}\textcolor[HTML]{000000}{(singleton}\textcolor[HTML]{000000}{\enskip{}}\textcolor[HTML]{000000}{\char36{}1))}\textcolor[HTML]{C000C0}{\char93{}}\\
\textcolor[HTML]{000000}{\enskip{}\enskip{}}\textcolor[HTML]{AF5F00}{-\enskip{}}\textcolor[HTML]{008080}{name}\textcolor[HTML]{C000C0}{:}\textcolor[HTML]{000000}{\enskip{}}\textcolor[HTML]{000000}{getsize}\\
\textcolor[HTML]{000000}{\enskip{}\enskip{}\enskip{}\enskip{}}\textcolor[HTML]{008080}{args}\textcolor[HTML]{C000C0}{:}\textcolor[HTML]{000000}{\enskip{}}\textcolor[HTML]{C000C0}{\char91{}\char93{}}\\
\textcolor[HTML]{000000}{\enskip{}\enskip{}\enskip{}\enskip{}}\textcolor[HTML]{008080}{return}\textcolor[HTML]{C000C0}{:}\\
\textcolor[HTML]{000000}{\enskip{}\enskip{}\enskip{}\enskip{}\enskip{}\enskip{}}\textcolor[HTML]{AF5F00}{-\enskip{}}\textcolor[HTML]{008080}{name}\textcolor[HTML]{C000C0}{:}\textcolor[HTML]{000000}{\enskip{}}\textcolor[HTML]{000000}{result}\\
\textcolor[HTML]{000000}{\enskip{}\enskip{}\enskip{}\enskip{}\enskip{}\enskip{}\enskip{}\enskip{}}\textcolor[HTML]{008080}{type}\textcolor[HTML]{C000C0}{:}\textcolor[HTML]{000000}{\enskip{}}\textcolor[HTML]{000000}{Int}\\
\textcolor[HTML]{000000}{\enskip{}\enskip{}\enskip{}\enskip{}}\textcolor[HTML]{008080}{requires}\textcolor[HTML]{C000C0}{:}\textcolor[HTML]{000000}{\enskip{}|}\\
\textcolor[HTML]{000000}{\enskip{}\enskip{}\enskip{}\enskip{}\enskip{}\enskip{}}\textcolor[HTML]{C00000}{true}\\
\textcolor[HTML]{000000}{\enskip{}\enskip{}\enskip{}\enskip{}}\textcolor[HTML]{008080}{ensures}\textcolor[HTML]{C000C0}{:}\textcolor[HTML]{000000}{\enskip{}|}\\
\textcolor[HTML]{000000}{\enskip{}\enskip{}\enskip{}\enskip{}\enskip{}\enskip{}}\textcolor[HTML]{000000}{(and}\textcolor[HTML]{000000}{\enskip{}}\textcolor[HTML]{000000}{(=}\textcolor[HTML]{000000}{\enskip{}}\textcolor[HTML]{000000}{S\char95{}new}\textcolor[HTML]{000000}{\enskip{}}\textcolor[HTML]{000000}{S)}\\
\textcolor[HTML]{000000}{\enskip{}\enskip{}\enskip{}\enskip{}\enskip{}\enskip{}\enskip{}\enskip{}\enskip{}\enskip{}\enskip{}}\textcolor[HTML]{000000}{(=}\textcolor[HTML]{000000}{\enskip{}}\textcolor[HTML]{000000}{size\char95{}new}\textcolor[HTML]{000000}{\enskip{}}\textcolor[HTML]{000000}{size)}\\
\textcolor[HTML]{000000}{\enskip{}\enskip{}\enskip{}\enskip{}\enskip{}\enskip{}\enskip{}\enskip{}\enskip{}\enskip{}\enskip{}}\textcolor[HTML]{000000}{(=}\textcolor[HTML]{000000}{\enskip{}}\textcolor[HTML]{000000}{size}\textcolor[HTML]{000000}{\enskip{}}\textcolor[HTML]{000000}{result))}\\
\textcolor[HTML]{000000}{\enskip{}\enskip{}\enskip{}\enskip{}}\textcolor[HTML]{008080}{terms}\textcolor[HTML]{C000C0}{:}\\
\textcolor[HTML]{000000}{\enskip{}\enskip{}\enskip{}\enskip{}\enskip{}\enskip{}}\textcolor[HTML]{008080}{Int}\textcolor[HTML]{C000C0}{:}\textcolor[HTML]{000000}{\enskip{}}\textcolor[HTML]{C000C0}{\char91{}}\textcolor[HTML]{000000}{size}\textcolor[HTML]{C000C0}{\char93{}}\\
\\
\textcolor[HTML]{008080}{predicates}\textcolor[HTML]{C000C0}{:}\\
\textcolor[HTML]{000000}{\enskip{}\enskip{}}\textcolor[HTML]{AF5F00}{-\enskip{}}\textcolor[HTML]{008080}{name}\textcolor[HTML]{C000C0}{:}\textcolor[HTML]{000000}{\enskip{}}\textcolor[HTML]{C00000}{"}\textcolor[HTML]{C00000}{=}\textcolor[HTML]{C00000}{"}\\
\textcolor[HTML]{000000}{\enskip{}\enskip{}\enskip{}\enskip{}}\textcolor[HTML]{008080}{type}\textcolor[HTML]{C000C0}{:}\textcolor[HTML]{000000}{\enskip{}}\textcolor[HTML]{C000C0}{\char91{}}\textcolor[HTML]{000000}{Int}\textcolor[HTML]{000000}{,\enskip{}}\textcolor[HTML]{000000}{Int}\textcolor[HTML]{C000C0}{\char93{}}\\
\textcolor[HTML]{000000}{\enskip{}\enskip{}}\textcolor[HTML]{AF5F00}{-\enskip{}}\textcolor[HTML]{008080}{name}\textcolor[HTML]{C000C0}{:}\textcolor[HTML]{000000}{\enskip{}}\textcolor[HTML]{C00000}{"}\textcolor[HTML]{C00000}{=}\textcolor[HTML]{C00000}{"}\\
\textcolor[HTML]{000000}{\enskip{}\enskip{}\enskip{}\enskip{}}\textcolor[HTML]{008080}{type}\textcolor[HTML]{C000C0}{:}\textcolor[HTML]{000000}{\enskip{}}\textcolor[HTML]{C000C0}{\char91{}}\textcolor[HTML]{000000}{E}\textcolor[HTML]{000000}{,\enskip{}}\textcolor[HTML]{000000}{E}\textcolor[HTML]{C000C0}{\char93{}}\\
\textcolor[HTML]{000000}{\enskip{}\enskip{}}\textcolor[HTML]{AF5F00}{-\enskip{}}\textcolor[HTML]{008080}{name}\textcolor[HTML]{C000C0}{:}\textcolor[HTML]{000000}{\enskip{}}\textcolor[HTML]{C00000}{"}\textcolor[HTML]{C00000}{=}\textcolor[HTML]{C00000}{"}\\
\textcolor[HTML]{000000}{\enskip{}\enskip{}\enskip{}\enskip{}}\textcolor[HTML]{008080}{type}\textcolor[HTML]{C000C0}{:}\textcolor[HTML]{000000}{\enskip{}}\textcolor[HTML]{C000C0}{\char91{}}\textcolor[HTML]{000000}{(Set}\textcolor[HTML]{000000}{\enskip{}}\textcolor[HTML]{000000}{E)}\textcolor[HTML]{000000}{,\enskip{}}\textcolor[HTML]{000000}{(Set}\textcolor[HTML]{000000}{\enskip{}}\textcolor[HTML]{000000}{E)}\textcolor[HTML]{C000C0}{\char93{}}\\
\textcolor[HTML]{000000}{\enskip{}\enskip{}}\textcolor[HTML]{AF5F00}{-\enskip{}}\textcolor[HTML]{008080}{name}\textcolor[HTML]{C000C0}{:}\textcolor[HTML]{000000}{\enskip{}}\textcolor[HTML]{C00000}{"}\textcolor[HTML]{C00000}{member}\textcolor[HTML]{C00000}{"}\\
\textcolor[HTML]{000000}{\enskip{}\enskip{}\enskip{}\enskip{}}\textcolor[HTML]{008080}{type}\textcolor[HTML]{C000C0}{:}\textcolor[HTML]{000000}{\enskip{}}\textcolor[HTML]{C000C0}{\char91{}}\textcolor[HTML]{000000}{E}\textcolor[HTML]{000000}{,\enskip{}}\textcolor[HTML]{000000}{(Set}\textcolor[HTML]{000000}{\enskip{}}\textcolor[HTML]{000000}{E)}\textcolor[HTML]{C000C0}{\char93{}}\\

}

\begin{itemize}
\item \CCmethod{add} \CCbowtie\ \CCmethod{add}

 Simple:

[\CCequal{\CCvar{y1}}{\CCvar{x1}} $\wedge$ \CCmember{\CCvar{y1}}{\CCvar{S}}]

 $\vee$ [\CCnot{\CCequal{\CCvar{y1}}{\CCvar{x1}}}]

 Poke:

[\CCequal{\CCvar{y1}}{\CCvar{x1}} $\wedge$ \CCmember{\CCvar{y1}}{\CCvar{S}}]

 $\vee$ [\CCnot{\CCequal{\CCvar{y1}}{\CCvar{x1}}}]

\item \CCmethod{add} \CCbowtie\ \CCmethod{contains}

 Simple:

[\CCequal{\CCvar{y1}}{\CCvar{x1}} $\wedge$ \CCmember{\CCvar{y1}}{\CCvar{S}}]

 $\vee$ [\CCnot{\CCequal{\CCvar{y1}}{\CCvar{x1}}}]

 Poke:

[\CCmember{\CCvar{x1}}{\CCvar{S}}]

 $\vee$ [\CCnot{\CCmember{\CCvar{x1}}{\CCvar{S}}} $\wedge$ \CCnot{\CCequal{\CCvar{y1}}{\CCvar{x1}}}]

\item \CCmethod{add} \CCbowtie\ \CCmethod{getsize}

 Simple:

\CCmember{\CCvar{x1}}{\CCvar{S}}

 Poke:

\CCmember{\CCvar{x1}}{\CCvar{S}}

\item \CCmethod{add} \CCbowtie\ \CCmethod{remove}

 Simple:

\CCnot{\CCequal{\CCvar{y1}}{\CCvar{x1}}}

 Poke:

\CCnot{\CCequal{\CCvar{y1}}{\CCvar{x1}}}

\item \CCmethod{contains} \CCbowtie\ \CCmethod{contains}

 Simple:

true

 Poke:

true

\item \CCmethod{contains} \CCbowtie\ \CCmethod{getsize}

 Simple:

true

 Poke:

true

\item \CCmethod{contains} \CCbowtie\ \CCmethod{remove}

 Simple:

[\CCequal{\CCvar{y1}}{\CCvar{x1}} $\wedge$ \CCequal{\CCvar{1}}{\CCvar{size}} $\wedge$ \CCnot{\CCmember{\CCvar{y1}}{\CCvar{S}}}]

 $\vee$ [\CCequal{\CCvar{y1}}{\CCvar{x1}} $\wedge$ \CCnot{\CCequal{\CCvar{1}}{\CCvar{size}}} $\wedge$ \CCnot{\CCmember{\CCvar{y1}}{\CCvar{S}}}]

 $\vee$ [\CCnot{\CCequal{\CCvar{y1}}{\CCvar{x1}}}]

 Poke:

[\CCequal{\CCsetminus{\CCvar{S}}{\CCsingleton{\CCvar{x1}}}}{\CCsingleton{\CCvar{y1}}}]

 $\vee$ [\CCnot{\CCequal{\CCsetminus{\CCvar{S}}{\CCsingleton{\CCvar{x1}}}}{\CCsingleton{\CCvar{y1}}}} $\wedge$ \CCmember{\CCvar{y1}}{\CCsingleton{\CCvar{x1}}} $\wedge$ \CCnot{\CCmember{\CCvar{y1}}{\CCvar{S}}}]

 $\vee$ [\CCnot{\CCequal{\CCsetminus{\CCvar{S}}{\CCsingleton{\CCvar{x1}}}}{\CCsingleton{\CCvar{y1}}}} $\wedge$ \CCnot{\CCmember{\CCvar{y1}}{\CCsingleton{\CCvar{x1}}}}]

\item \CCmethod{getsize} \CCbowtie\ \CCmethod{getsize}

 Simple:

true

 Poke:

true

\item \CCmethod{getsize} \CCbowtie\ \CCmethod{remove}

 Simple:

[\CCequal{\CCvar{1}}{\CCvar{size}} $\wedge$ \CCnot{\CCmember{\CCvar{y1}}{\CCvar{S}}}]

 $\vee$ [\CCnot{\CCequal{\CCvar{1}}{\CCvar{size}}} $\wedge$ \CCnot{\CCmember{\CCvar{y1}}{\CCvar{S}}}]

 Poke:

\CCnot{\CCmember{\CCvar{y1}}{\CCvar{S}}}

\item \CCmethod{remove} \CCbowtie\ \CCmethod{remove}

 Simple:

[\CCequal{\CCvar{1}}{\CCvar{size}} $\wedge$ \CCequal{\CCvar{y1}}{\CCvar{x1}} $\wedge$ \CCnot{\CCmember{\CCvar{y1}}{\CCvar{S}}}]

 $\vee$ [\CCequal{\CCvar{1}}{\CCvar{size}} $\wedge$ \CCnot{\CCequal{\CCvar{y1}}{\CCvar{x1}}}]

 $\vee$ [\CCnot{\CCequal{\CCvar{1}}{\CCvar{size}}} $\wedge$ \CCequal{\CCvar{y1}}{\CCvar{x1}} $\wedge$ \CCnot{\CCmember{\CCvar{y1}}{\CCvar{S}}}]

 $\vee$ [\CCnot{\CCequal{\CCvar{1}}{\CCvar{size}}} $\wedge$ \CCnot{\CCequal{\CCvar{y1}}{\CCvar{x1}}}]

 Poke:

[\CCequal{\CCsetminus{\CCvar{S}}{\CCsingleton{\CCvar{y1}}}}{\CCsingleton{\CCvar{x1}}}]

 $\vee$ [\CCnot{\CCequal{\CCsetminus{\CCvar{S}}{\CCsingleton{\CCvar{y1}}}}{\CCsingleton{\CCvar{x1}}}} $\wedge$ \CCmember{\CCvar{y1}}{\CCsingleton{\CCvar{x1}}} $\wedge$ \CCnot{\CCmember{\CCvar{y1}}{\CCvar{S}}}]

 $\vee$ [\CCnot{\CCequal{\CCsetminus{\CCvar{S}}{\CCsingleton{\CCvar{y1}}}}{\CCsingleton{\CCvar{x1}}}} $\wedge$ \CCnot{\CCmember{\CCvar{y1}}{\CCsingleton{\CCvar{x1}}}}]

\end{itemize}

\subsection{HashTable}
\scriptsize
\label{yml:hashtable}
{\ttfamily\noindent
\textcolor[HTML]{0000C0}{\char35{}\enskip{}Hash\enskip{}table\enskip{}data\enskip{}structure's\enskip{}abstract\enskip{}definition}\\
\\
\textcolor[HTML]{008080}{name}\textcolor[HTML]{C000C0}{:}\textcolor[HTML]{000000}{\enskip{}}\textcolor[HTML]{000000}{HashTable}\\
\\
\textcolor[HTML]{008080}{preamble}\textcolor[HTML]{C000C0}{:}\textcolor[HTML]{000000}{\enskip{}|}\\
\textcolor[HTML]{000000}{\enskip{}\enskip{}}\textcolor[HTML]{000000}{(declare-sort}\textcolor[HTML]{000000}{\enskip{}}\textcolor[HTML]{000000}{E}\textcolor[HTML]{000000}{\enskip{}}\textcolor[HTML]{000000}{0)}\\
\textcolor[HTML]{000000}{\enskip{}\enskip{}}\textcolor[HTML]{000000}{(declare-sort}\textcolor[HTML]{000000}{\enskip{}}\textcolor[HTML]{000000}{F}\textcolor[HTML]{000000}{\enskip{}}\textcolor[HTML]{000000}{0)}\\
\\
\textcolor[HTML]{008080}{state}\textcolor[HTML]{C000C0}{:}\\
\textcolor[HTML]{000000}{\enskip{}\enskip{}}\textcolor[HTML]{AF5F00}{-\enskip{}}\textcolor[HTML]{008080}{name}\textcolor[HTML]{C000C0}{:}\textcolor[HTML]{000000}{\enskip{}}\textcolor[HTML]{000000}{keys}\\
\textcolor[HTML]{000000}{\enskip{}\enskip{}\enskip{}\enskip{}}\textcolor[HTML]{008080}{type}\textcolor[HTML]{C000C0}{:}\textcolor[HTML]{000000}{\enskip{}}\textcolor[HTML]{000000}{(Set}\textcolor[HTML]{000000}{\enskip{}}\textcolor[HTML]{000000}{E)}\\
\textcolor[HTML]{000000}{\enskip{}\enskip{}}\textcolor[HTML]{AF5F00}{-\enskip{}}\textcolor[HTML]{008080}{name}\textcolor[HTML]{C000C0}{:}\textcolor[HTML]{000000}{\enskip{}}\textcolor[HTML]{000000}{H}\\
\textcolor[HTML]{000000}{\enskip{}\enskip{}\enskip{}\enskip{}}\textcolor[HTML]{008080}{type}\textcolor[HTML]{C000C0}{:}\textcolor[HTML]{000000}{\enskip{}}\textcolor[HTML]{000000}{(Array}\textcolor[HTML]{000000}{\enskip{}}\textcolor[HTML]{000000}{E}\textcolor[HTML]{000000}{\enskip{}}\textcolor[HTML]{000000}{F)}\\
\textcolor[HTML]{000000}{\enskip{}\enskip{}}\textcolor[HTML]{AF5F00}{-\enskip{}}\textcolor[HTML]{008080}{name}\textcolor[HTML]{C000C0}{:}\textcolor[HTML]{000000}{\enskip{}}\textcolor[HTML]{000000}{size}\\
\textcolor[HTML]{000000}{\enskip{}\enskip{}\enskip{}\enskip{}}\textcolor[HTML]{008080}{type}\textcolor[HTML]{C000C0}{:}\textcolor[HTML]{000000}{\enskip{}}\textcolor[HTML]{000000}{Int}\\
\\
\textcolor[HTML]{008080}{states\char95{}equal}\textcolor[HTML]{C000C0}{:}\\
\textcolor[HTML]{000000}{\enskip{}\enskip{}}\textcolor[HTML]{008080}{definition}\textcolor[HTML]{C000C0}{:}\textcolor[HTML]{000000}{\enskip{}|}\\
\textcolor[HTML]{000000}{\enskip{}\enskip{}\enskip{}\enskip{}}\textcolor[HTML]{000000}{(and}\textcolor[HTML]{000000}{\enskip{}}\textcolor[HTML]{000000}{(=}\textcolor[HTML]{000000}{\enskip{}}\textcolor[HTML]{000000}{keys\char95{}1}\textcolor[HTML]{000000}{\enskip{}}\textcolor[HTML]{000000}{keys\char95{}2)}\\
\textcolor[HTML]{000000}{\enskip{}\enskip{}\enskip{}\enskip{}\enskip{}\enskip{}\enskip{}\enskip{}\enskip{}}\textcolor[HTML]{000000}{(=}\textcolor[HTML]{000000}{\enskip{}}\textcolor[HTML]{000000}{H\char95{}1}\textcolor[HTML]{000000}{\enskip{}}\textcolor[HTML]{000000}{H\char95{}2)}\\
\textcolor[HTML]{000000}{\enskip{}\enskip{}\enskip{}\enskip{}\enskip{}\enskip{}\enskip{}\enskip{}\enskip{}}\textcolor[HTML]{000000}{(=}\textcolor[HTML]{000000}{\enskip{}}\textcolor[HTML]{000000}{size\char95{}1}\textcolor[HTML]{000000}{\enskip{}}\textcolor[HTML]{000000}{size\char95{}2))}\\
\\
\textcolor[HTML]{008080}{methods}\textcolor[HTML]{C000C0}{:}\\
\textcolor[HTML]{000000}{\enskip{}\enskip{}}\textcolor[HTML]{AF5F00}{-\enskip{}}\textcolor[HTML]{008080}{name}\textcolor[HTML]{C000C0}{:}\textcolor[HTML]{000000}{\enskip{}}\textcolor[HTML]{000000}{haskey}\\
\textcolor[HTML]{000000}{\enskip{}\enskip{}\enskip{}\enskip{}}\textcolor[HTML]{008080}{args}\textcolor[HTML]{C000C0}{:}\\
\textcolor[HTML]{000000}{\enskip{}\enskip{}\enskip{}\enskip{}\enskip{}\enskip{}}\textcolor[HTML]{AF5F00}{-\enskip{}}\textcolor[HTML]{008080}{name}\textcolor[HTML]{C000C0}{:}\textcolor[HTML]{000000}{\enskip{}}\textcolor[HTML]{000000}{k0}\\
\textcolor[HTML]{000000}{\enskip{}\enskip{}\enskip{}\enskip{}\enskip{}\enskip{}\enskip{}\enskip{}}\textcolor[HTML]{008080}{type}\textcolor[HTML]{C000C0}{:}\textcolor[HTML]{000000}{\enskip{}}\textcolor[HTML]{000000}{E}\\
\textcolor[HTML]{000000}{\enskip{}\enskip{}\enskip{}\enskip{}}\textcolor[HTML]{008080}{return}\textcolor[HTML]{C000C0}{:}\\
\textcolor[HTML]{000000}{\enskip{}\enskip{}\enskip{}\enskip{}\enskip{}\enskip{}}\textcolor[HTML]{AF5F00}{-\enskip{}}\textcolor[HTML]{008080}{name}\textcolor[HTML]{C000C0}{:}\textcolor[HTML]{000000}{\enskip{}}\textcolor[HTML]{000000}{result}\\
\textcolor[HTML]{000000}{\enskip{}\enskip{}\enskip{}\enskip{}\enskip{}\enskip{}\enskip{}\enskip{}}\textcolor[HTML]{008080}{type}\textcolor[HTML]{C000C0}{:}\textcolor[HTML]{000000}{\enskip{}}\textcolor[HTML]{000000}{Bool}\\
\textcolor[HTML]{000000}{\enskip{}\enskip{}\enskip{}\enskip{}}\textcolor[HTML]{008080}{requires}\textcolor[HTML]{C000C0}{:}\textcolor[HTML]{000000}{\enskip{}|}\\
\textcolor[HTML]{000000}{\enskip{}\enskip{}\enskip{}\enskip{}\enskip{}\enskip{}}\textcolor[HTML]{C00000}{true}\\
\textcolor[HTML]{000000}{\enskip{}\enskip{}\enskip{}\enskip{}}\textcolor[HTML]{008080}{ensures}\textcolor[HTML]{C000C0}{:}\textcolor[HTML]{000000}{\enskip{}|}\\
\textcolor[HTML]{000000}{\enskip{}\enskip{}\enskip{}\enskip{}\enskip{}\enskip{}}\textcolor[HTML]{000000}{(and}\textcolor[HTML]{000000}{\enskip{}}\textcolor[HTML]{000000}{(=}\textcolor[HTML]{000000}{\enskip{}}\textcolor[HTML]{000000}{keys\char95{}new}\textcolor[HTML]{000000}{\enskip{}}\textcolor[HTML]{000000}{keys)}\\
\textcolor[HTML]{000000}{\enskip{}\enskip{}\enskip{}\enskip{}\enskip{}\enskip{}\enskip{}\enskip{}\enskip{}\enskip{}\enskip{}}\textcolor[HTML]{000000}{(=}\textcolor[HTML]{000000}{\enskip{}}\textcolor[HTML]{000000}{H\char95{}new}\textcolor[HTML]{000000}{\enskip{}}\textcolor[HTML]{000000}{H)}\\
\textcolor[HTML]{000000}{\enskip{}\enskip{}\enskip{}\enskip{}\enskip{}\enskip{}\enskip{}\enskip{}\enskip{}\enskip{}\enskip{}}\textcolor[HTML]{000000}{(=}\textcolor[HTML]{000000}{\enskip{}}\textcolor[HTML]{000000}{size\char95{}new}\textcolor[HTML]{000000}{\enskip{}}\textcolor[HTML]{000000}{size)}\\
\textcolor[HTML]{000000}{\enskip{}\enskip{}\enskip{}\enskip{}\enskip{}\enskip{}\enskip{}\enskip{}\enskip{}\enskip{}\enskip{}}\textcolor[HTML]{000000}{(=}\textcolor[HTML]{000000}{\enskip{}}\textcolor[HTML]{000000}{(member}\textcolor[HTML]{000000}{\enskip{}}\textcolor[HTML]{000000}{k0}\textcolor[HTML]{000000}{\enskip{}}\textcolor[HTML]{000000}{keys)}\textcolor[HTML]{000000}{\enskip{}}\textcolor[HTML]{000000}{result)}\\
\textcolor[HTML]{000000}{\enskip{}\enskip{}\enskip{}\enskip{}\enskip{}\enskip{}\enskip{}}\textcolor[HTML]{000000}{)}\\
\textcolor[HTML]{000000}{\enskip{}\enskip{}\enskip{}\enskip{}}\textcolor[HTML]{008080}{terms}\textcolor[HTML]{C000C0}{:}\\
\textcolor[HTML]{000000}{\enskip{}\enskip{}\enskip{}\enskip{}\enskip{}\enskip{}}\textcolor[HTML]{008080}{Int}\textcolor[HTML]{C000C0}{:}\textcolor[HTML]{000000}{\enskip{}}\textcolor[HTML]{C000C0}{\char91{}}\textcolor[HTML]{000000}{size}\textcolor[HTML]{C000C0}{\char93{}}\\
\textcolor[HTML]{000000}{\enskip{}\enskip{}\enskip{}\enskip{}\enskip{}\enskip{}}\textcolor[HTML]{008080}{E}\textcolor[HTML]{C000C0}{:}\textcolor[HTML]{000000}{\enskip{}}\textcolor[HTML]{C000C0}{\char91{}}\textcolor[HTML]{000000}{\char36{}1}\textcolor[HTML]{C000C0}{\char93{}}\\
\textcolor[HTML]{000000}{\enskip{}\enskip{}\enskip{}\enskip{}\enskip{}\enskip{}}\textcolor[HTML]{000000}{(Set}\textcolor[HTML]{000000}{\enskip{}}\textcolor[HTML]{000000}{E)}\textcolor[HTML]{000000}{:\enskip{}}\textcolor[HTML]{C000C0}{\char91{}}\textcolor[HTML]{000000}{keys}\textcolor[HTML]{C000C0}{\char93{}}\\
\textcolor[HTML]{000000}{\enskip{}\enskip{}\enskip{}\enskip{}\enskip{}\enskip{}}\textcolor[HTML]{000000}{(Array}\textcolor[HTML]{000000}{\enskip{}}\textcolor[HTML]{000000}{E}\textcolor[HTML]{000000}{\enskip{}}\textcolor[HTML]{000000}{F)}\textcolor[HTML]{000000}{:\enskip{}}\textcolor[HTML]{C000C0}{\char91{}}\textcolor[HTML]{000000}{H}\textcolor[HTML]{C000C0}{\char93{}}\\
\textcolor[HTML]{000000}{\enskip{}\enskip{}}\textcolor[HTML]{AF5F00}{-\enskip{}}\textcolor[HTML]{008080}{name}\textcolor[HTML]{C000C0}{:}\textcolor[HTML]{000000}{\enskip{}}\textcolor[HTML]{000000}{remove}\\
\textcolor[HTML]{000000}{\enskip{}\enskip{}\enskip{}\enskip{}}\textcolor[HTML]{008080}{args}\textcolor[HTML]{C000C0}{:}\\
\textcolor[HTML]{000000}{\enskip{}\enskip{}\enskip{}\enskip{}\enskip{}\enskip{}}\textcolor[HTML]{AF5F00}{-\enskip{}}\textcolor[HTML]{008080}{name}\textcolor[HTML]{C000C0}{:}\textcolor[HTML]{000000}{\enskip{}}\textcolor[HTML]{000000}{v}\\
\textcolor[HTML]{000000}{\enskip{}\enskip{}\enskip{}\enskip{}\enskip{}\enskip{}\enskip{}\enskip{}}\textcolor[HTML]{008080}{type}\textcolor[HTML]{C000C0}{:}\textcolor[HTML]{000000}{\enskip{}}\textcolor[HTML]{000000}{E}\\
\textcolor[HTML]{000000}{\enskip{}\enskip{}\enskip{}\enskip{}}\textcolor[HTML]{008080}{return}\textcolor[HTML]{C000C0}{:}\\
\textcolor[HTML]{000000}{\enskip{}\enskip{}\enskip{}\enskip{}\enskip{}\enskip{}}\textcolor[HTML]{AF5F00}{-\enskip{}}\textcolor[HTML]{008080}{name}\textcolor[HTML]{C000C0}{:}\textcolor[HTML]{000000}{\enskip{}}\textcolor[HTML]{000000}{result}\\
\textcolor[HTML]{000000}{\enskip{}\enskip{}\enskip{}\enskip{}\enskip{}\enskip{}\enskip{}\enskip{}}\textcolor[HTML]{008080}{type}\textcolor[HTML]{C000C0}{:}\textcolor[HTML]{000000}{\enskip{}}\textcolor[HTML]{000000}{Bool}\\
\textcolor[HTML]{000000}{\enskip{}\enskip{}\enskip{}\enskip{}}\textcolor[HTML]{008080}{requires}\textcolor[HTML]{C000C0}{:}\textcolor[HTML]{000000}{\enskip{}|}\\
\textcolor[HTML]{000000}{\enskip{}\enskip{}\enskip{}\enskip{}\enskip{}\enskip{}}\textcolor[HTML]{C00000}{true}\\
\textcolor[HTML]{000000}{\enskip{}\enskip{}\enskip{}\enskip{}}\textcolor[HTML]{008080}{ensures}\textcolor[HTML]{C000C0}{:}\textcolor[HTML]{000000}{\enskip{}|}\\
\textcolor[HTML]{000000}{\enskip{}\enskip{}\enskip{}\enskip{}\enskip{}\enskip{}}\textcolor[HTML]{000000}{(ite}\textcolor[HTML]{000000}{\enskip{}}\textcolor[HTML]{000000}{(member}\textcolor[HTML]{000000}{\enskip{}}\textcolor[HTML]{000000}{v}\textcolor[HTML]{000000}{\enskip{}}\textcolor[HTML]{000000}{keys)}\\
\textcolor[HTML]{000000}{\enskip{}\enskip{}\enskip{}\enskip{}\enskip{}\enskip{}\enskip{}\enskip{}\enskip{}\enskip{}\enskip{}}\textcolor[HTML]{000000}{(and}\textcolor[HTML]{000000}{\enskip{}}\textcolor[HTML]{000000}{(=}\textcolor[HTML]{000000}{\enskip{}}\textcolor[HTML]{000000}{keys\char95{}new}\textcolor[HTML]{000000}{\enskip{}}\textcolor[HTML]{000000}{(setminus}\textcolor[HTML]{000000}{\enskip{}}\textcolor[HTML]{000000}{keys}\textcolor[HTML]{000000}{\enskip{}}\textcolor[HTML]{000000}{(singleton}\textcolor[HTML]{000000}{\enskip{}}\textcolor[HTML]{000000}{v)))}\\
\textcolor[HTML]{000000}{\enskip{}\enskip{}\enskip{}\enskip{}\enskip{}\enskip{}\enskip{}\enskip{}\enskip{}\enskip{}\enskip{}\enskip{}\enskip{}\enskip{}\enskip{}\enskip{}}\textcolor[HTML]{000000}{(=}\textcolor[HTML]{000000}{\enskip{}}\textcolor[HTML]{000000}{size\char95{}new}\textcolor[HTML]{000000}{\enskip{}}\textcolor[HTML]{000000}{(-}\textcolor[HTML]{000000}{\enskip{}}\textcolor[HTML]{000000}{size}\textcolor[HTML]{000000}{\enskip{}}\textcolor[HTML]{000000}{1))}\\
\textcolor[HTML]{000000}{\enskip{}\enskip{}\enskip{}\enskip{}\enskip{}\enskip{}\enskip{}\enskip{}\enskip{}\enskip{}\enskip{}\enskip{}\enskip{}\enskip{}\enskip{}\enskip{}}\textcolor[HTML]{000000}{(=}\textcolor[HTML]{000000}{\enskip{}}\textcolor[HTML]{000000}{H\char95{}new}\textcolor[HTML]{000000}{\enskip{}}\textcolor[HTML]{000000}{H)}\\
\textcolor[HTML]{000000}{\enskip{}\enskip{}\enskip{}\enskip{}\enskip{}\enskip{}\enskip{}\enskip{}\enskip{}\enskip{}\enskip{}\enskip{}\enskip{}\enskip{}\enskip{}\enskip{}}\textcolor[HTML]{000000}{result)}\\
\textcolor[HTML]{000000}{\enskip{}\enskip{}\enskip{}\enskip{}\enskip{}\enskip{}\enskip{}\enskip{}\enskip{}\enskip{}\enskip{}}\textcolor[HTML]{000000}{(and}\textcolor[HTML]{000000}{\enskip{}}\textcolor[HTML]{000000}{(=}\textcolor[HTML]{000000}{\enskip{}}\textcolor[HTML]{000000}{keys\char95{}new}\textcolor[HTML]{000000}{\enskip{}}\textcolor[HTML]{000000}{keys)}\\
\textcolor[HTML]{000000}{\enskip{}\enskip{}\enskip{}\enskip{}\enskip{}\enskip{}\enskip{}\enskip{}\enskip{}\enskip{}\enskip{}\enskip{}\enskip{}\enskip{}\enskip{}\enskip{}}\textcolor[HTML]{000000}{(=}\textcolor[HTML]{000000}{\enskip{}}\textcolor[HTML]{000000}{size\char95{}new}\textcolor[HTML]{000000}{\enskip{}}\textcolor[HTML]{000000}{size)}\\
\textcolor[HTML]{000000}{\enskip{}\enskip{}\enskip{}\enskip{}\enskip{}\enskip{}\enskip{}\enskip{}\enskip{}\enskip{}\enskip{}\enskip{}\enskip{}\enskip{}\enskip{}\enskip{}}\textcolor[HTML]{000000}{(=}\textcolor[HTML]{000000}{\enskip{}}\textcolor[HTML]{000000}{H\char95{}new}\textcolor[HTML]{000000}{\enskip{}}\textcolor[HTML]{000000}{H)}\\
\textcolor[HTML]{000000}{\enskip{}\enskip{}\enskip{}\enskip{}\enskip{}\enskip{}\enskip{}\enskip{}\enskip{}\enskip{}\enskip{}\enskip{}\enskip{}\enskip{}\enskip{}\enskip{}}\textcolor[HTML]{000000}{(not}\textcolor[HTML]{000000}{\enskip{}}\textcolor[HTML]{000000}{result)))}\\
\textcolor[HTML]{000000}{\enskip{}\enskip{}\enskip{}\enskip{}}\textcolor[HTML]{008080}{terms}\textcolor[HTML]{C000C0}{:}\\
\textcolor[HTML]{000000}{\enskip{}\enskip{}\enskip{}\enskip{}\enskip{}\enskip{}}\textcolor[HTML]{008080}{Int}\textcolor[HTML]{C000C0}{:}\textcolor[HTML]{000000}{\enskip{}}\textcolor[HTML]{C000C0}{\char91{}}\textcolor[HTML]{000000}{size}\textcolor[HTML]{000000}{,\enskip{}}\textcolor[HTML]{C00000}{1}\textcolor[HTML]{000000}{,\enskip{}}\textcolor[HTML]{000000}{(-}\textcolor[HTML]{000000}{\enskip{}}\textcolor[HTML]{000000}{size}\textcolor[HTML]{000000}{\enskip{}}\textcolor[HTML]{000000}{1)}\textcolor[HTML]{C000C0}{\char93{}}\\
\textcolor[HTML]{000000}{\enskip{}\enskip{}\enskip{}\enskip{}\enskip{}\enskip{}}\textcolor[HTML]{008080}{E}\textcolor[HTML]{C000C0}{:}\textcolor[HTML]{000000}{\enskip{}}\textcolor[HTML]{C000C0}{\char91{}}\textcolor[HTML]{000000}{\char36{}1}\textcolor[HTML]{C000C0}{\char93{}}\\
\textcolor[HTML]{000000}{\enskip{}\enskip{}\enskip{}\enskip{}\enskip{}\enskip{}}\textcolor[HTML]{000000}{(Set}\textcolor[HTML]{000000}{\enskip{}}\textcolor[HTML]{000000}{E)}\textcolor[HTML]{000000}{:\enskip{}}\textcolor[HTML]{C000C0}{\char91{}}\textcolor[HTML]{000000}{keys}\textcolor[HTML]{000000}{,\enskip{}}\textcolor[HTML]{000000}{(singleton}\textcolor[HTML]{000000}{\enskip{}}\textcolor[HTML]{000000}{\char36{}1)}\textcolor[HTML]{000000}{,\enskip{}}\textcolor[HTML]{000000}{(setminus}\textcolor[HTML]{000000}{\enskip{}}\textcolor[HTML]{000000}{keys}\textcolor[HTML]{000000}{\enskip{}}\textcolor[HTML]{000000}{(singleton}\textcolor[HTML]{000000}{\enskip{}}\textcolor[HTML]{000000}{\char36{}1))}\textcolor[HTML]{C000C0}{\char93{}}\\
\textcolor[HTML]{000000}{\enskip{}\enskip{}\enskip{}\enskip{}\enskip{}\enskip{}}\textcolor[HTML]{000000}{(Array}\textcolor[HTML]{000000}{\enskip{}}\textcolor[HTML]{000000}{E}\textcolor[HTML]{000000}{\enskip{}}\textcolor[HTML]{000000}{F)}\textcolor[HTML]{000000}{:\enskip{}}\textcolor[HTML]{C000C0}{\char91{}}\textcolor[HTML]{000000}{H}\textcolor[HTML]{C000C0}{\char93{}}\\
\textcolor[HTML]{000000}{\enskip{}\enskip{}}\textcolor[HTML]{AF5F00}{-\enskip{}}\textcolor[HTML]{008080}{name}\textcolor[HTML]{C000C0}{:}\textcolor[HTML]{000000}{\enskip{}}\textcolor[HTML]{000000}{put}\\
\textcolor[HTML]{000000}{\enskip{}\enskip{}\enskip{}\enskip{}}\textcolor[HTML]{008080}{args}\textcolor[HTML]{C000C0}{:}\\
\textcolor[HTML]{000000}{\enskip{}\enskip{}\enskip{}\enskip{}\enskip{}\enskip{}}\textcolor[HTML]{AF5F00}{-\enskip{}}\textcolor[HTML]{008080}{name}\textcolor[HTML]{C000C0}{:}\textcolor[HTML]{000000}{\enskip{}}\textcolor[HTML]{000000}{k0}\\
\textcolor[HTML]{000000}{\enskip{}\enskip{}\enskip{}\enskip{}\enskip{}\enskip{}\enskip{}\enskip{}}\textcolor[HTML]{008080}{type}\textcolor[HTML]{C000C0}{:}\textcolor[HTML]{000000}{\enskip{}}\textcolor[HTML]{000000}{E}\\
\textcolor[HTML]{000000}{\enskip{}\enskip{}\enskip{}\enskip{}\enskip{}\enskip{}}\textcolor[HTML]{AF5F00}{-\enskip{}}\textcolor[HTML]{008080}{name}\textcolor[HTML]{C000C0}{:}\textcolor[HTML]{000000}{\enskip{}}\textcolor[HTML]{000000}{v0}\\
\textcolor[HTML]{000000}{\enskip{}\enskip{}\enskip{}\enskip{}\enskip{}\enskip{}\enskip{}\enskip{}}\textcolor[HTML]{008080}{type}\textcolor[HTML]{C000C0}{:}\textcolor[HTML]{000000}{\enskip{}}\textcolor[HTML]{000000}{F}\\
\textcolor[HTML]{000000}{\enskip{}\enskip{}\enskip{}\enskip{}}\textcolor[HTML]{008080}{return}\textcolor[HTML]{C000C0}{:}\\
\textcolor[HTML]{000000}{\enskip{}\enskip{}\enskip{}\enskip{}\enskip{}\enskip{}}\textcolor[HTML]{AF5F00}{-\enskip{}}\textcolor[HTML]{008080}{name}\textcolor[HTML]{C000C0}{:}\textcolor[HTML]{000000}{\enskip{}}\textcolor[HTML]{000000}{result}\\
\textcolor[HTML]{000000}{\enskip{}\enskip{}\enskip{}\enskip{}\enskip{}\enskip{}\enskip{}\enskip{}}\textcolor[HTML]{008080}{type}\textcolor[HTML]{C000C0}{:}\textcolor[HTML]{000000}{\enskip{}}\textcolor[HTML]{000000}{Bool}\\
\textcolor[HTML]{000000}{\enskip{}\enskip{}\enskip{}\enskip{}}\textcolor[HTML]{008080}{requires}\textcolor[HTML]{C000C0}{:}\textcolor[HTML]{000000}{\enskip{}|}\\
\textcolor[HTML]{000000}{\enskip{}\enskip{}\enskip{}\enskip{}\enskip{}\enskip{}}\textcolor[HTML]{C00000}{true}\\
\textcolor[HTML]{000000}{\enskip{}\enskip{}\enskip{}\enskip{}}\textcolor[HTML]{008080}{ensures}\textcolor[HTML]{C000C0}{:}\textcolor[HTML]{000000}{\enskip{}|}\\
\textcolor[HTML]{000000}{\enskip{}\enskip{}\enskip{}\enskip{}\enskip{}\enskip{}}\textcolor[HTML]{000000}{(ite}\textcolor[HTML]{000000}{\enskip{}}\textcolor[HTML]{000000}{(member}\textcolor[HTML]{000000}{\enskip{}}\textcolor[HTML]{000000}{k0}\textcolor[HTML]{000000}{\enskip{}}\textcolor[HTML]{000000}{keys)}\\
\textcolor[HTML]{000000}{\enskip{}\enskip{}\enskip{}\enskip{}\enskip{}\enskip{}\enskip{}\enskip{}\enskip{}\enskip{}\enskip{}}\textcolor[HTML]{000000}{(and}\textcolor[HTML]{000000}{\enskip{}}\textcolor[HTML]{000000}{(=}\textcolor[HTML]{000000}{\enskip{}}\textcolor[HTML]{000000}{keys\char95{}new}\textcolor[HTML]{000000}{\enskip{}}\textcolor[HTML]{000000}{keys)}\\
\textcolor[HTML]{000000}{\enskip{}\enskip{}\enskip{}\enskip{}\enskip{}\enskip{}\enskip{}\enskip{}\enskip{}\enskip{}\enskip{}\enskip{}\enskip{}\enskip{}\enskip{}\enskip{}}\textcolor[HTML]{000000}{(=}\textcolor[HTML]{000000}{\enskip{}}\textcolor[HTML]{000000}{size\char95{}new}\textcolor[HTML]{000000}{\enskip{}}\textcolor[HTML]{000000}{size)}\\
\textcolor[HTML]{000000}{\enskip{}\enskip{}\enskip{}\enskip{}\enskip{}\enskip{}\enskip{}\enskip{}\enskip{}\enskip{}\enskip{}\enskip{}\enskip{}\enskip{}\enskip{}\enskip{}}\textcolor[HTML]{000000}{(ite}\textcolor[HTML]{000000}{\enskip{}}\textcolor[HTML]{000000}{(=}\textcolor[HTML]{000000}{\enskip{}}\textcolor[HTML]{000000}{v0}\textcolor[HTML]{000000}{\enskip{}}\textcolor[HTML]{000000}{(select}\textcolor[HTML]{000000}{\enskip{}}\textcolor[HTML]{000000}{H}\textcolor[HTML]{000000}{\enskip{}}\textcolor[HTML]{000000}{k0))}\\
\textcolor[HTML]{000000}{\enskip{}\enskip{}\enskip{}\enskip{}\enskip{}\enskip{}\enskip{}\enskip{}\enskip{}\enskip{}\enskip{}\enskip{}\enskip{}\enskip{}\enskip{}\enskip{}\enskip{}\enskip{}\enskip{}\enskip{}\enskip{}}\textcolor[HTML]{000000}{(and}\textcolor[HTML]{000000}{\enskip{}}\textcolor[HTML]{000000}{(not}\textcolor[HTML]{000000}{\enskip{}}\textcolor[HTML]{000000}{result)}\\
\textcolor[HTML]{000000}{\enskip{}\enskip{}\enskip{}\enskip{}\enskip{}\enskip{}\enskip{}\enskip{}\enskip{}\enskip{}\enskip{}\enskip{}\enskip{}\enskip{}\enskip{}\enskip{}\enskip{}\enskip{}\enskip{}\enskip{}\enskip{}\enskip{}\enskip{}\enskip{}\enskip{}\enskip{}}\textcolor[HTML]{000000}{(=}\textcolor[HTML]{000000}{\enskip{}}\textcolor[HTML]{000000}{H\char95{}new}\textcolor[HTML]{000000}{\enskip{}}\textcolor[HTML]{000000}{H))}\\
\textcolor[HTML]{000000}{\enskip{}\enskip{}\enskip{}\enskip{}\enskip{}\enskip{}\enskip{}\enskip{}\enskip{}\enskip{}\enskip{}\enskip{}\enskip{}\enskip{}\enskip{}\enskip{}\enskip{}\enskip{}\enskip{}\enskip{}\enskip{}}\textcolor[HTML]{000000}{(and}\textcolor[HTML]{000000}{\enskip{}}\textcolor[HTML]{000000}{result}\\
\textcolor[HTML]{000000}{\enskip{}\enskip{}\enskip{}\enskip{}\enskip{}\enskip{}\enskip{}\enskip{}\enskip{}\enskip{}\enskip{}\enskip{}\enskip{}\enskip{}\enskip{}\enskip{}\enskip{}\enskip{}\enskip{}\enskip{}\enskip{}\enskip{}\enskip{}\enskip{}\enskip{}\enskip{}}\textcolor[HTML]{000000}{(=}\textcolor[HTML]{000000}{\enskip{}}\textcolor[HTML]{000000}{H\char95{}new}\textcolor[HTML]{000000}{\enskip{}}\textcolor[HTML]{000000}{(store}\textcolor[HTML]{000000}{\enskip{}}\textcolor[HTML]{000000}{H}\textcolor[HTML]{000000}{\enskip{}}\textcolor[HTML]{000000}{k0}\textcolor[HTML]{000000}{\enskip{}}\textcolor[HTML]{000000}{v0)))))}\\
\textcolor[HTML]{000000}{\enskip{}\enskip{}\enskip{}\enskip{}\enskip{}\enskip{}\enskip{}\enskip{}\enskip{}\enskip{}\enskip{}}\textcolor[HTML]{000000}{(and}\textcolor[HTML]{000000}{\enskip{}}\textcolor[HTML]{000000}{(=}\textcolor[HTML]{000000}{\enskip{}}\textcolor[HTML]{000000}{keys\char95{}new}\textcolor[HTML]{000000}{\enskip{}}\textcolor[HTML]{000000}{(insert}\textcolor[HTML]{000000}{\enskip{}}\textcolor[HTML]{000000}{k0}\textcolor[HTML]{000000}{\enskip{}}\textcolor[HTML]{000000}{keys))}\\
\textcolor[HTML]{000000}{\enskip{}\enskip{}\enskip{}\enskip{}\enskip{}\enskip{}\enskip{}\enskip{}\enskip{}\enskip{}\enskip{}\enskip{}\enskip{}\enskip{}\enskip{}\enskip{}}\textcolor[HTML]{000000}{(=}\textcolor[HTML]{000000}{\enskip{}}\textcolor[HTML]{000000}{size\char95{}new}\textcolor[HTML]{000000}{\enskip{}}\textcolor[HTML]{000000}{(+}\textcolor[HTML]{000000}{\enskip{}}\textcolor[HTML]{000000}{size}\textcolor[HTML]{000000}{\enskip{}}\textcolor[HTML]{000000}{1))}\\
\textcolor[HTML]{000000}{\enskip{}\enskip{}\enskip{}\enskip{}\enskip{}\enskip{}\enskip{}\enskip{}\enskip{}\enskip{}\enskip{}\enskip{}\enskip{}\enskip{}\enskip{}\enskip{}}\textcolor[HTML]{000000}{result}\\
\textcolor[HTML]{000000}{\enskip{}\enskip{}\enskip{}\enskip{}\enskip{}\enskip{}\enskip{}\enskip{}\enskip{}\enskip{}\enskip{}\enskip{}\enskip{}\enskip{}\enskip{}\enskip{}}\textcolor[HTML]{000000}{(=}\textcolor[HTML]{000000}{\enskip{}}\textcolor[HTML]{000000}{H\char95{}new}\textcolor[HTML]{000000}{\enskip{}}\textcolor[HTML]{000000}{(store}\textcolor[HTML]{000000}{\enskip{}}\textcolor[HTML]{000000}{H}\textcolor[HTML]{000000}{\enskip{}}\textcolor[HTML]{000000}{k0}\textcolor[HTML]{000000}{\enskip{}}\textcolor[HTML]{000000}{v0))))}\\
\textcolor[HTML]{000000}{\enskip{}\enskip{}\enskip{}\enskip{}}\textcolor[HTML]{008080}{terms}\textcolor[HTML]{C000C0}{:}\\
\textcolor[HTML]{000000}{\enskip{}\enskip{}\enskip{}\enskip{}\enskip{}\enskip{}}\textcolor[HTML]{008080}{Int}\textcolor[HTML]{C000C0}{:}\textcolor[HTML]{000000}{\enskip{}}\textcolor[HTML]{C000C0}{\char91{}}\textcolor[HTML]{000000}{size}\textcolor[HTML]{000000}{,\enskip{}}\textcolor[HTML]{C00000}{1}\textcolor[HTML]{000000}{,\enskip{}}\textcolor[HTML]{000000}{(+}\textcolor[HTML]{000000}{\enskip{}}\textcolor[HTML]{000000}{size}\textcolor[HTML]{000000}{\enskip{}}\textcolor[HTML]{000000}{1)}\textcolor[HTML]{C000C0}{\char93{}}\\
\textcolor[HTML]{000000}{\enskip{}\enskip{}\enskip{}\enskip{}\enskip{}\enskip{}}\textcolor[HTML]{008080}{E}\textcolor[HTML]{C000C0}{:}\textcolor[HTML]{000000}{\enskip{}}\textcolor[HTML]{C000C0}{\char91{}}\textcolor[HTML]{000000}{\char36{}1}\textcolor[HTML]{C000C0}{\char93{}}\\
\textcolor[HTML]{000000}{\enskip{}\enskip{}\enskip{}\enskip{}\enskip{}\enskip{}}\textcolor[HTML]{008080}{F}\textcolor[HTML]{C000C0}{:}\textcolor[HTML]{000000}{\enskip{}}\textcolor[HTML]{C000C0}{\char91{}}\textcolor[HTML]{000000}{\char36{}2}\textcolor[HTML]{000000}{,\enskip{}}\textcolor[HTML]{000000}{(select}\textcolor[HTML]{000000}{\enskip{}}\textcolor[HTML]{000000}{H}\textcolor[HTML]{000000}{\enskip{}}\textcolor[HTML]{000000}{\char36{}1)}\textcolor[HTML]{000000}{,\enskip{}}\textcolor[HTML]{C000C0}{\char93{}}\\
\textcolor[HTML]{000000}{\enskip{}\enskip{}\enskip{}\enskip{}\enskip{}\enskip{}}\textcolor[HTML]{000000}{(Set}\textcolor[HTML]{000000}{\enskip{}}\textcolor[HTML]{000000}{E)}\textcolor[HTML]{000000}{:\enskip{}}\textcolor[HTML]{C000C0}{\char91{}}\textcolor[HTML]{000000}{keys}\textcolor[HTML]{000000}{,\enskip{}}\textcolor[HTML]{000000}{(insert}\textcolor[HTML]{000000}{\enskip{}}\textcolor[HTML]{000000}{\char36{}1}\textcolor[HTML]{000000}{\enskip{}}\textcolor[HTML]{000000}{keys)}\textcolor[HTML]{C000C0}{\char93{}}\\
\textcolor[HTML]{000000}{\enskip{}\enskip{}\enskip{}\enskip{}\enskip{}\enskip{}}\textcolor[HTML]{000000}{(Array}\textcolor[HTML]{000000}{\enskip{}}\textcolor[HTML]{000000}{E}\textcolor[HTML]{000000}{\enskip{}}\textcolor[HTML]{000000}{F)}\textcolor[HTML]{000000}{:\enskip{}}\textcolor[HTML]{C000C0}{\char91{}}\textcolor[HTML]{000000}{H}\textcolor[HTML]{000000}{,\enskip{}}\textcolor[HTML]{000000}{(store}\textcolor[HTML]{000000}{\enskip{}}\textcolor[HTML]{000000}{H}\textcolor[HTML]{000000}{\enskip{}}\textcolor[HTML]{000000}{\char36{}1}\textcolor[HTML]{000000}{\enskip{}}\textcolor[HTML]{000000}{\char36{}2)}\textcolor[HTML]{C000C0}{\char93{}}\\
\textcolor[HTML]{000000}{\enskip{}\enskip{}}\textcolor[HTML]{AF5F00}{-\enskip{}}\textcolor[HTML]{008080}{name}\textcolor[HTML]{C000C0}{:}\textcolor[HTML]{000000}{\enskip{}}\textcolor[HTML]{000000}{get}\\
\textcolor[HTML]{000000}{\enskip{}\enskip{}\enskip{}\enskip{}}\textcolor[HTML]{008080}{args}\textcolor[HTML]{C000C0}{:}\\
\textcolor[HTML]{000000}{\enskip{}\enskip{}\enskip{}\enskip{}\enskip{}\enskip{}}\textcolor[HTML]{AF5F00}{-\enskip{}}\textcolor[HTML]{008080}{name}\textcolor[HTML]{C000C0}{:}\textcolor[HTML]{000000}{\enskip{}}\textcolor[HTML]{000000}{k0}\\
\textcolor[HTML]{000000}{\enskip{}\enskip{}\enskip{}\enskip{}\enskip{}\enskip{}\enskip{}\enskip{}}\textcolor[HTML]{008080}{type}\textcolor[HTML]{C000C0}{:}\textcolor[HTML]{000000}{\enskip{}}\textcolor[HTML]{000000}{E}\\
\textcolor[HTML]{000000}{\enskip{}\enskip{}\enskip{}\enskip{}}\textcolor[HTML]{008080}{return}\textcolor[HTML]{C000C0}{:}\\
\textcolor[HTML]{000000}{\enskip{}\enskip{}\enskip{}\enskip{}\enskip{}\enskip{}}\textcolor[HTML]{AF5F00}{-\enskip{}}\textcolor[HTML]{008080}{name}\textcolor[HTML]{C000C0}{:}\textcolor[HTML]{000000}{\enskip{}}\textcolor[HTML]{000000}{result}\\
\textcolor[HTML]{000000}{\enskip{}\enskip{}\enskip{}\enskip{}\enskip{}\enskip{}\enskip{}\enskip{}}\textcolor[HTML]{008080}{type}\textcolor[HTML]{C000C0}{:}\textcolor[HTML]{000000}{\enskip{}}\textcolor[HTML]{000000}{F}\\
\textcolor[HTML]{000000}{\enskip{}\enskip{}\enskip{}\enskip{}}\textcolor[HTML]{008080}{requires}\textcolor[HTML]{C000C0}{:}\textcolor[HTML]{000000}{\enskip{}|}\\
\textcolor[HTML]{000000}{\enskip{}\enskip{}\enskip{}\enskip{}\enskip{}\enskip{}}\textcolor[HTML]{000000}{(member}\textcolor[HTML]{000000}{\enskip{}}\textcolor[HTML]{000000}{k0}\textcolor[HTML]{000000}{\enskip{}}\textcolor[HTML]{000000}{keys)}\\
\textcolor[HTML]{000000}{\enskip{}\enskip{}\enskip{}\enskip{}}\textcolor[HTML]{008080}{ensures}\textcolor[HTML]{C000C0}{:}\textcolor[HTML]{000000}{\enskip{}|}\\
\textcolor[HTML]{000000}{\enskip{}\enskip{}\enskip{}\enskip{}\enskip{}\enskip{}}\textcolor[HTML]{000000}{(and}\textcolor[HTML]{000000}{\enskip{}}\textcolor[HTML]{000000}{(=}\textcolor[HTML]{000000}{\enskip{}}\textcolor[HTML]{000000}{keys\char95{}new}\textcolor[HTML]{000000}{\enskip{}}\textcolor[HTML]{000000}{keys)}\\
\textcolor[HTML]{000000}{\enskip{}\enskip{}\enskip{}\enskip{}\enskip{}\enskip{}\enskip{}\enskip{}\enskip{}\enskip{}\enskip{}}\textcolor[HTML]{000000}{(=}\textcolor[HTML]{000000}{\enskip{}}\textcolor[HTML]{000000}{H\char95{}new}\textcolor[HTML]{000000}{\enskip{}}\textcolor[HTML]{000000}{H)}\\
\textcolor[HTML]{000000}{\enskip{}\enskip{}\enskip{}\enskip{}\enskip{}\enskip{}\enskip{}\enskip{}\enskip{}\enskip{}\enskip{}}\textcolor[HTML]{000000}{(=}\textcolor[HTML]{000000}{\enskip{}}\textcolor[HTML]{000000}{size\char95{}new}\textcolor[HTML]{000000}{\enskip{}}\textcolor[HTML]{000000}{size)}\\
\textcolor[HTML]{000000}{\enskip{}\enskip{}\enskip{}\enskip{}\enskip{}\enskip{}\enskip{}\enskip{}\enskip{}\enskip{}\enskip{}}\textcolor[HTML]{000000}{(=}\textcolor[HTML]{000000}{\enskip{}}\textcolor[HTML]{000000}{(select}\textcolor[HTML]{000000}{\enskip{}}\textcolor[HTML]{000000}{H}\textcolor[HTML]{000000}{\enskip{}}\textcolor[HTML]{000000}{k0)}\textcolor[HTML]{000000}{\enskip{}}\textcolor[HTML]{000000}{result)}\\
\textcolor[HTML]{000000}{\enskip{}\enskip{}\enskip{}\enskip{}\enskip{}\enskip{}\enskip{}\enskip{}\enskip{}\enskip{}\enskip{}}\textcolor[HTML]{000000}{)}\\
\textcolor[HTML]{000000}{\enskip{}\enskip{}\enskip{}\enskip{}}\textcolor[HTML]{008080}{terms}\textcolor[HTML]{C000C0}{:}\\
\textcolor[HTML]{000000}{\enskip{}\enskip{}\enskip{}\enskip{}\enskip{}\enskip{}}\textcolor[HTML]{008080}{Int}\textcolor[HTML]{C000C0}{:}\textcolor[HTML]{000000}{\enskip{}}\textcolor[HTML]{C000C0}{\char91{}}\textcolor[HTML]{000000}{size}\textcolor[HTML]{C000C0}{\char93{}}\\
\textcolor[HTML]{000000}{\enskip{}\enskip{}\enskip{}\enskip{}\enskip{}\enskip{}}\textcolor[HTML]{008080}{E}\textcolor[HTML]{C000C0}{:}\textcolor[HTML]{000000}{\enskip{}}\textcolor[HTML]{C000C0}{\char91{}}\textcolor[HTML]{000000}{\char36{}1}\textcolor[HTML]{C000C0}{\char93{}}\\
\textcolor[HTML]{000000}{\enskip{}\enskip{}\enskip{}\enskip{}\enskip{}\enskip{}}\textcolor[HTML]{008080}{F}\textcolor[HTML]{C000C0}{:}\textcolor[HTML]{000000}{\enskip{}}\textcolor[HTML]{C000C0}{\char91{}}\textcolor[HTML]{000000}{(select}\textcolor[HTML]{000000}{\enskip{}}\textcolor[HTML]{000000}{H}\textcolor[HTML]{000000}{\enskip{}}\textcolor[HTML]{000000}{\char36{}1)}\textcolor[HTML]{C000C0}{\char93{}}\\
\textcolor[HTML]{000000}{\enskip{}\enskip{}\enskip{}\enskip{}\enskip{}\enskip{}}\textcolor[HTML]{000000}{(Set}\textcolor[HTML]{000000}{\enskip{}}\textcolor[HTML]{000000}{E)}\textcolor[HTML]{000000}{:\enskip{}}\textcolor[HTML]{C000C0}{\char91{}}\textcolor[HTML]{000000}{keys}\textcolor[HTML]{C000C0}{\char93{}}\\
\textcolor[HTML]{000000}{\enskip{}\enskip{}\enskip{}\enskip{}\enskip{}\enskip{}}\textcolor[HTML]{000000}{(Array}\textcolor[HTML]{000000}{\enskip{}}\textcolor[HTML]{000000}{E}\textcolor[HTML]{000000}{\enskip{}}\textcolor[HTML]{000000}{F)}\textcolor[HTML]{000000}{:\enskip{}}\textcolor[HTML]{C000C0}{\char91{}}\textcolor[HTML]{000000}{H}\textcolor[HTML]{C000C0}{\char93{}}\\
\textcolor[HTML]{000000}{\enskip{}\enskip{}}\textcolor[HTML]{AF5F00}{-\enskip{}}\textcolor[HTML]{008080}{name}\textcolor[HTML]{C000C0}{:}\textcolor[HTML]{000000}{\enskip{}}\textcolor[HTML]{000000}{size}\\
\textcolor[HTML]{000000}{\enskip{}\enskip{}\enskip{}\enskip{}}\textcolor[HTML]{008080}{args}\textcolor[HTML]{C000C0}{:}\textcolor[HTML]{000000}{\enskip{}}\textcolor[HTML]{C000C0}{\char91{}\char93{}}\\
\textcolor[HTML]{000000}{\enskip{}\enskip{}\enskip{}\enskip{}}\textcolor[HTML]{008080}{return}\textcolor[HTML]{C000C0}{:}\\
\textcolor[HTML]{000000}{\enskip{}\enskip{}\enskip{}\enskip{}\enskip{}\enskip{}}\textcolor[HTML]{AF5F00}{-\enskip{}}\textcolor[HTML]{008080}{name}\textcolor[HTML]{C000C0}{:}\textcolor[HTML]{000000}{\enskip{}}\textcolor[HTML]{000000}{result}\\
\textcolor[HTML]{000000}{\enskip{}\enskip{}\enskip{}\enskip{}\enskip{}\enskip{}\enskip{}\enskip{}}\textcolor[HTML]{008080}{type}\textcolor[HTML]{C000C0}{:}\textcolor[HTML]{000000}{\enskip{}}\textcolor[HTML]{000000}{Int}\\
\textcolor[HTML]{000000}{\enskip{}\enskip{}\enskip{}\enskip{}}\textcolor[HTML]{008080}{requires}\textcolor[HTML]{C000C0}{:}\textcolor[HTML]{000000}{\enskip{}|}\\
\textcolor[HTML]{000000}{\enskip{}\enskip{}\enskip{}\enskip{}\enskip{}\enskip{}}\textcolor[HTML]{C00000}{true}\\
\textcolor[HTML]{000000}{\enskip{}\enskip{}\enskip{}\enskip{}}\textcolor[HTML]{008080}{ensures}\textcolor[HTML]{C000C0}{:}\textcolor[HTML]{000000}{\enskip{}|}\\
\textcolor[HTML]{000000}{\enskip{}\enskip{}\enskip{}\enskip{}\enskip{}\enskip{}}\textcolor[HTML]{000000}{(and}\textcolor[HTML]{000000}{\enskip{}}\textcolor[HTML]{000000}{(=}\textcolor[HTML]{000000}{\enskip{}}\textcolor[HTML]{000000}{keys\char95{}new}\textcolor[HTML]{000000}{\enskip{}}\textcolor[HTML]{000000}{keys)}\\
\textcolor[HTML]{000000}{\enskip{}\enskip{}\enskip{}\enskip{}\enskip{}\enskip{}\enskip{}\enskip{}\enskip{}\enskip{}\enskip{}}\textcolor[HTML]{000000}{(=}\textcolor[HTML]{000000}{\enskip{}}\textcolor[HTML]{000000}{H\char95{}new}\textcolor[HTML]{000000}{\enskip{}}\textcolor[HTML]{000000}{H)}\\
\textcolor[HTML]{000000}{\enskip{}\enskip{}\enskip{}\enskip{}\enskip{}\enskip{}\enskip{}\enskip{}\enskip{}\enskip{}\enskip{}}\textcolor[HTML]{000000}{(=}\textcolor[HTML]{000000}{\enskip{}}\textcolor[HTML]{000000}{size\char95{}new}\textcolor[HTML]{000000}{\enskip{}}\textcolor[HTML]{000000}{size)}\\
\textcolor[HTML]{000000}{\enskip{}\enskip{}\enskip{}\enskip{}\enskip{}\enskip{}\enskip{}\enskip{}\enskip{}\enskip{}\enskip{}}\textcolor[HTML]{000000}{(=}\textcolor[HTML]{000000}{\enskip{}}\textcolor[HTML]{000000}{size}\textcolor[HTML]{000000}{\enskip{}}\textcolor[HTML]{000000}{result))}\\
\textcolor[HTML]{000000}{\enskip{}\enskip{}\enskip{}\enskip{}}\textcolor[HTML]{008080}{terms}\textcolor[HTML]{C000C0}{:}\\
\textcolor[HTML]{000000}{\enskip{}\enskip{}\enskip{}\enskip{}\enskip{}\enskip{}}\textcolor[HTML]{008080}{Int}\textcolor[HTML]{C000C0}{:}\textcolor[HTML]{000000}{\enskip{}}\textcolor[HTML]{C000C0}{\char91{}}\textcolor[HTML]{000000}{size}\textcolor[HTML]{C000C0}{\char93{}}\\
\textcolor[HTML]{000000}{\enskip{}\enskip{}\enskip{}\enskip{}\enskip{}\enskip{}}\textcolor[HTML]{000000}{(Set}\textcolor[HTML]{000000}{\enskip{}}\textcolor[HTML]{000000}{E)}\textcolor[HTML]{000000}{:\enskip{}}\textcolor[HTML]{C000C0}{\char91{}}\textcolor[HTML]{000000}{keys}\textcolor[HTML]{C000C0}{\char93{}}\\
\textcolor[HTML]{000000}{\enskip{}\enskip{}\enskip{}\enskip{}\enskip{}\enskip{}}\textcolor[HTML]{000000}{(Array}\textcolor[HTML]{000000}{\enskip{}}\textcolor[HTML]{000000}{E}\textcolor[HTML]{000000}{\enskip{}}\textcolor[HTML]{000000}{F)}\textcolor[HTML]{000000}{:\enskip{}}\textcolor[HTML]{C000C0}{\char91{}}\textcolor[HTML]{000000}{H}\textcolor[HTML]{C000C0}{\char93{}}\\
\\
\textcolor[HTML]{008080}{predicates}\textcolor[HTML]{C000C0}{:}\\
\textcolor[HTML]{000000}{\enskip{}\enskip{}}\textcolor[HTML]{AF5F00}{-\enskip{}}\textcolor[HTML]{008080}{name}\textcolor[HTML]{C000C0}{:}\textcolor[HTML]{000000}{\enskip{}}\textcolor[HTML]{C00000}{"}\textcolor[HTML]{C00000}{=}\textcolor[HTML]{C00000}{"}\\
\textcolor[HTML]{000000}{\enskip{}\enskip{}\enskip{}\enskip{}}\textcolor[HTML]{008080}{type}\textcolor[HTML]{C000C0}{:}\textcolor[HTML]{000000}{\enskip{}}\textcolor[HTML]{C000C0}{\char91{}}\textcolor[HTML]{000000}{Int}\textcolor[HTML]{000000}{,\enskip{}}\textcolor[HTML]{000000}{Int}\textcolor[HTML]{C000C0}{\char93{}}\\
\textcolor[HTML]{000000}{\enskip{}\enskip{}}\textcolor[HTML]{AF5F00}{-\enskip{}}\textcolor[HTML]{008080}{name}\textcolor[HTML]{C000C0}{:}\textcolor[HTML]{000000}{\enskip{}}\textcolor[HTML]{C00000}{"}\textcolor[HTML]{C00000}{=}\textcolor[HTML]{C00000}{"}\\
\textcolor[HTML]{000000}{\enskip{}\enskip{}\enskip{}\enskip{}}\textcolor[HTML]{008080}{type}\textcolor[HTML]{C000C0}{:}\textcolor[HTML]{000000}{\enskip{}}\textcolor[HTML]{C000C0}{\char91{}}\textcolor[HTML]{000000}{E}\textcolor[HTML]{000000}{,\enskip{}}\textcolor[HTML]{000000}{E}\textcolor[HTML]{C000C0}{\char93{}}\\
\textcolor[HTML]{000000}{\enskip{}\enskip{}}\textcolor[HTML]{AF5F00}{-\enskip{}}\textcolor[HTML]{008080}{name}\textcolor[HTML]{C000C0}{:}\textcolor[HTML]{000000}{\enskip{}}\textcolor[HTML]{C00000}{"}\textcolor[HTML]{C00000}{=}\textcolor[HTML]{C00000}{"}\\
\textcolor[HTML]{000000}{\enskip{}\enskip{}\enskip{}\enskip{}}\textcolor[HTML]{008080}{type}\textcolor[HTML]{C000C0}{:}\textcolor[HTML]{000000}{\enskip{}}\textcolor[HTML]{C000C0}{\char91{}}\textcolor[HTML]{000000}{F}\textcolor[HTML]{000000}{,\enskip{}}\textcolor[HTML]{000000}{F}\textcolor[HTML]{C000C0}{\char93{}}\\
\textcolor[HTML]{000000}{\enskip{}\enskip{}}\textcolor[HTML]{AF5F00}{-\enskip{}}\textcolor[HTML]{008080}{name}\textcolor[HTML]{C000C0}{:}\textcolor[HTML]{000000}{\enskip{}}\textcolor[HTML]{C00000}{"}\textcolor[HTML]{C00000}{=}\textcolor[HTML]{C00000}{"}\\
\textcolor[HTML]{000000}{\enskip{}\enskip{}\enskip{}\enskip{}}\textcolor[HTML]{008080}{type}\textcolor[HTML]{C000C0}{:}\textcolor[HTML]{000000}{\enskip{}}\textcolor[HTML]{C000C0}{\char91{}}\textcolor[HTML]{000000}{(Set}\textcolor[HTML]{000000}{\enskip{}}\textcolor[HTML]{000000}{E)}\textcolor[HTML]{000000}{,\enskip{}}\textcolor[HTML]{000000}{(Set}\textcolor[HTML]{000000}{\enskip{}}\textcolor[HTML]{000000}{E)}\textcolor[HTML]{C000C0}{\char93{}}\\
\textcolor[HTML]{000000}{\enskip{}\enskip{}}\textcolor[HTML]{AF5F00}{-\enskip{}}\textcolor[HTML]{008080}{name}\textcolor[HTML]{C000C0}{:}\textcolor[HTML]{000000}{\enskip{}}\textcolor[HTML]{C00000}{"}\textcolor[HTML]{C00000}{=}\textcolor[HTML]{C00000}{"}\\
\textcolor[HTML]{000000}{\enskip{}\enskip{}\enskip{}\enskip{}}\textcolor[HTML]{008080}{type}\textcolor[HTML]{C000C0}{:}\textcolor[HTML]{000000}{\enskip{}}\textcolor[HTML]{C000C0}{\char91{}}\textcolor[HTML]{000000}{(Array}\textcolor[HTML]{000000}{\enskip{}}\textcolor[HTML]{000000}{E}\textcolor[HTML]{000000}{\enskip{}}\textcolor[HTML]{000000}{F)}\textcolor[HTML]{000000}{,\enskip{}}\textcolor[HTML]{000000}{(Array}\textcolor[HTML]{000000}{\enskip{}}\textcolor[HTML]{000000}{E}\textcolor[HTML]{000000}{\enskip{}}\textcolor[HTML]{000000}{F)}\textcolor[HTML]{C000C0}{\char93{}}\\
\textcolor[HTML]{000000}{\enskip{}\enskip{}}\textcolor[HTML]{AF5F00}{-\enskip{}}\textcolor[HTML]{008080}{name}\textcolor[HTML]{C000C0}{:}\textcolor[HTML]{000000}{\enskip{}}\textcolor[HTML]{C00000}{"}\textcolor[HTML]{C00000}{member}\textcolor[HTML]{C00000}{"}\\
\textcolor[HTML]{000000}{\enskip{}\enskip{}\enskip{}\enskip{}}\textcolor[HTML]{008080}{type}\textcolor[HTML]{C000C0}{:}\textcolor[HTML]{000000}{\enskip{}}\textcolor[HTML]{C000C0}{\char91{}}\textcolor[HTML]{000000}{E}\textcolor[HTML]{000000}{,\enskip{}}\textcolor[HTML]{000000}{(Set}\textcolor[HTML]{000000}{\enskip{}}\textcolor[HTML]{000000}{E)}\textcolor[HTML]{C000C0}{\char93{}}\\
\\

}

\begin{itemize}
\item \CCmethod{get} \CCbowtie\ \CCmethod{get}

 Simple:

true

 Poke:

true

\item \CCmethod{get} \CCbowtie\ \CCmethod{haskey}

 Simple:

true

 Poke:

true

\item \CCmethod{put} \CCrightmover\ \CCmethod{get}

 Simple:

[\CCequal{\CCvar{x2}}{\CCselect{\CCvar{H}}{\CCvar{y1}}} $\wedge$ \CCmember{\CCvar{y1}}{\CCvar{keys}}]

 $\vee$ [\CCnot{\CCequal{\CCvar{x2}}{\CCselect{\CCvar{H}}{\CCvar{y1}}}} $\wedge$ \CCnot{\CCequal{\CCvar{y1}}{\CCvar{x1}}}]

 Poke:

[\CCequal{\CCstore{\CCvar{H}}{\CCvar{x1}}{\CCvar{x2}}}{\CCvar{H}} $\wedge$ \CCmember{\CCvar{y1}}{\CCvar{keys}}]

 $\vee$ [\CCnot{\CCequal{\CCstore{\CCvar{H}}{\CCvar{x1}}{\CCvar{x2}}}{\CCvar{H}}} $\wedge$ \CCnot{\CCequal{\CCvar{y1}}{\CCvar{x1}}}]

\item \CCmethod{get} \CCrightmover\ \CCmethod{put}

 Simple:

[\CCequal{\CCselect{\CCvar{H}}{\CCvar{y1}}}{\CCvar{y2}}]

 $\vee$ [\CCnot{\CCequal{\CCselect{\CCvar{H}}{\CCvar{y1}}}{\CCvar{y2}}} $\wedge$ \CCnot{\CCequal{\CCvar{y1}}{\CCvar{x1}}}]

 Poke:

[\CCequal{\CCselect{\CCvar{H}}{\CCvar{y1}}}{\CCvar{y2}}]

 $\vee$ [\CCnot{\CCequal{\CCselect{\CCvar{H}}{\CCvar{y1}}}{\CCvar{y2}}} $\wedge$ \CCnot{\CCequal{\CCvar{y1}}{\CCvar{x1}}}]

\item \CCmethod{remove} \CCrightmover\ \CCmethod{get}

 Simple:

true

 Poke:

true

\item \CCmethod{get} \CCrightmover\ \CCmethod{remove}

 Simple:

[\CCequal{\CCvar{1}}{\CCvar{size}} $\wedge$ \CCnot{\CCequal{\CCvar{y1}}{\CCvar{x1}}}]

 $\vee$ [\CCnot{\CCequal{\CCvar{1}}{\CCvar{size}}} $\wedge$ \CCnot{\CCequal{\CCvar{y1}}{\CCvar{x1}}}]

 Poke:

\CCnot{\CCequal{\CCvar{y1}}{\CCvar{x1}}}

\item \CCmethod{get} \CCbowtie\ \CCmethod{size}

 Simple:

true

 Poke:

true

\item \CCmethod{haskey} \CCbowtie\ \CCmethod{haskey}

 Simple:

true

 Poke:

true

\item \CCmethod{haskey} \CCbowtie\ \CCmethod{put}

 Simple:

[\CCequal{\CCvar{y1}}{\CCvar{x1}} $\wedge$ \CCmember{\CCvar{y1}}{\CCvar{keys}}]

 $\vee$ [\CCnot{\CCequal{\CCvar{y1}}{\CCvar{x1}}}]

 Poke:

[\CCmember{\CCvar{y1}}{\CCvar{keys}}]

 $\vee$ [\CCnot{\CCmember{\CCvar{y1}}{\CCvar{keys}}} $\wedge$ \CCnot{\CCequal{\CCvar{y1}}{\CCvar{x1}}}]

\item \CCmethod{haskey} \CCbowtie\ \CCmethod{remove}

 Simple:

[\CCequal{\CCvar{y1}}{\CCvar{x1}} $\wedge$ \CCequal{\CCvar{1}}{\CCvar{size}} $\wedge$ \CCnot{\CCmember{\CCvar{y1}}{\CCvar{keys}}}]

 $\vee$ [\CCequal{\CCvar{y1}}{\CCvar{x1}} $\wedge$ \CCnot{\CCequal{\CCvar{1}}{\CCvar{size}}} $\wedge$ \CCnot{\CCmember{\CCvar{y1}}{\CCvar{keys}}}]

 $\vee$ [\CCnot{\CCequal{\CCvar{y1}}{\CCvar{x1}}}]

 Poke:

[\CCmember{\CCvar{x1}}{\CCvar{keys}} $\wedge$ \CCnot{\CCequal{\CCvar{y1}}{\CCvar{x1}}}]

 $\vee$ [\CCnot{\CCmember{\CCvar{x1}}{\CCvar{keys}}}]

\item \CCmethod{haskey} \CCbowtie\ \CCmethod{size}

 Simple:

true

 Poke:

true

\item \CCmethod{put} \CCbowtie\ \CCmethod{put}

 Simple:

[\CCequal{\CCvar{x2}}{\CCvar{y2}} $\wedge$ \CCequal{\CCvar{x2}}{\CCselect{\CCvar{H}}{\CCvar{y1}}} $\wedge$ \CCmember{\CCvar{y1}}{\CCvar{keys}}]

 $\vee$ [\CCequal{\CCvar{x2}}{\CCvar{y2}} $\wedge$ \CCequal{\CCvar{x2}}{\CCselect{\CCvar{H}}{\CCvar{y1}}} $\wedge$ \CCnot{\CCmember{\CCvar{y1}}{\CCvar{keys}}} $\wedge$ \CCnot{\CCequal{\CCvar{y1}}{\CCvar{x1}}}]

 $\vee$ [\CCequal{\CCvar{x2}}{\CCvar{y2}} $\wedge$ \CCnot{\CCequal{\CCvar{x2}}{\CCselect{\CCvar{H}}{\CCvar{y1}}}} $\wedge$ \CCnot{\CCequal{\CCvar{y1}}{\CCvar{x1}}}]

 $\vee$ [\CCnot{\CCequal{\CCvar{x2}}{\CCvar{y2}}} $\wedge$ \CCnot{\CCequal{\CCvar{y1}}{\CCvar{x1}}}]

 Poke:

[\CCequal{\CCselect{\CCvar{H}}{\CCvar{y1}}}{\CCvar{y2}} $\wedge$ \CCequal{\CCvar{x2}}{\CCselect{\CCvar{H}}{\CCvar{x1}}} $\wedge$ \CCequal{\CCplus{\CCvar{size}}{\CCvar{1}}}{\CCvar{1}} $\wedge$ \CCmember{\CCvar{y1}}{\CCvar{keys}}]

 $\vee$ [\CCequal{\CCselect{\CCvar{H}}{\CCvar{y1}}}{\CCvar{y2}} $\wedge$ \CCequal{\CCvar{x2}}{\CCselect{\CCvar{H}}{\CCvar{x1}}} $\wedge$ \CCequal{\CCplus{\CCvar{size}}{\CCvar{1}}}{\CCvar{1}} $\wedge$ \CCnot{\CCmember{\CCvar{y1}}{\CCvar{keys}}} $\wedge$ \CCnot{\CCequal{\CCvar{y1}}{\CCvar{x1}}}]

 $\vee$ [\CCequal{\CCselect{\CCvar{H}}{\CCvar{y1}}}{\CCvar{y2}} $\wedge$ \CCequal{\CCvar{x2}}{\CCselect{\CCvar{H}}{\CCvar{x1}}} $\wedge$ \CCnot{\CCequal{\CCplus{\CCvar{size}}{\CCvar{1}}}{\CCvar{1}}} $\wedge$ \CCmember{\CCvar{x1}}{\CCvar{keys}}]

 $\vee$ [\CCequal{\CCselect{\CCvar{H}}{\CCvar{y1}}}{\CCvar{y2}} $\wedge$ \CCequal{\CCvar{x2}}{\CCselect{\CCvar{H}}{\CCvar{x1}}} $\wedge$ \CCnot{\CCequal{\CCplus{\CCvar{size}}{\CCvar{1}}}{\CCvar{1}}} $\wedge$ \CCnot{\CCmember{\CCvar{x1}}{\CCvar{keys}}} $\wedge$ \CCnot{\CCequal{\CCvar{y1}}{\CCvar{x1}}}]

 $\vee$ [\CCequal{\CCselect{\CCvar{H}}{\CCvar{y1}}}{\CCvar{y2}} $\wedge$ \CCnot{\CCequal{\CCvar{x2}}{\CCselect{\CCvar{H}}{\CCvar{x1}}}} $\wedge$ \CCnot{\CCequal{\CCvar{y1}}{\CCvar{x1}}}]

 $\vee$ [\CCnot{\CCequal{\CCselect{\CCvar{H}}{\CCvar{y1}}}{\CCvar{y2}}} $\wedge$ \CCnot{\CCequal{\CCvar{y1}}{\CCvar{x1}}}]

\item \CCmethod{put} \CCbowtie\ \CCmethod{remove}

 Simple:

\CCnot{\CCequal{\CCvar{y1}}{\CCvar{x1}}}

 Poke:

\CCnot{\CCequal{\CCvar{y1}}{\CCvar{x1}}}

\item \CCmethod{put} \CCbowtie\ \CCmethod{size}

 Simple:

\CCmember{\CCvar{x1}}{\CCvar{keys}}

 Poke:

\CCmember{\CCvar{x1}}{\CCvar{keys}}

\item \CCmethod{remove} \CCbowtie\ \CCmethod{remove}

 Simple:

[\CCequal{\CCvar{1}}{\CCvar{size}} $\wedge$ \CCequal{\CCvar{y1}}{\CCvar{x1}} $\wedge$ \CCnot{\CCmember{\CCvar{y1}}{\CCvar{keys}}}]

 $\vee$ [\CCequal{\CCvar{1}}{\CCvar{size}} $\wedge$ \CCnot{\CCequal{\CCvar{y1}}{\CCvar{x1}}}]

 $\vee$ [\CCnot{\CCequal{\CCvar{1}}{\CCvar{size}}} $\wedge$ \CCequal{\CCvar{y1}}{\CCvar{x1}} $\wedge$ \CCnot{\CCmember{\CCvar{y1}}{\CCvar{keys}}}]

 $\vee$ [\CCnot{\CCequal{\CCvar{1}}{\CCvar{size}}} $\wedge$ \CCnot{\CCequal{\CCvar{y1}}{\CCvar{x1}}}]

 Poke:

[\CCequal{\CCsetminus{\CCvar{keys}}{\CCsingleton{\CCvar{x1}}}}{\CCsingleton{\CCvar{y1}}}]

 $\vee$ [\CCnot{\CCequal{\CCsetminus{\CCvar{keys}}{\CCsingleton{\CCvar{x1}}}}{\CCsingleton{\CCvar{y1}}}} $\wedge$ \CCmember{\CCvar{y1}}{\CCsingleton{\CCvar{x1}}} $\wedge$ \CCnot{\CCmember{\CCvar{y1}}{\CCvar{keys}}}]

 $\vee$ [\CCnot{\CCequal{\CCsetminus{\CCvar{keys}}{\CCsingleton{\CCvar{x1}}}}{\CCsingleton{\CCvar{y1}}}} $\wedge$ \CCnot{\CCmember{\CCvar{y1}}{\CCsingleton{\CCvar{x1}}}}]

\item \CCmethod{remove} \CCbowtie\ \CCmethod{size}

 Simple:

[\CCequal{\CCvar{1}}{\CCvar{size}} $\wedge$ \CCnot{\CCmember{\CCvar{x1}}{\CCvar{keys}}}]

 $\vee$ [\CCnot{\CCequal{\CCvar{1}}{\CCvar{size}}} $\wedge$ \CCnot{\CCmember{\CCvar{x1}}{\CCvar{keys}}}]

 Poke:

\CCnot{\CCmember{\CCvar{x1}}{\CCvar{keys}}}

\item \CCmethod{size} \CCbowtie\ \CCmethod{size}

 Simple:

true

 Poke:

true

\end{itemize}

\subsection{Stack}
\scriptsize
\label{yml:stack}
{\ttfamily\noindent
\\
\textcolor[HTML]{0000C0}{\char35{}\enskip{}Stack\enskip{}definition}\\
\\
\textcolor[HTML]{008080}{name}\textcolor[HTML]{C000C0}{:}\textcolor[HTML]{000000}{\enskip{}}\textcolor[HTML]{000000}{stack}\\
\\
\textcolor[HTML]{008080}{preamble}\textcolor[HTML]{C000C0}{:}\textcolor[HTML]{000000}{\enskip{}|}\\
\textcolor[HTML]{000000}{\enskip{}\enskip{}}\textcolor[HTML]{000000}{(declare-sort}\textcolor[HTML]{000000}{\enskip{}}\textcolor[HTML]{000000}{E}\textcolor[HTML]{000000}{\enskip{}}\textcolor[HTML]{000000}{0)}\\
\\
\textcolor[HTML]{008080}{state}\textcolor[HTML]{C000C0}{:}\\
\textcolor[HTML]{000000}{\enskip{}\enskip{}}\textcolor[HTML]{AF5F00}{-\enskip{}}\textcolor[HTML]{008080}{name}\textcolor[HTML]{C000C0}{:}\textcolor[HTML]{000000}{\enskip{}}\textcolor[HTML]{000000}{size}\\
\textcolor[HTML]{000000}{\enskip{}\enskip{}\enskip{}\enskip{}}\textcolor[HTML]{008080}{type}\textcolor[HTML]{C000C0}{:}\textcolor[HTML]{000000}{\enskip{}}\textcolor[HTML]{000000}{Int}\\
\textcolor[HTML]{000000}{\enskip{}\enskip{}}\textcolor[HTML]{AF5F00}{-\enskip{}}\textcolor[HTML]{008080}{name}\textcolor[HTML]{C000C0}{:}\textcolor[HTML]{000000}{\enskip{}}\textcolor[HTML]{000000}{top}\\
\textcolor[HTML]{000000}{\enskip{}\enskip{}\enskip{}\enskip{}}\textcolor[HTML]{008080}{type}\textcolor[HTML]{C000C0}{:}\textcolor[HTML]{000000}{\enskip{}}\textcolor[HTML]{000000}{E}\\
\textcolor[HTML]{000000}{\enskip{}\enskip{}}\textcolor[HTML]{AF5F00}{-\enskip{}}\textcolor[HTML]{008080}{name}\textcolor[HTML]{C000C0}{:}\textcolor[HTML]{000000}{\enskip{}}\textcolor[HTML]{000000}{nextToTop}\\
\textcolor[HTML]{000000}{\enskip{}\enskip{}\enskip{}\enskip{}}\textcolor[HTML]{008080}{type}\textcolor[HTML]{C000C0}{:}\textcolor[HTML]{000000}{\enskip{}}\textcolor[HTML]{000000}{E}\\
\textcolor[HTML]{000000}{\enskip{}\enskip{}}\textcolor[HTML]{AF5F00}{-\enskip{}}\textcolor[HTML]{008080}{name}\textcolor[HTML]{C000C0}{:}\textcolor[HTML]{000000}{\enskip{}}\textcolor[HTML]{000000}{secondToTop}\\
\textcolor[HTML]{000000}{\enskip{}\enskip{}\enskip{}\enskip{}}\textcolor[HTML]{008080}{type}\textcolor[HTML]{C000C0}{:}\textcolor[HTML]{000000}{\enskip{}}\textcolor[HTML]{000000}{E}\\
\textcolor[HTML]{000000}{\enskip{}\enskip{}}\textcolor[HTML]{AF5F00}{-\enskip{}}\textcolor[HTML]{008080}{name}\textcolor[HTML]{C000C0}{:}\textcolor[HTML]{000000}{\enskip{}}\textcolor[HTML]{000000}{thirdToTop}\\
\textcolor[HTML]{000000}{\enskip{}\enskip{}\enskip{}\enskip{}}\textcolor[HTML]{008080}{type}\textcolor[HTML]{C000C0}{:}\textcolor[HTML]{000000}{\enskip{}}\textcolor[HTML]{000000}{E}\\
\\
\textcolor[HTML]{008080}{states\char95{}equal}\textcolor[HTML]{C000C0}{:}\\
\textcolor[HTML]{000000}{\enskip{}\enskip{}}\textcolor[HTML]{008080}{definition}\textcolor[HTML]{C000C0}{:}\\
\textcolor[HTML]{000000}{\enskip{}\enskip{}\enskip{}\enskip{}}\textcolor[HTML]{000000}{(and}\textcolor[HTML]{000000}{\enskip{}}\textcolor[HTML]{000000}{(=}\textcolor[HTML]{000000}{\enskip{}}\textcolor[HTML]{000000}{size\char95{}1}\textcolor[HTML]{000000}{\enskip{}}\textcolor[HTML]{000000}{size\char95{}2)}\\
\textcolor[HTML]{000000}{\enskip{}\enskip{}\enskip{}\enskip{}\enskip{}\enskip{}\enskip{}\enskip{}\enskip{}}\textcolor[HTML]{000000}{(or}\textcolor[HTML]{000000}{\enskip{}}\textcolor[HTML]{000000}{(=}\textcolor[HTML]{000000}{\enskip{}}\textcolor[HTML]{000000}{size\char95{}1}\textcolor[HTML]{000000}{\enskip{}}\textcolor[HTML]{000000}{0)}\\
\textcolor[HTML]{000000}{\enskip{}\enskip{}\enskip{}\enskip{}\enskip{}\enskip{}\enskip{}\enskip{}\enskip{}\enskip{}\enskip{}\enskip{}\enskip{}}\textcolor[HTML]{000000}{(and}\textcolor[HTML]{000000}{\enskip{}}\textcolor[HTML]{000000}{(=}\textcolor[HTML]{000000}{\enskip{}}\textcolor[HTML]{000000}{size\char95{}1}\textcolor[HTML]{000000}{\enskip{}}\textcolor[HTML]{000000}{1)}\textcolor[HTML]{000000}{\enskip{}}\textcolor[HTML]{000000}{(=}\textcolor[HTML]{000000}{\enskip{}}\textcolor[HTML]{000000}{top\char95{}1}\textcolor[HTML]{000000}{\enskip{}}\textcolor[HTML]{000000}{top\char95{}2))}\\
\textcolor[HTML]{000000}{\enskip{}\enskip{}\enskip{}\enskip{}\enskip{}\enskip{}\enskip{}\enskip{}\enskip{}\enskip{}\enskip{}\enskip{}\enskip{}}\textcolor[HTML]{000000}{(and}\textcolor[HTML]{000000}{\enskip{}}\textcolor[HTML]{000000}{(=}\textcolor[HTML]{000000}{\enskip{}}\textcolor[HTML]{000000}{top\char95{}1}\textcolor[HTML]{000000}{\enskip{}}\textcolor[HTML]{000000}{top\char95{}2)}\textcolor[HTML]{000000}{\enskip{}}\textcolor[HTML]{000000}{(=}\textcolor[HTML]{000000}{\enskip{}}\textcolor[HTML]{000000}{nextToTop\char95{}1}\textcolor[HTML]{000000}{\enskip{}}\textcolor[HTML]{000000}{nextToTop\char95{}2))))}\\
\\
\textcolor[HTML]{008080}{methods}\textcolor[HTML]{C000C0}{:}\\
\textcolor[HTML]{000000}{\enskip{}\enskip{}}\textcolor[HTML]{AF5F00}{-\enskip{}}\textcolor[HTML]{008080}{name}\textcolor[HTML]{C000C0}{:}\textcolor[HTML]{000000}{\enskip{}}\textcolor[HTML]{000000}{push}\\
\textcolor[HTML]{000000}{\enskip{}\enskip{}\enskip{}\enskip{}}\textcolor[HTML]{008080}{args}\textcolor[HTML]{C000C0}{:}\\
\textcolor[HTML]{000000}{\enskip{}\enskip{}\enskip{}\enskip{}\enskip{}\enskip{}}\textcolor[HTML]{AF5F00}{-\enskip{}}\textcolor[HTML]{008080}{name}\textcolor[HTML]{C000C0}{:}\textcolor[HTML]{000000}{\enskip{}}\textcolor[HTML]{000000}{v}\\
\textcolor[HTML]{000000}{\enskip{}\enskip{}\enskip{}\enskip{}\enskip{}\enskip{}\enskip{}\enskip{}}\textcolor[HTML]{008080}{type}\textcolor[HTML]{C000C0}{:}\textcolor[HTML]{000000}{\enskip{}}\textcolor[HTML]{000000}{E}\\
\textcolor[HTML]{000000}{\enskip{}\enskip{}\enskip{}\enskip{}}\textcolor[HTML]{008080}{return}\textcolor[HTML]{C000C0}{:}\\
\textcolor[HTML]{000000}{\enskip{}\enskip{}\enskip{}\enskip{}\enskip{}\enskip{}}\textcolor[HTML]{AF5F00}{-\enskip{}}\textcolor[HTML]{008080}{name}\textcolor[HTML]{C000C0}{:}\textcolor[HTML]{000000}{\enskip{}}\textcolor[HTML]{000000}{result}\\
\textcolor[HTML]{000000}{\enskip{}\enskip{}\enskip{}\enskip{}\enskip{}\enskip{}\enskip{}\enskip{}}\textcolor[HTML]{008080}{type}\textcolor[HTML]{C000C0}{:}\textcolor[HTML]{000000}{\enskip{}}\textcolor[HTML]{000000}{Bool}\\
\textcolor[HTML]{000000}{\enskip{}\enskip{}\enskip{}\enskip{}}\textcolor[HTML]{008080}{requires}\textcolor[HTML]{C000C0}{:}\textcolor[HTML]{000000}{\enskip{}|}\\
\textcolor[HTML]{000000}{\enskip{}\enskip{}\enskip{}\enskip{}\enskip{}\enskip{}}\textcolor[HTML]{000000}{(>=}\textcolor[HTML]{000000}{\enskip{}}\textcolor[HTML]{000000}{size}\textcolor[HTML]{000000}{\enskip{}}\textcolor[HTML]{000000}{0)}\\
\textcolor[HTML]{000000}{\enskip{}\enskip{}\enskip{}\enskip{}}\textcolor[HTML]{008080}{ensures}\textcolor[HTML]{C000C0}{:}\textcolor[HTML]{000000}{\enskip{}|}\\
\textcolor[HTML]{000000}{\enskip{}\enskip{}\enskip{}\enskip{}\enskip{}\enskip{}}\textcolor[HTML]{000000}{(and}\textcolor[HTML]{000000}{\enskip{}}\textcolor[HTML]{000000}{(=}\textcolor[HTML]{000000}{\enskip{}}\textcolor[HTML]{000000}{size\char95{}new}\textcolor[HTML]{000000}{\enskip{}}\textcolor[HTML]{000000}{(+}\textcolor[HTML]{000000}{\enskip{}}\textcolor[HTML]{000000}{size}\textcolor[HTML]{000000}{\enskip{}}\textcolor[HTML]{000000}{1))}\\
\textcolor[HTML]{000000}{\enskip{}\enskip{}\enskip{}\enskip{}\enskip{}\enskip{}\enskip{}\enskip{}\enskip{}\enskip{}\enskip{}}\textcolor[HTML]{000000}{(=}\textcolor[HTML]{000000}{\enskip{}}\textcolor[HTML]{000000}{top\char95{}new}\textcolor[HTML]{000000}{\enskip{}}\textcolor[HTML]{000000}{v)}\\
\textcolor[HTML]{000000}{\enskip{}\enskip{}\enskip{}\enskip{}\enskip{}\enskip{}\enskip{}\enskip{}\enskip{}\enskip{}\enskip{}}\textcolor[HTML]{000000}{(=}\textcolor[HTML]{000000}{\enskip{}}\textcolor[HTML]{000000}{nextToTop\char95{}new}\textcolor[HTML]{000000}{\enskip{}}\textcolor[HTML]{000000}{top)}\\
\textcolor[HTML]{000000}{\enskip{}\enskip{}\enskip{}\enskip{}\enskip{}\enskip{}\enskip{}\enskip{}\enskip{}\enskip{}\enskip{}}\textcolor[HTML]{000000}{(=}\textcolor[HTML]{000000}{\enskip{}}\textcolor[HTML]{000000}{secondToTop\char95{}new}\textcolor[HTML]{000000}{\enskip{}}\textcolor[HTML]{000000}{nextToTop)}\\
\textcolor[HTML]{000000}{\enskip{}\enskip{}\enskip{}\enskip{}\enskip{}\enskip{}\enskip{}\enskip{}\enskip{}\enskip{}\enskip{}}\textcolor[HTML]{000000}{(=}\textcolor[HTML]{000000}{\enskip{}}\textcolor[HTML]{000000}{thirdToTop\char95{}new}\textcolor[HTML]{000000}{\enskip{}}\textcolor[HTML]{000000}{secondToTop)}\\
\textcolor[HTML]{000000}{\enskip{}\enskip{}\enskip{}\enskip{}\enskip{}\enskip{}\enskip{}\enskip{}\enskip{}\enskip{}\enskip{}}\textcolor[HTML]{000000}{(=}\textcolor[HTML]{000000}{\enskip{}}\textcolor[HTML]{000000}{result}\textcolor[HTML]{000000}{\enskip{}}\textcolor[HTML]{C00000}{true}\textcolor[HTML]{000000}{))}\\
\textcolor[HTML]{000000}{\enskip{}\enskip{}\enskip{}\enskip{}}\textcolor[HTML]{008080}{terms}\textcolor[HTML]{C000C0}{:}\\
\textcolor[HTML]{000000}{\enskip{}\enskip{}\enskip{}\enskip{}\enskip{}\enskip{}}\textcolor[HTML]{008080}{Int}\textcolor[HTML]{C000C0}{:}\textcolor[HTML]{000000}{\enskip{}}\textcolor[HTML]{C000C0}{\char91{}}\textcolor[HTML]{000000}{size}\textcolor[HTML]{000000}{,\enskip{}}\textcolor[HTML]{C00000}{1}\textcolor[HTML]{000000}{,\enskip{}}\textcolor[HTML]{000000}{(+}\textcolor[HTML]{000000}{\enskip{}}\textcolor[HTML]{000000}{size}\textcolor[HTML]{000000}{\enskip{}}\textcolor[HTML]{000000}{1)}\textcolor[HTML]{C000C0}{\char93{}}\\
\textcolor[HTML]{000000}{\enskip{}\enskip{}\enskip{}\enskip{}\enskip{}\enskip{}}\textcolor[HTML]{008080}{E}\textcolor[HTML]{C000C0}{:}\textcolor[HTML]{000000}{\enskip{}}\textcolor[HTML]{C000C0}{\char91{}}\textcolor[HTML]{000000}{top}\textcolor[HTML]{000000}{,\enskip{}}\textcolor[HTML]{000000}{nextToTop}\textcolor[HTML]{000000}{,\enskip{}}\textcolor[HTML]{000000}{secondToTop}\textcolor[HTML]{000000}{,\enskip{}}\textcolor[HTML]{000000}{thirdToTop}\textcolor[HTML]{000000}{,\enskip{}}\textcolor[HTML]{000000}{\char36{}1}\textcolor[HTML]{C000C0}{\char93{}}\\
\textcolor[HTML]{000000}{\enskip{}\enskip{}}\textcolor[HTML]{AF5F00}{-\enskip{}}\textcolor[HTML]{008080}{name}\textcolor[HTML]{C000C0}{:}\textcolor[HTML]{000000}{\enskip{}}\textcolor[HTML]{000000}{pop}\\
\textcolor[HTML]{000000}{\enskip{}\enskip{}\enskip{}\enskip{}}\textcolor[HTML]{008080}{args}\textcolor[HTML]{C000C0}{:}\textcolor[HTML]{000000}{\enskip{}}\textcolor[HTML]{C000C0}{\char91{}\char93{}}\\
\textcolor[HTML]{000000}{\enskip{}\enskip{}\enskip{}\enskip{}}\textcolor[HTML]{008080}{return}\textcolor[HTML]{C000C0}{:}\\
\textcolor[HTML]{000000}{\enskip{}\enskip{}\enskip{}\enskip{}\enskip{}\enskip{}}\textcolor[HTML]{AF5F00}{-\enskip{}}\textcolor[HTML]{008080}{name}\textcolor[HTML]{C000C0}{:}\textcolor[HTML]{000000}{\enskip{}}\textcolor[HTML]{000000}{result}\\
\textcolor[HTML]{000000}{\enskip{}\enskip{}\enskip{}\enskip{}\enskip{}\enskip{}\enskip{}\enskip{}}\textcolor[HTML]{008080}{type}\textcolor[HTML]{C000C0}{:}\textcolor[HTML]{000000}{\enskip{}}\textcolor[HTML]{000000}{E}\\
\textcolor[HTML]{000000}{\enskip{}\enskip{}\enskip{}\enskip{}}\textcolor[HTML]{008080}{requires}\textcolor[HTML]{C000C0}{:}\textcolor[HTML]{000000}{\enskip{}|}\\
\textcolor[HTML]{000000}{\enskip{}\enskip{}\enskip{}\enskip{}\enskip{}\enskip{}}\textcolor[HTML]{000000}{(>=}\textcolor[HTML]{000000}{\enskip{}}\textcolor[HTML]{000000}{size}\textcolor[HTML]{000000}{\enskip{}}\textcolor[HTML]{000000}{1)}\\
\textcolor[HTML]{000000}{\enskip{}\enskip{}\enskip{}\enskip{}}\textcolor[HTML]{008080}{ensures}\textcolor[HTML]{C000C0}{:}\textcolor[HTML]{000000}{\enskip{}|}\\
\textcolor[HTML]{000000}{\enskip{}\enskip{}\enskip{}\enskip{}\enskip{}\enskip{}}\textcolor[HTML]{000000}{(and}\textcolor[HTML]{000000}{\enskip{}}\textcolor[HTML]{000000}{(=}\textcolor[HTML]{000000}{\enskip{}}\textcolor[HTML]{000000}{size\char95{}new}\textcolor[HTML]{000000}{\enskip{}}\textcolor[HTML]{000000}{(-}\textcolor[HTML]{000000}{\enskip{}}\textcolor[HTML]{000000}{size}\textcolor[HTML]{000000}{\enskip{}}\textcolor[HTML]{000000}{1))}\\
\textcolor[HTML]{000000}{\enskip{}\enskip{}\enskip{}\enskip{}\enskip{}\enskip{}\enskip{}\enskip{}\enskip{}\enskip{}\enskip{}}\textcolor[HTML]{000000}{(=}\textcolor[HTML]{000000}{\enskip{}}\textcolor[HTML]{000000}{result}\textcolor[HTML]{000000}{\enskip{}}\textcolor[HTML]{000000}{top)}\\
\textcolor[HTML]{000000}{\enskip{}\enskip{}\enskip{}\enskip{}\enskip{}\enskip{}\enskip{}\enskip{}\enskip{}\enskip{}\enskip{}}\textcolor[HTML]{000000}{(=}\textcolor[HTML]{000000}{\enskip{}}\textcolor[HTML]{000000}{top\char95{}new}\textcolor[HTML]{000000}{\enskip{}}\textcolor[HTML]{000000}{nextToTop)}\\
\textcolor[HTML]{000000}{\enskip{}\enskip{}\enskip{}\enskip{}\enskip{}\enskip{}\enskip{}\enskip{}\enskip{}\enskip{}\enskip{}}\textcolor[HTML]{000000}{(=}\textcolor[HTML]{000000}{\enskip{}}\textcolor[HTML]{000000}{nextToTop\char95{}new}\textcolor[HTML]{000000}{\enskip{}}\textcolor[HTML]{000000}{secondToTop)}\\
\textcolor[HTML]{000000}{\enskip{}\enskip{}\enskip{}\enskip{}\enskip{}\enskip{}\enskip{}\enskip{}\enskip{}\enskip{}\enskip{}}\textcolor[HTML]{000000}{(=}\textcolor[HTML]{000000}{\enskip{}}\textcolor[HTML]{000000}{secondToTop\char95{}new}\textcolor[HTML]{000000}{\enskip{}}\textcolor[HTML]{000000}{thirdToTop))}\\
\textcolor[HTML]{000000}{\enskip{}\enskip{}\enskip{}\enskip{}}\textcolor[HTML]{008080}{terms}\textcolor[HTML]{C000C0}{:}\\
\textcolor[HTML]{000000}{\enskip{}\enskip{}\enskip{}\enskip{}\enskip{}\enskip{}}\textcolor[HTML]{008080}{Int}\textcolor[HTML]{C000C0}{:}\textcolor[HTML]{000000}{\enskip{}}\textcolor[HTML]{C000C0}{\char91{}}\textcolor[HTML]{000000}{size}\textcolor[HTML]{000000}{,\enskip{}}\textcolor[HTML]{C00000}{1}\textcolor[HTML]{000000}{,\enskip{}}\textcolor[HTML]{000000}{(-}\textcolor[HTML]{000000}{\enskip{}}\textcolor[HTML]{000000}{size}\textcolor[HTML]{000000}{\enskip{}}\textcolor[HTML]{000000}{1)}\textcolor[HTML]{000000}{,\enskip{}}\textcolor[HTML]{C00000}{0}\textcolor[HTML]{C000C0}{\char93{}}\\
\textcolor[HTML]{000000}{\enskip{}\enskip{}\enskip{}\enskip{}\enskip{}\enskip{}}\textcolor[HTML]{008080}{E}\textcolor[HTML]{C000C0}{:}\textcolor[HTML]{000000}{\enskip{}}\textcolor[HTML]{C000C0}{\char91{}}\textcolor[HTML]{000000}{top}\textcolor[HTML]{000000}{,\enskip{}}\textcolor[HTML]{000000}{nextToTop}\textcolor[HTML]{000000}{,\enskip{}}\textcolor[HTML]{000000}{secondToTop}\textcolor[HTML]{000000}{,\enskip{}}\textcolor[HTML]{000000}{thirdToTop}\textcolor[HTML]{C000C0}{\char93{}}\\
\textcolor[HTML]{000000}{\enskip{}\enskip{}}\textcolor[HTML]{AF5F00}{-\enskip{}}\textcolor[HTML]{008080}{name}\textcolor[HTML]{C000C0}{:}\textcolor[HTML]{000000}{\enskip{}}\textcolor[HTML]{000000}{clear}\\
\textcolor[HTML]{000000}{\enskip{}\enskip{}\enskip{}\enskip{}}\textcolor[HTML]{008080}{args}\textcolor[HTML]{C000C0}{:}\textcolor[HTML]{000000}{\enskip{}}\textcolor[HTML]{C000C0}{\char91{}\char93{}}\\
\textcolor[HTML]{000000}{\enskip{}\enskip{}\enskip{}\enskip{}}\textcolor[HTML]{008080}{return}\textcolor[HTML]{C000C0}{:}\\
\textcolor[HTML]{000000}{\enskip{}\enskip{}\enskip{}\enskip{}\enskip{}\enskip{}}\textcolor[HTML]{AF5F00}{-\enskip{}}\textcolor[HTML]{008080}{name}\textcolor[HTML]{C000C0}{:}\textcolor[HTML]{000000}{\enskip{}}\textcolor[HTML]{000000}{result}\\
\textcolor[HTML]{000000}{\enskip{}\enskip{}\enskip{}\enskip{}\enskip{}\enskip{}\enskip{}\enskip{}}\textcolor[HTML]{008080}{type}\textcolor[HTML]{C000C0}{:}\textcolor[HTML]{000000}{\enskip{}}\textcolor[HTML]{000000}{Bool}\\
\textcolor[HTML]{000000}{\enskip{}\enskip{}\enskip{}\enskip{}}\textcolor[HTML]{008080}{requires}\textcolor[HTML]{C000C0}{:}\textcolor[HTML]{000000}{\enskip{}|}\\
\textcolor[HTML]{000000}{\enskip{}\enskip{}\enskip{}\enskip{}\enskip{}\enskip{}}\textcolor[HTML]{000000}{(>=}\textcolor[HTML]{000000}{\enskip{}}\textcolor[HTML]{000000}{size}\textcolor[HTML]{000000}{\enskip{}}\textcolor[HTML]{000000}{0)}\\
\textcolor[HTML]{000000}{\enskip{}\enskip{}\enskip{}\enskip{}}\textcolor[HTML]{008080}{ensures}\textcolor[HTML]{C000C0}{:}\textcolor[HTML]{000000}{\enskip{}|}\\
\textcolor[HTML]{000000}{\enskip{}\enskip{}\enskip{}\enskip{}\enskip{}\enskip{}}\textcolor[HTML]{000000}{(and}\textcolor[HTML]{000000}{\enskip{}}\textcolor[HTML]{000000}{(=}\textcolor[HTML]{000000}{\enskip{}}\textcolor[HTML]{000000}{size\char95{}new}\textcolor[HTML]{000000}{\enskip{}}\textcolor[HTML]{000000}{0)}\\
\textcolor[HTML]{000000}{\enskip{}\enskip{}\enskip{}\enskip{}\enskip{}\enskip{}\enskip{}\enskip{}\enskip{}\enskip{}\enskip{}}\textcolor[HTML]{000000}{(=}\textcolor[HTML]{000000}{\enskip{}}\textcolor[HTML]{000000}{result}\textcolor[HTML]{000000}{\enskip{}}\textcolor[HTML]{C00000}{true}\textcolor[HTML]{000000}{))}\\
\textcolor[HTML]{000000}{\enskip{}\enskip{}\enskip{}\enskip{}}\textcolor[HTML]{008080}{terms}\textcolor[HTML]{C000C0}{:}\\
\textcolor[HTML]{000000}{\enskip{}\enskip{}\enskip{}\enskip{}\enskip{}\enskip{}}\textcolor[HTML]{008080}{Int}\textcolor[HTML]{C000C0}{:}\textcolor[HTML]{000000}{\enskip{}}\textcolor[HTML]{C000C0}{\char91{}}\textcolor[HTML]{000000}{size}\textcolor[HTML]{000000}{,\enskip{}}\textcolor[HTML]{C00000}{0}\textcolor[HTML]{C000C0}{\char93{}}\\
\textcolor[HTML]{000000}{\enskip{}\enskip{}\enskip{}\enskip{}\enskip{}\enskip{}}\textcolor[HTML]{008080}{E}\textcolor[HTML]{C000C0}{:}\textcolor[HTML]{000000}{\enskip{}}\textcolor[HTML]{C000C0}{\char91{}}\textcolor[HTML]{000000}{top}\textcolor[HTML]{000000}{,\enskip{}}\textcolor[HTML]{000000}{nextToTop}\textcolor[HTML]{000000}{,\enskip{}}\textcolor[HTML]{000000}{secondToTop}\textcolor[HTML]{000000}{,\enskip{}}\textcolor[HTML]{000000}{thirdToTop}\textcolor[HTML]{C000C0}{\char93{}}\\
\\
\textcolor[HTML]{008080}{predicates}\textcolor[HTML]{C000C0}{:}\\
\textcolor[HTML]{000000}{\enskip{}\enskip{}}\textcolor[HTML]{AF5F00}{-\enskip{}}\textcolor[HTML]{008080}{name}\textcolor[HTML]{C000C0}{:}\textcolor[HTML]{000000}{\enskip{}}\textcolor[HTML]{C00000}{"}\textcolor[HTML]{C00000}{=}\textcolor[HTML]{C00000}{"}\\
\textcolor[HTML]{000000}{\enskip{}\enskip{}\enskip{}\enskip{}}\textcolor[HTML]{008080}{type}\textcolor[HTML]{C000C0}{:}\textcolor[HTML]{000000}{\enskip{}}\textcolor[HTML]{C000C0}{\char91{}}\textcolor[HTML]{000000}{Int}\textcolor[HTML]{000000}{,\enskip{}}\textcolor[HTML]{000000}{Int}\textcolor[HTML]{C000C0}{\char93{}}\\
\textcolor[HTML]{000000}{\enskip{}\enskip{}}\textcolor[HTML]{AF5F00}{-\enskip{}}\textcolor[HTML]{008080}{name}\textcolor[HTML]{C000C0}{:}\textcolor[HTML]{000000}{\enskip{}}\textcolor[HTML]{C00000}{"}\textcolor[HTML]{C00000}{=}\textcolor[HTML]{C00000}{"}\\
\textcolor[HTML]{000000}{\enskip{}\enskip{}\enskip{}\enskip{}}\textcolor[HTML]{008080}{type}\textcolor[HTML]{C000C0}{:}\textcolor[HTML]{000000}{\enskip{}}\textcolor[HTML]{C000C0}{\char91{}}\textcolor[HTML]{000000}{E}\textcolor[HTML]{000000}{,\enskip{}}\textcolor[HTML]{000000}{E}\textcolor[HTML]{C000C0}{\char93{}}\\

}

\begin{itemize}
\item \CCmethod{clear} \CCbowtie\ \CCmethod{clear}

 Simple:

true

 Poke:

true

\item \CCmethod{clear} \CCbowtie\ \CCmethod{pop}

 Simple:

false

 Poke:

false

\item \CCmethod{clear} \CCbowtie\ \CCmethod{push}

 Simple:

false

 Poke:

false

\item \CCmethod{pop} \CCbowtie\ \CCmethod{pop}

 Simple:

\CCequal{\CCvar{nextToTop}}{\CCvar{top}}

 Poke:

\CCequal{\CCvar{nextToTop}}{\CCvar{top}}

\item \CCmethod{push} \CCrightmover\ \CCmethod{pop}

 Simple:

[\CCequal{\CCvar{1}}{\CCvar{size}} $\wedge$ \CCequal{\CCvar{nextToTop}}{\CCvar{top}} $\wedge$ \CCequal{\CCvar{nextToTop}}{\CCvar{thirdToTop}} $\wedge$ \CCequal{\CCvar{nextToTop}}{\CCvar{x1}}]

 $\vee$ [\CCequal{\CCvar{1}}{\CCvar{size}} $\wedge$ \CCequal{\CCvar{nextToTop}}{\CCvar{top}} $\wedge$ \CCnot{\CCequal{\CCvar{nextToTop}}{\CCvar{thirdToTop}}} $\wedge$ \CCequal{\CCvar{nextToTop}}{\CCvar{x1}}]

 $\vee$ [\CCequal{\CCvar{1}}{\CCvar{size}} $\wedge$ \CCnot{\CCequal{\CCvar{nextToTop}}{\CCvar{top}}} $\wedge$ \CCequal{\CCvar{nextToTop}}{\CCvar{thirdToTop}} $\wedge$ \CCequal{\CCvar{nextToTop}}{\CCvar{secondToTop}} $\wedge$ \CCequal{\CCvar{top}}{\CCvar{x1}}]

 $\vee$ [\CCequal{\CCvar{1}}{\CCvar{size}} $\wedge$ \CCnot{\CCequal{\CCvar{nextToTop}}{\CCvar{top}}} $\wedge$ \CCequal{\CCvar{nextToTop}}{\CCvar{thirdToTop}} $\wedge$ \CCnot{\CCequal{\CCvar{nextToTop}}{\CCvar{secondToTop}}} $\wedge$ \CCequal{\CCvar{top}}{\CCvar{x1}}]

 $\vee$ [\CCequal{\CCvar{1}}{\CCvar{size}} $\wedge$ \CCnot{\CCequal{\CCvar{nextToTop}}{\CCvar{top}}} $\wedge$ \CCnot{\CCequal{\CCvar{nextToTop}}{\CCvar{thirdToTop}}} $\wedge$ \CCequal{\CCvar{nextToTop}}{\CCvar{secondToTop}} $\wedge$ \CCequal{\CCvar{top}}{\CCvar{x1}}]

 $\vee$ [\CCequal{\CCvar{1}}{\CCvar{size}} $\wedge$ \CCnot{\CCequal{\CCvar{nextToTop}}{\CCvar{top}}} $\wedge$ \CCnot{\CCequal{\CCvar{nextToTop}}{\CCvar{thirdToTop}}} $\wedge$ \CCnot{\CCequal{\CCvar{nextToTop}}{\CCvar{secondToTop}}} $\wedge$ \CCequal{\CCvar{top}}{\CCvar{x1}}]

 $\vee$ [\CCnot{\CCequal{\CCvar{1}}{\CCvar{size}}} $\wedge$ \CCnot{\CCequal{\CCvar{0}}{\CCvar{size}}} $\wedge$ \CCequal{\CCvar{nextToTop}}{\CCvar{thirdToTop}} $\wedge$ \CCequal{\CCvar{nextToTop}}{\CCvar{secondToTop}} $\wedge$ \CCequal{\CCvar{top}}{\CCvar{x1}}]

 $\vee$ [\CCnot{\CCequal{\CCvar{1}}{\CCvar{size}}} $\wedge$ \CCnot{\CCequal{\CCvar{0}}{\CCvar{size}}} $\wedge$ \CCequal{\CCvar{nextToTop}}{\CCvar{thirdToTop}} $\wedge$ \CCnot{\CCequal{\CCvar{nextToTop}}{\CCvar{secondToTop}}} $\wedge$ \CCequal{\CCvar{top}}{\CCvar{x1}}]

 $\vee$ [\CCnot{\CCequal{\CCvar{1}}{\CCvar{size}}} $\wedge$ \CCnot{\CCequal{\CCvar{0}}{\CCvar{size}}} $\wedge$ \CCnot{\CCequal{\CCvar{nextToTop}}{\CCvar{thirdToTop}}} $\wedge$ \CCequal{\CCvar{nextToTop}}{\CCvar{secondToTop}} $\wedge$ \CCequal{\CCvar{top}}{\CCvar{x1}}]

 $\vee$ [\CCnot{\CCequal{\CCvar{1}}{\CCvar{size}}} $\wedge$ \CCnot{\CCequal{\CCvar{0}}{\CCvar{size}}} $\wedge$ \CCnot{\CCequal{\CCvar{nextToTop}}{\CCvar{thirdToTop}}} $\wedge$ \CCnot{\CCequal{\CCvar{nextToTop}}{\CCvar{secondToTop}}} $\wedge$ \CCequal{\CCvar{top}}{\CCvar{x1}}]

 Poke:

\CCnot{\CCequal{\CCvar{0}}{\CCvar{size}}} $\wedge$ \CCequal{\CCvar{top}}{\CCvar{x1}}

\item \CCmethod{pop} \CCrightmover\ \CCmethod{push}

 Simple:

[\CCequal{\CCvar{nextToTop}}{\CCvar{y1}} $\wedge$ \CCequal{\CCvar{nextToTop}}{\CCvar{top}}]

 $\vee$ [\CCnot{\CCequal{\CCvar{nextToTop}}{\CCvar{y1}}} $\wedge$ \CCequal{\CCvar{nextToTop}}{\CCvar{thirdToTop}} $\wedge$ \CCequal{\CCvar{nextToTop}}{\CCvar{secondToTop}} $\wedge$ \CCequal{\CCvar{y1}}{\CCvar{top}}]

 $\vee$ [\CCnot{\CCequal{\CCvar{nextToTop}}{\CCvar{y1}}} $\wedge$ \CCequal{\CCvar{nextToTop}}{\CCvar{thirdToTop}} $\wedge$ \CCnot{\CCequal{\CCvar{nextToTop}}{\CCvar{secondToTop}}} $\wedge$ \CCequal{\CCvar{y1}}{\CCvar{top}}]

 $\vee$ [\CCnot{\CCequal{\CCvar{nextToTop}}{\CCvar{y1}}} $\wedge$ \CCnot{\CCequal{\CCvar{nextToTop}}{\CCvar{thirdToTop}}} $\wedge$ \CCequal{\CCvar{nextToTop}}{\CCvar{secondToTop}} $\wedge$ \CCequal{\CCvar{y1}}{\CCvar{top}}]

 $\vee$ [\CCnot{\CCequal{\CCvar{nextToTop}}{\CCvar{y1}}} $\wedge$ \CCnot{\CCequal{\CCvar{nextToTop}}{\CCvar{thirdToTop}}} $\wedge$ \CCnot{\CCequal{\CCvar{nextToTop}}{\CCvar{secondToTop}}} $\wedge$ \CCequal{\CCvar{y1}}{\CCvar{top}}]

 Poke:

\CCequal{\CCvar{y1}}{\CCvar{top}}

\item \CCmethod{push} \CCbowtie\ \CCmethod{push}

 Simple:

[\CCequal{\CCvar{thirdToTop}}{\CCvar{y1}} $\wedge$ \CCequal{\CCvar{thirdToTop}}{\CCvar{x1}}]

 $\vee$ [\CCnot{\CCequal{\CCvar{thirdToTop}}{\CCvar{y1}}} $\wedge$ \CCequal{\CCvar{y1}}{\CCvar{x1}}]

 Poke:

\CCequal{\CCvar{y1}}{\CCvar{x1}}

\end{itemize}

\section{BlockKing: YML representation}
\label{apx:blockking}
{\ttfamily\noindent
\textcolor[HTML]{008080}{name}\textcolor[HTML]{C000C0}{:}\textcolor[HTML]{000000}{\enskip{}}\textcolor[HTML]{000000}{blockking}\\
\\
\textcolor[HTML]{008080}{preamble}\textcolor[HTML]{C000C0}{:}\textcolor[HTML]{000000}{\enskip{}|}\\
\textcolor[HTML]{000000}{\enskip{}\enskip{}}\textcolor[HTML]{000000}{(declare-fun}\textcolor[HTML]{000000}{\enskip{}}\textcolor[HTML]{000000}{modFn}\textcolor[HTML]{000000}{\enskip{}}\textcolor[HTML]{000000}{(Int)}\textcolor[HTML]{000000}{\enskip{}}\textcolor[HTML]{000000}{Int)}\\
\\
\textcolor[HTML]{008080}{state}\textcolor[HTML]{C000C0}{:}\\
\textcolor[HTML]{000000}{\enskip{}\enskip{}}\textcolor[HTML]{AF5F00}{-\enskip{}}\textcolor[HTML]{008080}{name}\textcolor[HTML]{C000C0}{:}\textcolor[HTML]{000000}{\enskip{}}\textcolor[HTML]{000000}{warrior}\\
\textcolor[HTML]{000000}{\enskip{}\enskip{}\enskip{}\enskip{}}\textcolor[HTML]{008080}{type}\textcolor[HTML]{C000C0}{:}\textcolor[HTML]{000000}{\enskip{}}\textcolor[HTML]{000000}{Int}\\
\textcolor[HTML]{000000}{\enskip{}\enskip{}}\textcolor[HTML]{AF5F00}{-\enskip{}}\textcolor[HTML]{008080}{name}\textcolor[HTML]{C000C0}{:}\textcolor[HTML]{000000}{\enskip{}}\textcolor[HTML]{000000}{warriorGold}\\
\textcolor[HTML]{000000}{\enskip{}\enskip{}\enskip{}\enskip{}}\textcolor[HTML]{008080}{type}\textcolor[HTML]{C000C0}{:}\textcolor[HTML]{000000}{\enskip{}}\textcolor[HTML]{000000}{Int}\\
\textcolor[HTML]{000000}{\enskip{}\enskip{}}\textcolor[HTML]{AF5F00}{-\enskip{}}\textcolor[HTML]{008080}{name}\textcolor[HTML]{C000C0}{:}\textcolor[HTML]{000000}{\enskip{}}\textcolor[HTML]{000000}{warriorBlock}\\
\textcolor[HTML]{000000}{\enskip{}\enskip{}\enskip{}\enskip{}}\textcolor[HTML]{008080}{type}\textcolor[HTML]{C000C0}{:}\textcolor[HTML]{000000}{\enskip{}}\textcolor[HTML]{000000}{Int}\\
\textcolor[HTML]{000000}{\enskip{}\enskip{}}\textcolor[HTML]{AF5F00}{-\enskip{}}\textcolor[HTML]{008080}{name}\textcolor[HTML]{C000C0}{:}\textcolor[HTML]{000000}{\enskip{}}\textcolor[HTML]{000000}{king}\\
\textcolor[HTML]{000000}{\enskip{}\enskip{}\enskip{}\enskip{}}\textcolor[HTML]{008080}{type}\textcolor[HTML]{C000C0}{:}\textcolor[HTML]{000000}{\enskip{}}\textcolor[HTML]{000000}{Int}\\
\textcolor[HTML]{000000}{\enskip{}\enskip{}}\textcolor[HTML]{AF5F00}{-\enskip{}}\textcolor[HTML]{008080}{name}\textcolor[HTML]{C000C0}{:}\textcolor[HTML]{000000}{\enskip{}}\textcolor[HTML]{000000}{kingBlock}\\
\textcolor[HTML]{000000}{\enskip{}\enskip{}\enskip{}\enskip{}}\textcolor[HTML]{008080}{type}\textcolor[HTML]{C000C0}{:}\textcolor[HTML]{000000}{\enskip{}}\textcolor[HTML]{000000}{Int}\\
\\
\textcolor[HTML]{008080}{methods}\textcolor[HTML]{C000C0}{:}\\
\textcolor[HTML]{000000}{\enskip{}\enskip{}}\textcolor[HTML]{AF5F00}{-\enskip{}}\textcolor[HTML]{008080}{name}\textcolor[HTML]{C000C0}{:}\textcolor[HTML]{000000}{\enskip{}}\textcolor[HTML]{000000}{enter}\\
\textcolor[HTML]{000000}{\enskip{}\enskip{}\enskip{}\enskip{}}\textcolor[HTML]{008080}{args}\textcolor[HTML]{C000C0}{:}\\
\textcolor[HTML]{000000}{\enskip{}\enskip{}\enskip{}\enskip{}\enskip{}\enskip{}}\textcolor[HTML]{AF5F00}{-\enskip{}}\textcolor[HTML]{008080}{name}\textcolor[HTML]{C000C0}{:}\textcolor[HTML]{000000}{\enskip{}}\textcolor[HTML]{000000}{msg\char95{}value}\\
\textcolor[HTML]{000000}{\enskip{}\enskip{}\enskip{}\enskip{}\enskip{}\enskip{}\enskip{}\enskip{}}\textcolor[HTML]{008080}{type}\textcolor[HTML]{C000C0}{:}\textcolor[HTML]{000000}{\enskip{}}\textcolor[HTML]{000000}{Int}\\
\textcolor[HTML]{000000}{\enskip{}\enskip{}\enskip{}\enskip{}\enskip{}\enskip{}}\textcolor[HTML]{AF5F00}{-\enskip{}}\textcolor[HTML]{008080}{name}\textcolor[HTML]{C000C0}{:}\textcolor[HTML]{000000}{\enskip{}}\textcolor[HTML]{000000}{msg\char95{}sender}\\
\textcolor[HTML]{000000}{\enskip{}\enskip{}\enskip{}\enskip{}\enskip{}\enskip{}\enskip{}\enskip{}}\textcolor[HTML]{008080}{type}\textcolor[HTML]{C000C0}{:}\textcolor[HTML]{000000}{\enskip{}}\textcolor[HTML]{000000}{Int}\\
\textcolor[HTML]{000000}{\enskip{}\enskip{}\enskip{}\enskip{}\enskip{}\enskip{}}\textcolor[HTML]{AF5F00}{-\enskip{}}\textcolor[HTML]{008080}{name}\textcolor[HTML]{C000C0}{:}\textcolor[HTML]{000000}{\enskip{}}\textcolor[HTML]{000000}{block\char95{}number}\\
\textcolor[HTML]{000000}{\enskip{}\enskip{}\enskip{}\enskip{}\enskip{}\enskip{}\enskip{}\enskip{}}\textcolor[HTML]{008080}{type}\textcolor[HTML]{C000C0}{:}\textcolor[HTML]{000000}{\enskip{}}\textcolor[HTML]{000000}{Int}\\
\textcolor[HTML]{000000}{\enskip{}\enskip{}\enskip{}\enskip{}\enskip{}\enskip{}}\textcolor[HTML]{AF5F00}{-\enskip{}}\textcolor[HTML]{008080}{name}\textcolor[HTML]{C000C0}{:}\textcolor[HTML]{000000}{\enskip{}}\textcolor[HTML]{000000}{random}\\
\textcolor[HTML]{000000}{\enskip{}\enskip{}\enskip{}\enskip{}\enskip{}\enskip{}\enskip{}\enskip{}}\textcolor[HTML]{008080}{type}\textcolor[HTML]{C000C0}{:}\textcolor[HTML]{000000}{\enskip{}}\textcolor[HTML]{000000}{Int}\\
\textcolor[HTML]{000000}{\enskip{}\enskip{}\enskip{}\enskip{}}\textcolor[HTML]{008080}{return}\textcolor[HTML]{C000C0}{:}\\
\textcolor[HTML]{000000}{\enskip{}\enskip{}\enskip{}\enskip{}\enskip{}\enskip{}}\textcolor[HTML]{AF5F00}{-\enskip{}}\textcolor[HTML]{008080}{name}\textcolor[HTML]{C000C0}{:}\textcolor[HTML]{000000}{\enskip{}}\textcolor[HTML]{000000}{result}\\
\textcolor[HTML]{000000}{\enskip{}\enskip{}\enskip{}\enskip{}\enskip{}\enskip{}\enskip{}\enskip{}}\textcolor[HTML]{008080}{type}\textcolor[HTML]{C000C0}{:}\textcolor[HTML]{000000}{\enskip{}}\textcolor[HTML]{000000}{Bool}\\
\textcolor[HTML]{000000}{\enskip{}\enskip{}\enskip{}\enskip{}}\textcolor[HTML]{008080}{requires}\textcolor[HTML]{C000C0}{:}\textcolor[HTML]{000000}{\enskip{}|}\\
\textcolor[HTML]{000000}{\enskip{}\enskip{}\enskip{}\enskip{}\enskip{}\enskip{}}\textcolor[HTML]{C00000}{true}\\
\textcolor[HTML]{000000}{\enskip{}\enskip{}\enskip{}\enskip{}}\textcolor[HTML]{008080}{ensures}\textcolor[HTML]{C000C0}{:}\textcolor[HTML]{000000}{\enskip{}|}\\
\textcolor[HTML]{000000}{\enskip{}\enskip{}\enskip{}\enskip{}\enskip{}\enskip{}}\textcolor[HTML]{000000}{(and}\textcolor[HTML]{000000}{\enskip{}}\textcolor[HTML]{000000}{result}\\
\textcolor[HTML]{000000}{\enskip{}\enskip{}\enskip{}\enskip{}\enskip{}\enskip{}}\textcolor[HTML]{000000}{(ite}\textcolor[HTML]{000000}{\enskip{}}\textcolor[HTML]{000000}{(<}\textcolor[HTML]{000000}{\enskip{}}\textcolor[HTML]{000000}{msg\char95{}value}\textcolor[HTML]{000000}{\enskip{}}\textcolor[HTML]{000000}{50)}\\
\textcolor[HTML]{000000}{\enskip{}\enskip{}\enskip{}\enskip{}\enskip{}\enskip{}\enskip{}\enskip{}\enskip{}\enskip{}\enskip{}}\textcolor[HTML]{000000}{(states\char95{}equal}\textcolor[HTML]{000000}{\enskip{}}\textcolor[HTML]{000000}{warrior}\textcolor[HTML]{000000}{\enskip{}}\textcolor[HTML]{000000}{warriorGold}\textcolor[HTML]{000000}{\enskip{}}\textcolor[HTML]{000000}{warriorBlock}\\
\textcolor[HTML]{000000}{\enskip{}\enskip{}\enskip{}\enskip{}\enskip{}\enskip{}\enskip{}\enskip{}\enskip{}\enskip{}\enskip{}\enskip{}\enskip{}\enskip{}\enskip{}\enskip{}\enskip{}\enskip{}\enskip{}\enskip{}\enskip{}\enskip{}\enskip{}\enskip{}\enskip{}}\textcolor[HTML]{000000}{king}\textcolor[HTML]{000000}{\enskip{}}\textcolor[HTML]{000000}{kingBlock}\textcolor[HTML]{000000}{\enskip{}}\textcolor[HTML]{000000}{err}\\
\textcolor[HTML]{000000}{\enskip{}\enskip{}\enskip{}\enskip{}\enskip{}\enskip{}\enskip{}\enskip{}\enskip{}\enskip{}\enskip{}\enskip{}\enskip{}\enskip{}\enskip{}\enskip{}\enskip{}\enskip{}\enskip{}\enskip{}\enskip{}\enskip{}\enskip{}\enskip{}\enskip{}}\textcolor[HTML]{000000}{warrior\char95{}new}\textcolor[HTML]{000000}{\enskip{}}\textcolor[HTML]{000000}{warriorGold\char95{}new}\textcolor[HTML]{000000}{\enskip{}}\textcolor[HTML]{000000}{warriorBlock\char95{}new}\\
\textcolor[HTML]{000000}{\enskip{}\enskip{}\enskip{}\enskip{}\enskip{}\enskip{}\enskip{}\enskip{}\enskip{}\enskip{}\enskip{}\enskip{}\enskip{}\enskip{}\enskip{}\enskip{}\enskip{}\enskip{}\enskip{}\enskip{}\enskip{}\enskip{}\enskip{}\enskip{}\enskip{}}\textcolor[HTML]{000000}{king\char95{}new}\textcolor[HTML]{000000}{\enskip{}}\textcolor[HTML]{000000}{kingBlock\char95{}new}\textcolor[HTML]{000000}{\enskip{}}\textcolor[HTML]{000000}{err\char95{}new)}\\
\textcolor[HTML]{000000}{\enskip{}\enskip{}\enskip{}\enskip{}\enskip{}\enskip{}\enskip{}\enskip{}\enskip{}\enskip{}\enskip{}}\textcolor[HTML]{000000}{(and}\textcolor[HTML]{000000}{\enskip{}}\textcolor[HTML]{000000}{(=}\textcolor[HTML]{000000}{\enskip{}}\textcolor[HTML]{000000}{warrior\char95{}new}\textcolor[HTML]{000000}{\enskip{}}\textcolor[HTML]{000000}{msg\char95{}sender)}\\
\textcolor[HTML]{000000}{\enskip{}\enskip{}\enskip{}\enskip{}\enskip{}\enskip{}\enskip{}\enskip{}\enskip{}\enskip{}\enskip{}\enskip{}\enskip{}\enskip{}\enskip{}\enskip{}}\textcolor[HTML]{000000}{(=}\textcolor[HTML]{000000}{\enskip{}}\textcolor[HTML]{000000}{warriorGold\char95{}new}\textcolor[HTML]{000000}{\enskip{}}\textcolor[HTML]{000000}{msg\char95{}value)}\\
\textcolor[HTML]{000000}{\enskip{}\enskip{}\enskip{}\enskip{}\enskip{}\enskip{}\enskip{}\enskip{}\enskip{}\enskip{}\enskip{}\enskip{}\enskip{}\enskip{}\enskip{}\enskip{}}\textcolor[HTML]{000000}{(=}\textcolor[HTML]{000000}{\enskip{}}\textcolor[HTML]{000000}{warriorBlock\char95{}new}\textcolor[HTML]{000000}{\enskip{}}\textcolor[HTML]{000000}{block\char95{}number)}\\
\textcolor[HTML]{000000}{\enskip{}\enskip{}\enskip{}\enskip{}\enskip{}\enskip{}\enskip{}\enskip{}\enskip{}\enskip{}\enskip{}\enskip{}\enskip{}\enskip{}\enskip{}\enskip{}}\textcolor[HTML]{000000}{(ite}\textcolor[HTML]{000000}{\enskip{}}\textcolor[HTML]{000000}{(=}\textcolor[HTML]{000000}{\enskip{}}\textcolor[HTML]{000000}{(modFn}\textcolor[HTML]{000000}{\enskip{}}\textcolor[HTML]{000000}{warriorBlock\char95{}new)}\textcolor[HTML]{000000}{\enskip{}}\textcolor[HTML]{000000}{random)}\\
\textcolor[HTML]{000000}{\enskip{}\enskip{}\enskip{}\enskip{}\enskip{}\enskip{}\enskip{}\enskip{}\enskip{}\enskip{}\enskip{}\enskip{}\enskip{}\enskip{}\enskip{}\enskip{}\enskip{}\enskip{}\enskip{}\enskip{}\enskip{}}\textcolor[HTML]{000000}{(and}\textcolor[HTML]{000000}{\enskip{}}\textcolor[HTML]{000000}{(=}\textcolor[HTML]{000000}{\enskip{}}\textcolor[HTML]{000000}{king\char95{}new}\textcolor[HTML]{000000}{\enskip{}}\textcolor[HTML]{000000}{warrior\char95{}new)}\\
\textcolor[HTML]{000000}{\enskip{}\enskip{}\enskip{}\enskip{}\enskip{}\enskip{}\enskip{}\enskip{}\enskip{}\enskip{}\enskip{}\enskip{}\enskip{}\enskip{}\enskip{}\enskip{}\enskip{}\enskip{}\enskip{}\enskip{}\enskip{}\enskip{}\enskip{}\enskip{}\enskip{}\enskip{}}\textcolor[HTML]{000000}{(=}\textcolor[HTML]{000000}{\enskip{}}\textcolor[HTML]{000000}{kingBlock\char95{}new}\textcolor[HTML]{000000}{\enskip{}}\textcolor[HTML]{000000}{warriorBlock\char95{}new))}\\
\textcolor[HTML]{000000}{\enskip{}\enskip{}\enskip{}\enskip{}\enskip{}\enskip{}\enskip{}\enskip{}\enskip{}\enskip{}\enskip{}\enskip{}\enskip{}\enskip{}\enskip{}\enskip{}\enskip{}\enskip{}\enskip{}\enskip{}\enskip{}}\textcolor[HTML]{000000}{(and}\textcolor[HTML]{000000}{\enskip{}}\textcolor[HTML]{000000}{(=}\textcolor[HTML]{000000}{\enskip{}}\textcolor[HTML]{000000}{king\char95{}new}\textcolor[HTML]{000000}{\enskip{}}\textcolor[HTML]{000000}{king)}\\
\textcolor[HTML]{000000}{\enskip{}\enskip{}\enskip{}\enskip{}\enskip{}\enskip{}\enskip{}\enskip{}\enskip{}\enskip{}\enskip{}\enskip{}\enskip{}\enskip{}\enskip{}\enskip{}\enskip{}\enskip{}\enskip{}\enskip{}\enskip{}\enskip{}\enskip{}\enskip{}\enskip{}\enskip{}}\textcolor[HTML]{000000}{(=}\textcolor[HTML]{000000}{\enskip{}}\textcolor[HTML]{000000}{kingBlock\char95{}new}\textcolor[HTML]{000000}{\enskip{}}\textcolor[HTML]{000000}{kingBlock))}\\
\textcolor[HTML]{000000}{\enskip{}\enskip{}\enskip{}\enskip{}\enskip{}\enskip{}\enskip{}\enskip{}\enskip{}\enskip{}\enskip{}\enskip{}\enskip{}\enskip{}\enskip{}\enskip{}}\textcolor[HTML]{000000}{)}\\
\textcolor[HTML]{000000}{\enskip{}\enskip{}\enskip{}\enskip{}\enskip{}\enskip{}\enskip{}\enskip{}\enskip{}\enskip{}\enskip{}}\textcolor[HTML]{000000}{)}\\
\textcolor[HTML]{000000}{\enskip{}\enskip{}\enskip{}\enskip{}\enskip{}\enskip{}}\textcolor[HTML]{000000}{)}\\
\textcolor[HTML]{000000}{\enskip{}\enskip{}\enskip{}\enskip{}\enskip{}\enskip{}}\textcolor[HTML]{000000}{)}\\
\\

}
Predicates:
\begin{verbatim}
(= x1 y1)
(= x2 y2)
(= x3 y3)
(= (modFn x3) x4)
(= (modFn y3) y4)
(< x1 50)
(< y1 50)
\end{verbatim}

\section{BlockKing Fixed: YML representation}
\label{apx:blockking_fixed}
{\ttfamily\noindent
\textcolor[HTML]{008080}{name}\textcolor[HTML]{C000C0}{:}\textcolor[HTML]{000000}{\enskip{}}\textcolor[HTML]{000000}{blockking\char95{}fixed}\\
\\
\textcolor[HTML]{008080}{preamble}\textcolor[HTML]{C000C0}{:}\textcolor[HTML]{000000}{\enskip{}|}\\
\textcolor[HTML]{000000}{\enskip{}\enskip{}}\textcolor[HTML]{000000}{(declare-fun}\textcolor[HTML]{000000}{\enskip{}}\textcolor[HTML]{000000}{modFn}\textcolor[HTML]{000000}{\enskip{}}\textcolor[HTML]{000000}{(Int)}\textcolor[HTML]{000000}{\enskip{}}\textcolor[HTML]{000000}{Int)}\\
\\
\textcolor[HTML]{008080}{state}\textcolor[HTML]{C000C0}{:}\\
\textcolor[HTML]{000000}{\enskip{}\enskip{}}\textcolor[HTML]{AF5F00}{-\enskip{}}\textcolor[HTML]{008080}{name}\textcolor[HTML]{C000C0}{:}\textcolor[HTML]{000000}{\enskip{}}\textcolor[HTML]{000000}{warrior}\\
\textcolor[HTML]{000000}{\enskip{}\enskip{}\enskip{}\enskip{}}\textcolor[HTML]{008080}{type}\textcolor[HTML]{C000C0}{:}\textcolor[HTML]{000000}{\enskip{}}\textcolor[HTML]{000000}{(Array}\textcolor[HTML]{000000}{\enskip{}}\textcolor[HTML]{000000}{Int}\textcolor[HTML]{000000}{\enskip{}}\textcolor[HTML]{000000}{Int)}\\
\textcolor[HTML]{000000}{\enskip{}\enskip{}}\textcolor[HTML]{AF5F00}{-\enskip{}}\textcolor[HTML]{008080}{name}\textcolor[HTML]{C000C0}{:}\textcolor[HTML]{000000}{\enskip{}}\textcolor[HTML]{000000}{warriorGold}\\
\textcolor[HTML]{000000}{\enskip{}\enskip{}\enskip{}\enskip{}}\textcolor[HTML]{008080}{type}\textcolor[HTML]{C000C0}{:}\textcolor[HTML]{000000}{\enskip{}}\textcolor[HTML]{000000}{(Array}\textcolor[HTML]{000000}{\enskip{}}\textcolor[HTML]{000000}{Int}\textcolor[HTML]{000000}{\enskip{}}\textcolor[HTML]{000000}{Int)}\\
\textcolor[HTML]{000000}{\enskip{}\enskip{}}\textcolor[HTML]{AF5F00}{-\enskip{}}\textcolor[HTML]{008080}{name}\textcolor[HTML]{C000C0}{:}\textcolor[HTML]{000000}{\enskip{}}\textcolor[HTML]{000000}{warriorBlock}\\
\textcolor[HTML]{000000}{\enskip{}\enskip{}\enskip{}\enskip{}}\textcolor[HTML]{008080}{type}\textcolor[HTML]{C000C0}{:}\textcolor[HTML]{000000}{\enskip{}}\textcolor[HTML]{000000}{(Array}\textcolor[HTML]{000000}{\enskip{}}\textcolor[HTML]{000000}{Int}\textcolor[HTML]{000000}{\enskip{}}\textcolor[HTML]{000000}{Int)}\\
\textcolor[HTML]{000000}{\enskip{}\enskip{}}\textcolor[HTML]{AF5F00}{-\enskip{}}\textcolor[HTML]{008080}{name}\textcolor[HTML]{C000C0}{:}\textcolor[HTML]{000000}{\enskip{}}\textcolor[HTML]{000000}{king}\\
\textcolor[HTML]{000000}{\enskip{}\enskip{}\enskip{}\enskip{}}\textcolor[HTML]{008080}{type}\textcolor[HTML]{C000C0}{:}\textcolor[HTML]{000000}{\enskip{}}\textcolor[HTML]{000000}{Int}\\
\textcolor[HTML]{000000}{\enskip{}\enskip{}}\textcolor[HTML]{AF5F00}{-\enskip{}}\textcolor[HTML]{008080}{name}\textcolor[HTML]{C000C0}{:}\textcolor[HTML]{000000}{\enskip{}}\textcolor[HTML]{000000}{kingBlock}\\
\textcolor[HTML]{000000}{\enskip{}\enskip{}\enskip{}\enskip{}}\textcolor[HTML]{008080}{type}\textcolor[HTML]{C000C0}{:}\textcolor[HTML]{000000}{\enskip{}}\textcolor[HTML]{000000}{Int}\\
\\
\textcolor[HTML]{008080}{methods}\textcolor[HTML]{C000C0}{:}\\
\textcolor[HTML]{000000}{\enskip{}\enskip{}}\textcolor[HTML]{AF5F00}{-\enskip{}}\textcolor[HTML]{008080}{name}\textcolor[HTML]{C000C0}{:}\textcolor[HTML]{000000}{\enskip{}}\textcolor[HTML]{000000}{enter}\\
\textcolor[HTML]{000000}{\enskip{}\enskip{}\enskip{}\enskip{}}\textcolor[HTML]{008080}{args}\textcolor[HTML]{C000C0}{:}\\
\textcolor[HTML]{000000}{\enskip{}\enskip{}\enskip{}\enskip{}\enskip{}\enskip{}}\textcolor[HTML]{AF5F00}{-\enskip{}}\textcolor[HTML]{008080}{name}\textcolor[HTML]{C000C0}{:}\textcolor[HTML]{000000}{\enskip{}}\textcolor[HTML]{000000}{msg\char95{}value}\\
\textcolor[HTML]{000000}{\enskip{}\enskip{}\enskip{}\enskip{}\enskip{}\enskip{}\enskip{}\enskip{}}\textcolor[HTML]{008080}{type}\textcolor[HTML]{C000C0}{:}\textcolor[HTML]{000000}{\enskip{}}\textcolor[HTML]{000000}{Int}\\
\textcolor[HTML]{000000}{\enskip{}\enskip{}\enskip{}\enskip{}\enskip{}\enskip{}}\textcolor[HTML]{AF5F00}{-\enskip{}}\textcolor[HTML]{008080}{name}\textcolor[HTML]{C000C0}{:}\textcolor[HTML]{000000}{\enskip{}}\textcolor[HTML]{000000}{msg\char95{}sender}\\
\textcolor[HTML]{000000}{\enskip{}\enskip{}\enskip{}\enskip{}\enskip{}\enskip{}\enskip{}\enskip{}}\textcolor[HTML]{008080}{type}\textcolor[HTML]{C000C0}{:}\textcolor[HTML]{000000}{\enskip{}}\textcolor[HTML]{000000}{Int}\\
\textcolor[HTML]{000000}{\enskip{}\enskip{}\enskip{}\enskip{}\enskip{}\enskip{}}\textcolor[HTML]{AF5F00}{-\enskip{}}\textcolor[HTML]{008080}{name}\textcolor[HTML]{C000C0}{:}\textcolor[HTML]{000000}{\enskip{}}\textcolor[HTML]{000000}{block\char95{}number}\\
\textcolor[HTML]{000000}{\enskip{}\enskip{}\enskip{}\enskip{}\enskip{}\enskip{}\enskip{}\enskip{}}\textcolor[HTML]{008080}{type}\textcolor[HTML]{C000C0}{:}\textcolor[HTML]{000000}{\enskip{}}\textcolor[HTML]{000000}{Int}\\
\textcolor[HTML]{000000}{\enskip{}\enskip{}\enskip{}\enskip{}\enskip{}\enskip{}}\textcolor[HTML]{AF5F00}{-\enskip{}}\textcolor[HTML]{008080}{name}\textcolor[HTML]{C000C0}{:}\textcolor[HTML]{000000}{\enskip{}}\textcolor[HTML]{000000}{random}\\
\textcolor[HTML]{000000}{\enskip{}\enskip{}\enskip{}\enskip{}\enskip{}\enskip{}\enskip{}\enskip{}}\textcolor[HTML]{008080}{type}\textcolor[HTML]{C000C0}{:}\textcolor[HTML]{000000}{\enskip{}}\textcolor[HTML]{000000}{Int}\\
\textcolor[HTML]{000000}{\enskip{}\enskip{}\enskip{}\enskip{}}\textcolor[HTML]{008080}{return}\textcolor[HTML]{C000C0}{:}\\
\textcolor[HTML]{000000}{\enskip{}\enskip{}\enskip{}\enskip{}\enskip{}\enskip{}}\textcolor[HTML]{AF5F00}{-\enskip{}}\textcolor[HTML]{008080}{name}\textcolor[HTML]{C000C0}{:}\textcolor[HTML]{000000}{\enskip{}}\textcolor[HTML]{000000}{result}\\
\textcolor[HTML]{000000}{\enskip{}\enskip{}\enskip{}\enskip{}\enskip{}\enskip{}\enskip{}\enskip{}}\textcolor[HTML]{008080}{type}\textcolor[HTML]{C000C0}{:}\textcolor[HTML]{000000}{\enskip{}}\textcolor[HTML]{000000}{Bool}\\
\textcolor[HTML]{000000}{\enskip{}\enskip{}\enskip{}\enskip{}}\textcolor[HTML]{008080}{requires}\textcolor[HTML]{C000C0}{:}\textcolor[HTML]{000000}{\enskip{}|}\\
\textcolor[HTML]{000000}{\enskip{}\enskip{}\enskip{}\enskip{}\enskip{}\enskip{}}\textcolor[HTML]{C00000}{true}\\
\textcolor[HTML]{000000}{\enskip{}\enskip{}\enskip{}\enskip{}}\textcolor[HTML]{008080}{ensures}\textcolor[HTML]{C000C0}{:}\textcolor[HTML]{000000}{\enskip{}|}\\
\textcolor[HTML]{000000}{\enskip{}\enskip{}\enskip{}\enskip{}\enskip{}\enskip{}}\textcolor[HTML]{000000}{(and}\textcolor[HTML]{000000}{\enskip{}}\textcolor[HTML]{000000}{result}\\
\textcolor[HTML]{000000}{\enskip{}\enskip{}\enskip{}\enskip{}\enskip{}\enskip{}}\textcolor[HTML]{000000}{(ite}\textcolor[HTML]{000000}{\enskip{}}\textcolor[HTML]{000000}{(<}\textcolor[HTML]{000000}{\enskip{}}\textcolor[HTML]{000000}{msg\char95{}value}\textcolor[HTML]{000000}{\enskip{}}\textcolor[HTML]{000000}{50)}\\
\textcolor[HTML]{000000}{\enskip{}\enskip{}\enskip{}\enskip{}\enskip{}\enskip{}\enskip{}\enskip{}\enskip{}\enskip{}\enskip{}}\textcolor[HTML]{000000}{(states\char95{}equal}\textcolor[HTML]{000000}{\enskip{}}\textcolor[HTML]{000000}{warrior}\textcolor[HTML]{000000}{\enskip{}}\textcolor[HTML]{000000}{warriorGold}\textcolor[HTML]{000000}{\enskip{}}\textcolor[HTML]{000000}{warriorBlock}\\
\textcolor[HTML]{000000}{\enskip{}\enskip{}\enskip{}\enskip{}\enskip{}\enskip{}\enskip{}\enskip{}\enskip{}\enskip{}\enskip{}\enskip{}\enskip{}\enskip{}\enskip{}\enskip{}\enskip{}\enskip{}\enskip{}\enskip{}\enskip{}\enskip{}\enskip{}\enskip{}\enskip{}}\textcolor[HTML]{000000}{king}\textcolor[HTML]{000000}{\enskip{}}\textcolor[HTML]{000000}{kingBlock}\textcolor[HTML]{000000}{\enskip{}}\textcolor[HTML]{000000}{err}\\
\textcolor[HTML]{000000}{\enskip{}\enskip{}\enskip{}\enskip{}\enskip{}\enskip{}\enskip{}\enskip{}\enskip{}\enskip{}\enskip{}\enskip{}\enskip{}\enskip{}\enskip{}\enskip{}\enskip{}\enskip{}\enskip{}\enskip{}\enskip{}\enskip{}\enskip{}\enskip{}\enskip{}}\textcolor[HTML]{000000}{warrior\char95{}new}\textcolor[HTML]{000000}{\enskip{}}\textcolor[HTML]{000000}{warriorGold\char95{}new}\textcolor[HTML]{000000}{\enskip{}}\textcolor[HTML]{000000}{warriorBlock\char95{}new}\\
\textcolor[HTML]{000000}{\enskip{}\enskip{}\enskip{}\enskip{}\enskip{}\enskip{}\enskip{}\enskip{}\enskip{}\enskip{}\enskip{}\enskip{}\enskip{}\enskip{}\enskip{}\enskip{}\enskip{}\enskip{}\enskip{}\enskip{}\enskip{}\enskip{}\enskip{}\enskip{}\enskip{}}\textcolor[HTML]{000000}{king\char95{}new}\textcolor[HTML]{000000}{\enskip{}}\textcolor[HTML]{000000}{kingBlock\char95{}new}\textcolor[HTML]{000000}{\enskip{}}\textcolor[HTML]{000000}{err\char95{}new)}\\
\textcolor[HTML]{000000}{\enskip{}\enskip{}\enskip{}\enskip{}\enskip{}\enskip{}\enskip{}\enskip{}\enskip{}\enskip{}\enskip{}}\textcolor[HTML]{000000}{(and}\textcolor[HTML]{000000}{\enskip{}}\textcolor[HTML]{000000}{(=}\textcolor[HTML]{000000}{\enskip{}}\textcolor[HTML]{000000}{warrior\char95{}new}\textcolor[HTML]{000000}{\enskip{}}\textcolor[HTML]{000000}{(store}\textcolor[HTML]{000000}{\enskip{}}\textcolor[HTML]{000000}{warrior}\textcolor[HTML]{000000}{\enskip{}}\textcolor[HTML]{000000}{msg\char95{}sender}\textcolor[HTML]{000000}{\enskip{}}\textcolor[HTML]{000000}{msg\char95{}sender))}\\
\textcolor[HTML]{000000}{\enskip{}\enskip{}\enskip{}\enskip{}\enskip{}\enskip{}\enskip{}\enskip{}\enskip{}\enskip{}\enskip{}\enskip{}\enskip{}\enskip{}\enskip{}\enskip{}}\textcolor[HTML]{000000}{(=}\textcolor[HTML]{000000}{\enskip{}}\textcolor[HTML]{000000}{warriorGold\char95{}new}\textcolor[HTML]{000000}{\enskip{}}\textcolor[HTML]{000000}{(store}\textcolor[HTML]{000000}{\enskip{}}\textcolor[HTML]{000000}{warriorGold}\textcolor[HTML]{000000}{\enskip{}}\textcolor[HTML]{000000}{msg\char95{}sender}\textcolor[HTML]{000000}{\enskip{}}\textcolor[HTML]{000000}{msg\char95{}value))}\\
\textcolor[HTML]{000000}{\enskip{}\enskip{}\enskip{}\enskip{}\enskip{}\enskip{}\enskip{}\enskip{}\enskip{}\enskip{}\enskip{}\enskip{}\enskip{}\enskip{}\enskip{}\enskip{}}\textcolor[HTML]{000000}{(=}\textcolor[HTML]{000000}{\enskip{}}\textcolor[HTML]{000000}{warriorBlock\char95{}new}\textcolor[HTML]{000000}{\enskip{}}\textcolor[HTML]{000000}{(store}\textcolor[HTML]{000000}{\enskip{}}\textcolor[HTML]{000000}{warriorBlock}\textcolor[HTML]{000000}{\enskip{}}\textcolor[HTML]{000000}{msg\char95{}sender}\textcolor[HTML]{000000}{\enskip{}}\textcolor[HTML]{000000}{block\char95{}number))}\\
\textcolor[HTML]{000000}{\enskip{}\enskip{}\enskip{}\enskip{}\enskip{}\enskip{}\enskip{}\enskip{}\enskip{}\enskip{}\enskip{}\enskip{}\enskip{}\enskip{}\enskip{}\enskip{}}\textcolor[HTML]{000000}{(ite}\textcolor[HTML]{000000}{\enskip{}}\textcolor[HTML]{000000}{(=}\textcolor[HTML]{000000}{\enskip{}}\textcolor[HTML]{000000}{(modFn}\textcolor[HTML]{000000}{\enskip{}}\textcolor[HTML]{000000}{(select}\textcolor[HTML]{000000}{\enskip{}}\textcolor[HTML]{000000}{warriorBlock\char95{}new}\textcolor[HTML]{000000}{\enskip{}}\textcolor[HTML]{000000}{msg\char95{}sender))}\textcolor[HTML]{000000}{\enskip{}}\textcolor[HTML]{000000}{random)}\\
\textcolor[HTML]{000000}{\enskip{}\enskip{}\enskip{}\enskip{}\enskip{}\enskip{}\enskip{}\enskip{}\enskip{}\enskip{}\enskip{}\enskip{}\enskip{}\enskip{}\enskip{}\enskip{}\enskip{}\enskip{}\enskip{}\enskip{}\enskip{}}\textcolor[HTML]{000000}{(and}\textcolor[HTML]{000000}{\enskip{}}\textcolor[HTML]{000000}{(=}\textcolor[HTML]{000000}{\enskip{}}\textcolor[HTML]{000000}{king\char95{}new}\textcolor[HTML]{000000}{\enskip{}}\textcolor[HTML]{000000}{(select}\textcolor[HTML]{000000}{\enskip{}}\textcolor[HTML]{000000}{warrior\char95{}new}\textcolor[HTML]{000000}{\enskip{}}\textcolor[HTML]{000000}{msg\char95{}sender))}\\
\textcolor[HTML]{000000}{\enskip{}\enskip{}\enskip{}\enskip{}\enskip{}\enskip{}\enskip{}\enskip{}\enskip{}\enskip{}\enskip{}\enskip{}\enskip{}\enskip{}\enskip{}\enskip{}\enskip{}\enskip{}\enskip{}\enskip{}\enskip{}\enskip{}\enskip{}\enskip{}\enskip{}\enskip{}}\textcolor[HTML]{000000}{(=}\textcolor[HTML]{000000}{\enskip{}}\textcolor[HTML]{000000}{kingBlock\char95{}new}\textcolor[HTML]{000000}{\enskip{}}\textcolor[HTML]{000000}{(select}\textcolor[HTML]{000000}{\enskip{}}\textcolor[HTML]{000000}{warriorBlock\char95{}new}\textcolor[HTML]{000000}{\enskip{}}\textcolor[HTML]{000000}{msg\char95{}sender)))}\\
\textcolor[HTML]{000000}{\enskip{}\enskip{}\enskip{}\enskip{}\enskip{}\enskip{}\enskip{}\enskip{}\enskip{}\enskip{}\enskip{}\enskip{}\enskip{}\enskip{}\enskip{}\enskip{}\enskip{}\enskip{}\enskip{}\enskip{}\enskip{}}\textcolor[HTML]{000000}{(and}\textcolor[HTML]{000000}{\enskip{}}\textcolor[HTML]{000000}{(=}\textcolor[HTML]{000000}{\enskip{}}\textcolor[HTML]{000000}{king\char95{}new}\textcolor[HTML]{000000}{\enskip{}}\textcolor[HTML]{000000}{king)}\\
\textcolor[HTML]{000000}{\enskip{}\enskip{}\enskip{}\enskip{}\enskip{}\enskip{}\enskip{}\enskip{}\enskip{}\enskip{}\enskip{}\enskip{}\enskip{}\enskip{}\enskip{}\enskip{}\enskip{}\enskip{}\enskip{}\enskip{}\enskip{}\enskip{}\enskip{}\enskip{}\enskip{}\enskip{}}\textcolor[HTML]{000000}{(=}\textcolor[HTML]{000000}{\enskip{}}\textcolor[HTML]{000000}{kingBlock\char95{}new}\textcolor[HTML]{000000}{\enskip{}}\textcolor[HTML]{000000}{kingBlock))}\\
\textcolor[HTML]{000000}{\enskip{}\enskip{}\enskip{}\enskip{}\enskip{}\enskip{}\enskip{}\enskip{}\enskip{}\enskip{}\enskip{}\enskip{}\enskip{}\enskip{}\enskip{}\enskip{}}\textcolor[HTML]{000000}{)}\\
\textcolor[HTML]{000000}{\enskip{}\enskip{}\enskip{}\enskip{}\enskip{}\enskip{}\enskip{}\enskip{}\enskip{}\enskip{}\enskip{}}\textcolor[HTML]{000000}{)}\\
\textcolor[HTML]{000000}{\enskip{}\enskip{}\enskip{}\enskip{}\enskip{}\enskip{}}\textcolor[HTML]{000000}{)}\\
\textcolor[HTML]{000000}{\enskip{}\enskip{}\enskip{}\enskip{}\enskip{}\enskip{}}\textcolor[HTML]{000000}{)}\\
\\

}
Predicates:
\begin{verbatim}
(= x1 y1)
(= x2 y2)
(= x3 y3)
(= (modFn x3) x4)
(= (modFn y3) y4)
(< x1 50)
(< y1 50)
\end{verbatim}

\end{document}